\documentclass[11pt]{article}
\usepackage{graphicx}
\usepackage{amsmath}
\usepackage{amssymb}
\usepackage{natbib}

\title{Wavelet-Based Cascade Model for Intermittent Structure in Terrestrial Environments}
\author{D. Keith Wilson \\
U.S. Army Engineer Research and Development Center\\
Hanover, NH 03755
\and Chris L. Pettit \\
Aerospace Engineering Department\\
U.S. Naval Academy\\
Annapolis, MD, 21402
\and Sergey N. Vecherin \\
U.S. Army Engineer Research and Development Center\\
Hanover, NH 03755}

\date{7 Nov.\ 2013}

\begin{document}

\maketitle

\begin{abstract}

A wavelet-like model for distributions of objects in natural and man-made terrestrial environments is developed. The model is constructed in a self-similar fashion, with the sizes, amplitudes, and numbers of objects occurring at a constant ratios between parent and offspring objects. The objects are randomly distributed in space according to a Poisson process. Fractal supports and a cascade model are used to organize objects intermittently in space. In its basic form, the model is for continuously varying random fields, although a level-cut is introduced to model two-phase random media. The report begins with a description of relevant concepts from fractal theory, and then progresses through static (time-invariant), steady-state, and non-steady models. The results can be applied to such diverse phenomena as turbulence, geologic distributions, urban buildings, vegetation, and arctic ice floes. The model can be used as a basis for synthesizing realistic terrestrial scenes, and for predicting the performance of sensing and communication systems in operating environments with complex, intermittent distributions of scattering objects.

\end{abstract}

\pagebreak

{\small \tableofcontents}

\pagebreak

\section{Introduction}

\subsection{Characteristics of Random Terrestrial Environments}
\label{sec:RHM}

\begin{figure}[ptb]
\begin{center}
\includegraphics[trim=150 150 150 150,clip,width=0.9\linewidth,natwidth=960,natheight=720]{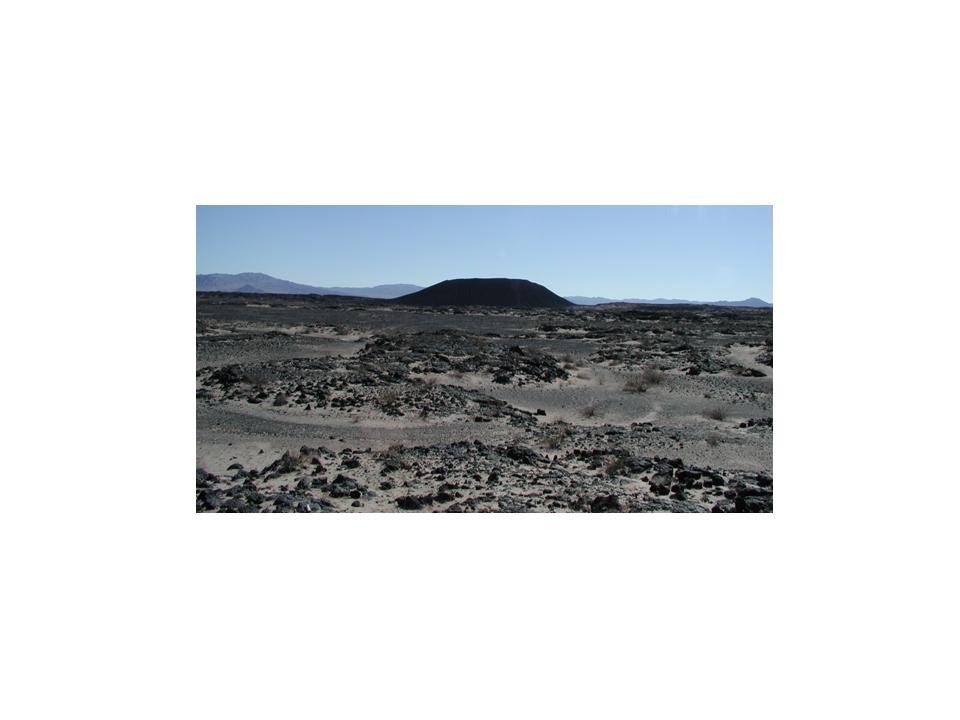}%
\caption{Volcanic crater and basaltic rock formations at the Amboy Crater, Mojave Desert, CA. The crater itself is visible on the horizon, in the middle of the picture. (Courtesy of David Finnegan, ERDC-CRREL.)}%
\label{fig:amboypic}%
\end{center}
\end{figure}

\begin{figure}[tbph]
\centering
\includegraphics[width=0.6\linewidth,natwidth=318,natheight=302]{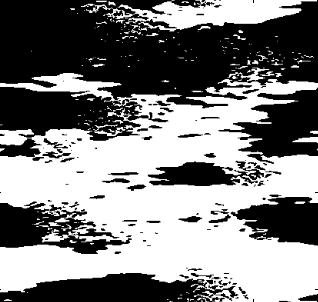}
\caption{Direct numerical simulation of turbulence in the stable (nocturnal) boundary layer. White areas represent relatively lower fluid density. (Courtesy of James Riley, University of Washington.)}
\label{fig:dns}
\end{figure}

\begin{figure}[tbph]
\centering
\includegraphics[width=0.8\linewidth,natwidth=547,natheight=345]{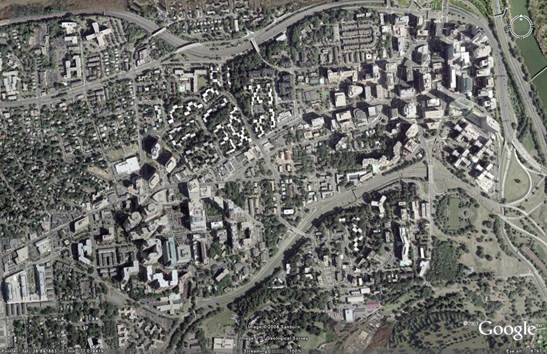}
\caption{Rosslyn-Arlington, Virginia, vicinity. (Imagery from Google Earth.)}
\label{fig:rosslyn}
\end{figure}

Terrestrial environments, both natural and man-made, often possess complex, random spatial distributions of objects. Figures~\ref{fig:amboypic}--\ref{fig:rosslyn} show some illustrative examples: a volcanic crater and rock formations, an up-close view of volcanic rock, arctic melt ponds, an arctic ice floe, a turbulence simulation, and urban/suburban topography, respectively. The visual similarities of these various \emph{random heterogeneous media} (RHM) are striking in several aspects, despite the very different phenomena underlying their creation. In particular, these RHM often share two traits which are focal points for this report: \emph{self-similarity} and \emph{intermittency}.

\subsubsection{Self-Similarity}

A \emph{self-similar} (or \emph{scale-invariant}) process appears the same regardless of the magnification. Given an image of such a process, we generally cannot discern the actual size of objects in the picture. In contrast, a scale-dependent process has one or more recognizable spatial (or time) scales associated with it. Many phenomena are self similar over some \emph{range} of scales. The largest scale present is called here the \emph{outer scale}, whereas the smallest scale is the \emph{inner scale}. Between these two scales lies the self-similarity range.

\begin{figure}[ptb]
\begin{center}
\includegraphics[width=0.7\linewidth,natwidth=543,natheight=408]{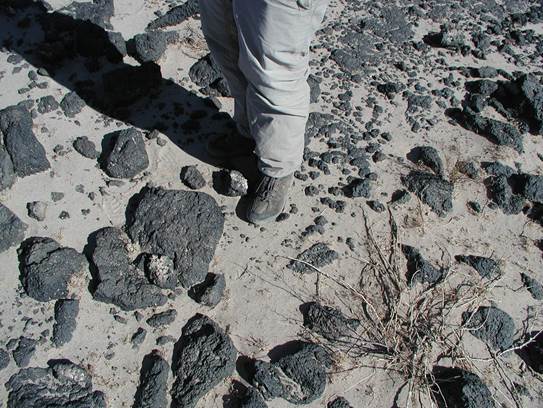}%
\caption{Close-up of volcanic (basaltic) rocks from the Amboy Crater vicinity. (Courtesy of David Finnegan, ERDC-CRREL.)}%
\label{fig:amboypic2}%
\end{center}
\end{figure}

\begin{figure}[tbph]
\centering
\includegraphics[width=0.8\linewidth,natwidth=450,natheight=300]{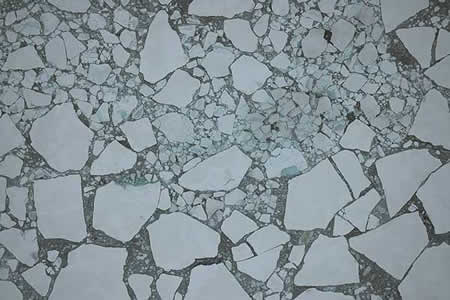}
\caption{Aerial image of an arctic ice floe. (Courtesy of Don Perovich, ERDC-CRREL.)}
\label{fig:icefloe}
\end{figure}

Figure~\ref{fig:icefloe}, which shows an ice floe, illustrates self similarity. From the image alone (without understanding the context in which it was produced), it is very difficult to determine the sizes of the floating ice blocks. If certain sections of the image are enlarged, they would appear very much like other sections. An outer scale is apparent, which corresponds to the largest dozen or so ice blocks. However, it is difficult to establish the presence of an inner scale without examination of higher-resolution imagery. Figure~\ref{fig:amboypic2}, showing volcanic rock in a desert, exhibits some of the same characteristics as Fig.~\ref{fig:icefloe}, although the spatial scale of the image is readily recognizable by the presence of a human foot. 

\begin{figure}[tbph]
\centering
\includegraphics[width=0.7\linewidth,natwidth=358,natheight=336]{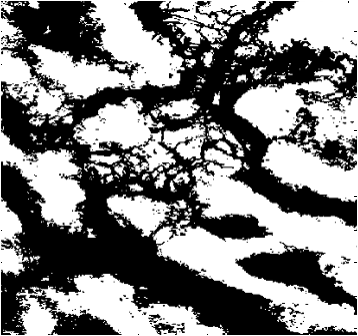}
\caption{Aerial image of arctic melt ponds. (Courtesy of Chris Polashenski, ERDC-CRREL.)}
\label{fig:melt_ponds}
\end{figure}

Self similarity can sometimes be associated with a \emph{cascade}, or iterative, process underlying the formation of the medium. The cascade consists of a sequence of reactions, by which larger objects, or \emph{parents}, break down into smaller objects, or \emph{offspring}. The cascade could also conceivably operate in the opposite direction, with smaller objects coalescing into larger ones. If the rules for the reaction are independent of scale, i.e., they are the same for all generations of the process, a self-similar structure results. For some real-world media, the cascade model may be closely connected to an identifiable physical processes, e.g., decay of turbulent eddies, weathering of rocks, or destruction of smaller buildings and replacement by larger ones. In other situations, the cascade may simply be a useful abstraction that leads to geometric characteristics similar to observations.

Self-similarity and iteration are also associated with \emph{fractals}, as conceptualized and popularized by Mandelbrot \citeyearpar{mandelbrot1977fractals}. A fractal is a set with a non-integer dimensional measure, which corresponds to a power-law dependence on wavenumber in an analysis of the spatial spectrum \citep{schroeder1991fractals}. Consider the arctic melt pond image, Fig.~\ref{fig:melt_ponds}, as an example. Progressively more fine structure would be revealed as the edges of the pond are magnified. Conceivably (although not obviously), the measured perimeter of the pond would increase as one tried to measure it with a smaller and smaller ruler, which responds to finer details in the edge of the ponds. Hence the perimeter of a melt pond apparently does not converge to a fixed value. In a sense, the perimeter exceeds the topological dimension of an ordinary line (i.e., one dimension).

\subsubsection{Intermittency}

\emph{Intermittency} means that the objects are not smoothly distributed in space and/or time; some synonyms are \emph{uneven}, \emph{irregular}, or \emph{sporadic}.\footnote{Most dictionaries list \emph{intermittency} and \emph{intermittence} as alternative spellings. The former is the more common spelling in the turbulence literature, and hence is adopted here.} The spatial distribution of volcanic rock outcrops in Fig.~\ref{fig:amboypic}, in the vicinity of the Amboy Crater, is an example. Some areas have abundant, coarse volcanic rock, whereas others are mostly sand or fine, powdery rock. Intermittency actually corresponds to a more ordered (organized) state of a random medium. When the positions of the objects are fully randomized, disorder and entropy are maximized. 

Mahrt \citeyearpar{mahrt1989intermittency} has proposed classifying intermittency into \emph{global} and \emph{intrinsic} types. Global intermittency is associated with the production of objects at the outer scale (the initial generation of the cascade process). In regions where such production is absent, activity at all scales is nearly totally lacking. This is clearly manifested by the turbulence simulation in Fig.~\ref{fig:dns} and by the individual rocky outcrops in Fig.~\ref{fig:amboypic}. Intrinsic intermittency, on the other hand, is associated with the cascade process. When the cascade process concentrates offspring into smaller regions of space relative to their parents, small objects tend to occur in pronounced bursts of activity. To some extent, this is visually evident in the turbulence simulation, Fig.~\ref{fig:dns}, in which fine structure occurs predominantly around the edges of certain large structures, and in the foreground of Fig.~\ref{fig:amboypic}, in which the small rocks appear to be relatively more isolated. 

When most authors use the term \emph{intermittency}, they are implicitly referring to the intrinsic type. For example, a landmark paper by Mandelbrot \citeyearpar{mandelbrot1974intermittent} describes intermittency in the dissipation rate of turbulence (the rate at which turbulent energy is converted to heat), as resulting from a turbulence cascade process. As the turbulence is examined at a finer scale, more regions become apparent that are not actively involved in the dissipation, hence resulting in a fractal dimension for the dissipation regions. This behavior is reflected by Fig.~\ref{fig:dns}. In analogy to cheesemaking, Mandelbrot termed the actively dissipating regions in turbulence the ``curds'' and the inactive regions the ``whey.'' He applied the same terminology to the clustering of matter in the universe and other phenomena. In the study of turbulence, intermittency of the dissipation rate has been an important and active topic of research for several decades \citep{kolmogorov1962refinement,obukhov1962some,sreenivasan1991fractals}.

Clustering of objects of a similar size might also be regarded as a manifestation of intermittency. We refer to this property as \emph{sorting by size}. Such sorting is evident in the construction pattern in Fig.~\ref{fig:rosslyn}, where large office buildings occur in certain regions, and smaller residential buildings in others. Sorting by size is also evident in each of the natural environments shown in Figs.~\ref{fig:amboypic}--\ref{fig:dns}. Sorting plausibly results from global intermittency and irregularities in the spatial and temporal production of objects.

To quantify the concept of intermittency, one can invoke a comparison to a normal, or Gaussian, probability distribution for the random fluctuations in a field. The normal distribution implies a certain frequency for extreme events; for example, there is approximately a $4.6\%$ chance of a fluctuation with an absolute value exceeding twice the standard deviation. Departure from the normal distribution is often quantified through the fourth moment, or \emph{kurtosis}. For a normal distribution, the kurtosis is exactly 3. A value larger than 3 indicates a higher probability of large fluctuations, which can be considered to be an indicator of an intermittent process. Further classification of intermittency into global, intrinsic, and size-sorting types would require more sophisticated analysis tools that account for the spatial scale of the fluctuations.

Although images were not specifically shown here, many other types of terrestrial environments, such as weather events, vegetation, populations, landscapes, surface elevations, can likewise be regarded as possessing self-similar and intermittent properties. Even extraterrestrial phenomena, such as asteroid sizes and the general distribution of matter in the universe, can be usefully modeled by such approaches.

\subsection{Motivation}

Realistic models of terrestrial environments have applications to many practical Army problems. In particular, this report was motivated by the desire to improve models for random atmospheric and terrain features of acoustic, seismic, and electromagnetic wave propagation calculations. The propagating waves are scattered by random inhomogeneities in the environment such as turbulence, dust, and rocks. Random fading and coherence loss in signals results, which degrades the performance of sensor systems used for detecting and localizing emitters, and for communications. The environmental modeling and wave propagation calculations may be done by theoretical \citep[e.g.,][]{TatarskiiVI:1971,ostashev1997acoustics} or numerical \citep[e.g.,][]{anderson2004tracked,wilson2005high} methods. In particular, the latter two studies demonstrated how the coupling of synthesized environmental scenes and high-fidelity wave propagation models running on parallel-processing supercomputers enables cost-effective, rapid virtual testing of new sensor systems and employment concepts.

\begin{figure}[tbph]
\centering
\includegraphics[width=0.8\linewidth,natwidth=966,natheight=449]{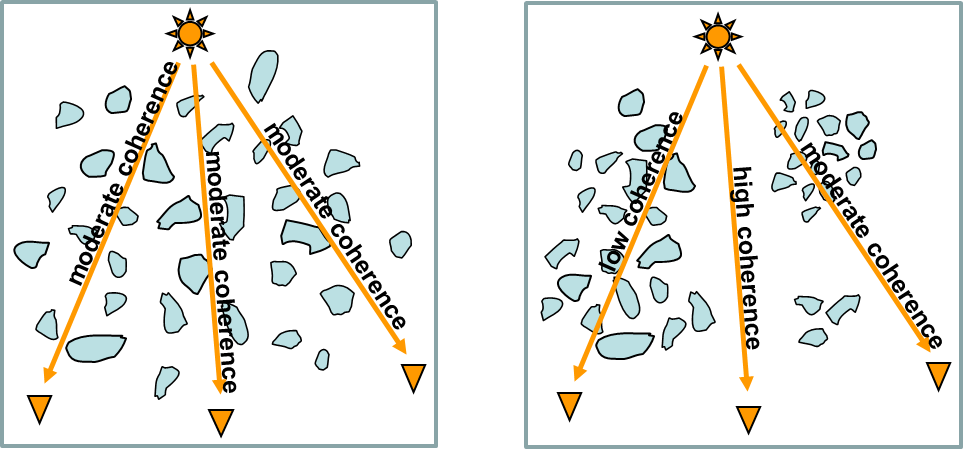}
\caption{Illustration of wave propagation without (left) and with (right) intermittency in the terrestrial environment. When intermittency is present, different propagation paths can experience drastically differing degrees of scattering and coherence loss.}
\label{fig:coherence}
\end{figure}

Conventional theoretical treatments of signal propagation \citep[e.g.,][]{TatarskiiVI:1971,rytov1989wave,ostashev1997acoustics} assume that heterogeneities (terrestrial objects) responsible for random scattering of the wavefield are evenly distributed in space. This is a significant shortcoming in existing capabilities for characterizing and predicting environmental effects on system performance. Figure~\ref{fig:coherence} depicts how an intermittent medium affects signal propagation. Along some paths, the waves encounter many scattering objects and signal coherence (a measure of signal randomness, which usually determines the sensing system performance) is strongly degraded. Other paths encounter relatively few scattering objects and thus have a high signal coherence. Conventional wave scattering theory assumes, in effect, that all paths experience similar scattering. It thus neglects the \emph{range} of possible outcomes and related uncertainty in system performance, which, depending on the environment, can be very significant. Urban areas provide a particularly important example: optical or RF communication systems may be rendered useless by obstructions in built-up sections, whereas signals may propagate very cleanly through open areas such as large lawns and parking lots.

To address such issues, some more recent papers on acoustic and electromagnetic propagation through turbulence have considered intermittency \citep[e.g.,][]{HentschelHGE:1983,TatarskiiVI:1985,GurvichAS:1986,WilsonDK:1996e}. Most of these studies were based on the log-normal model for turbulent intermittency, which is known to have substantial shortcomings \citep{mandelbrot1974intermittent}. Previous theoretical treatments are also usually restricted with regard to the length of the propagation path, size of the largest inhomogeneities, wavelength, and strength of the scattering. This report, while it does not address wave propagation directly, does provide new modeling approaches for self-similar, intermittent RHM that can subsequently provide a basis for improving theoretical treatments and numerical modeling, and thus lead to more realistic characterization of the performance of Army sensing and communication systems operating in complex, real-world environments.

\subsection{Structure of this Report}

This report explores how concepts for self-similar cascade processes can be incorporated into the previously developed \emph{quasi-wavelet} (QW) model for RHM, thus introducing realistic intermittency features into RHM modeling. QW models have been shown capable of synthesizing very realistic turbulence fields \citep[e.g.,][]{goedecke2006quasi,wilson2009quasi}, and were subsequently applied to surface geology \citep{wilson2008asc}. The approach in this report evolved out of an earlier paper presented at the Army Science Conference \citep{wilson2008asc}. One of the main improvements is that the three cascade reactions in the ASC paper have been replaced by a single, more general, reaction, which improves the analytical and stability properties of the model. The report also provides much additional background and a more systematic development of the model.

Section~\ref{sec:fractal} provides a conceptual introduction to fractals, which serves as a foundation for the subsequent treatment of cascade processes. Next, Sec.~\ref{sec:QW} introduces self-similar QW models. A systematic and simple approach is provided for describing the scale ratios, generations, and densification of scales. 

The following four sections describe the three basic model constructions: \emph{static}, \emph{steady}, and \emph{non-steady}. The static approach, described in Sec.~\ref{sec:staticmodel}, does not explicitly incorporate evolution in time; it describes a single ``snapshot'' of the random medium. Next, in Sec.~\ref{sec:steadycascade}, the steady-state cascade process model is introduced, in which creation and destruction of the QWs at each scale are in equilibrium. Sec.~\ref{sec:stat} derives formulas for determining model parameters from observed statistics of an RHM. It is shown that the static and steady-state models are indistinguishable, on the basis of second-order statistics, when only a single snapshot of the process is analyzed. Lastly, a modeling approach involving explicit simulation of the cascade process, which may be used for either a steady-state or non-steady-state model, is discussed in Sec.~\ref{sec:nonsteady}.

\section{Some Background on Fractals}
\label{sec:fractal}

The term \emph{fractal} was coined by \cite{mandelbrot1977fractals} to describe sets with non-integer dimensionality. A more precise, mathematical definition of a fractal is a set whose Hausdorff dimension exceeds its topological dimension. (The Hausdorff dimension will be defined and discussed shortly.) Such sets are generally constructed through recursive application of a self-similar construction rule, which makes them appear the same at all magnifications. Mandelbrot recognized that the geometry of many diverse phenomena, including coastlines, mountains, snowflakes, craters on the moon, turbulence, and clustering of matter in the universe, have fractal and self-similar properties.

As mentioned in Sec.~\ref{sec:RHM}, fractals may result from cascade processes, in which larger objects decompose into progressively smaller ones. The turbulence cascade, which involves stretching and shearing of larger eddies into smaller ones, and eventual dissipation of turbulent kinetic energy through viscous forces, is an example. Fracturing and weathering of rocks is another. Some cascade processes may actually operate in the reverse of this pattern, from smaller objects to larger ones. A cascade in this direction might be associated with storage, rather than dissipation, of energy.\footnote{This observation was made by Daniel Breton (CRREL) in a review of this report.} As a forest matures, it tends to support fewer but larger trees. Increasingly larger structures are typically built in a growing city, while smaller ones are torn down.

\subsection{Fractal Carpets}
\label{sec:carpet}

\begin{figure}[tbph]
\centering
\includegraphics[width=0.8\linewidth,natwidth=1200,natheight=900]{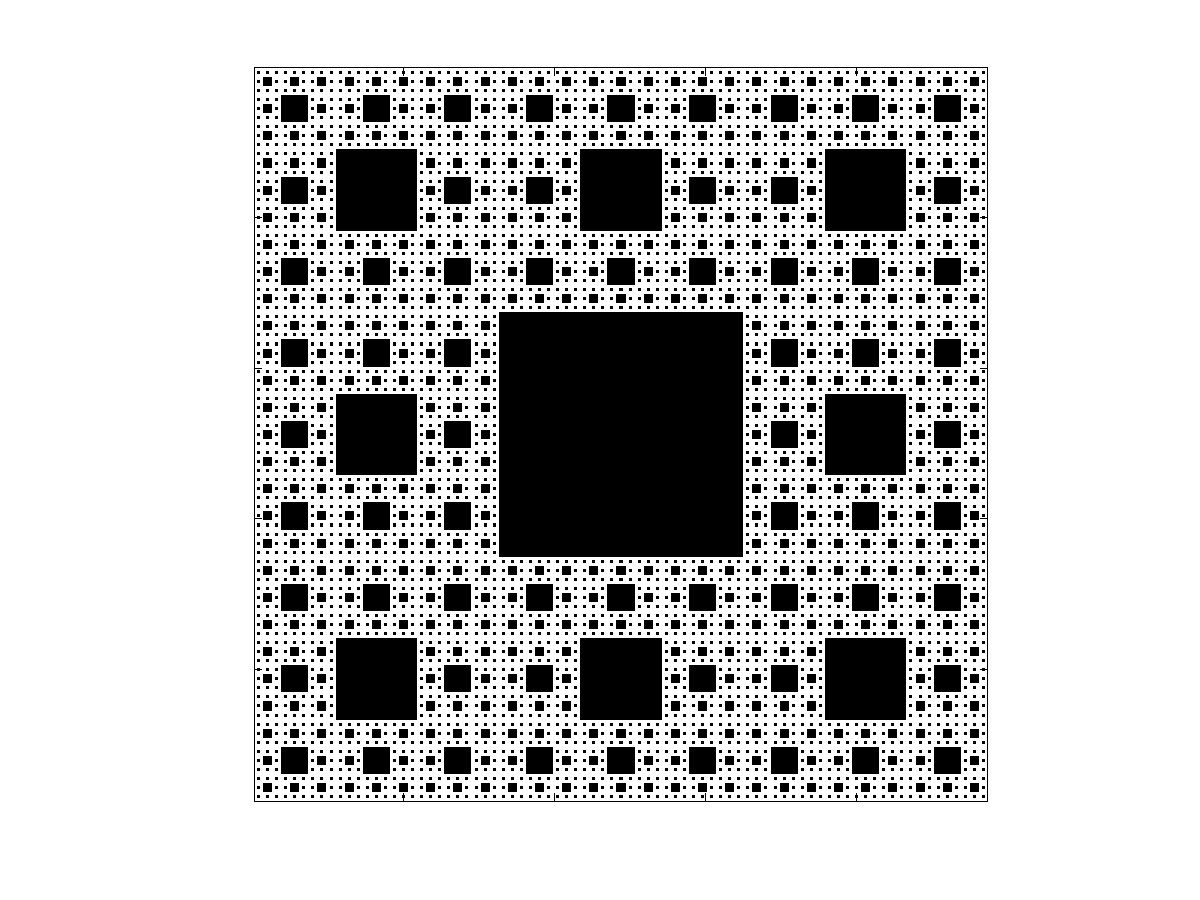}
\caption{The Sierpinski carpet, as constructed by removing the center square at each iteration.}
\label{fig:sierpinski}
\end{figure}

A famous example of a fractal is the Sierpinski carpet, shown in Fig.~\ref{fig:sierpinski}.\footnote{For reasons unknown to the present author, Mandelbrot~(\citeyear[p.\ 166]{mandelbrot1977fractals}) refers to the object in Fig.~\ref{fig:sierpinski} as a Sierpinski carpet, whereas Schroeder (\citeyear[p.\ 179]{schroeder1991fractals}) calls it a Cantor gasket. Mandelbrot's terminology is used here.}  The carpet is constructed from a square. A smaller square, with side length 1/3 of the original square, is then removed from the center. This leaves 8 squares of ``fabric,'' each with side length 1/3 of the original, around the edges. The process is then repeated by removing squares of fabric, with side length 1/9 of the original, from the centers of each of the remaining eight squares. After many iterations of this process, one has the carpet shown in Fig.~\ref{fig:sierpinski}.

The Hausdorff (fractal) dimension can be calculated from the formula $H=\log N/\log(1/\ell)$, where $N$ is the number of new objects created from the old one each time the ``measuring stick'' is decreased by a factor $\ell$. In the case of the Sierpinski carpet, $N=8$ and $\ell=1/3$. Hence $H=1.893\ldots$, which is less than the topographical dimension $D=2$. If no fabric were removed at each iteration ($N=9$ and $\ell=1/3$), we would have an integer Hausdorff dimension, $H=2$, as expected for a two-dimensional object such as a square.

\begin{figure}[tbph]
\centering
\includegraphics[width=0.8\linewidth,natwidth=1200,natheight=900]{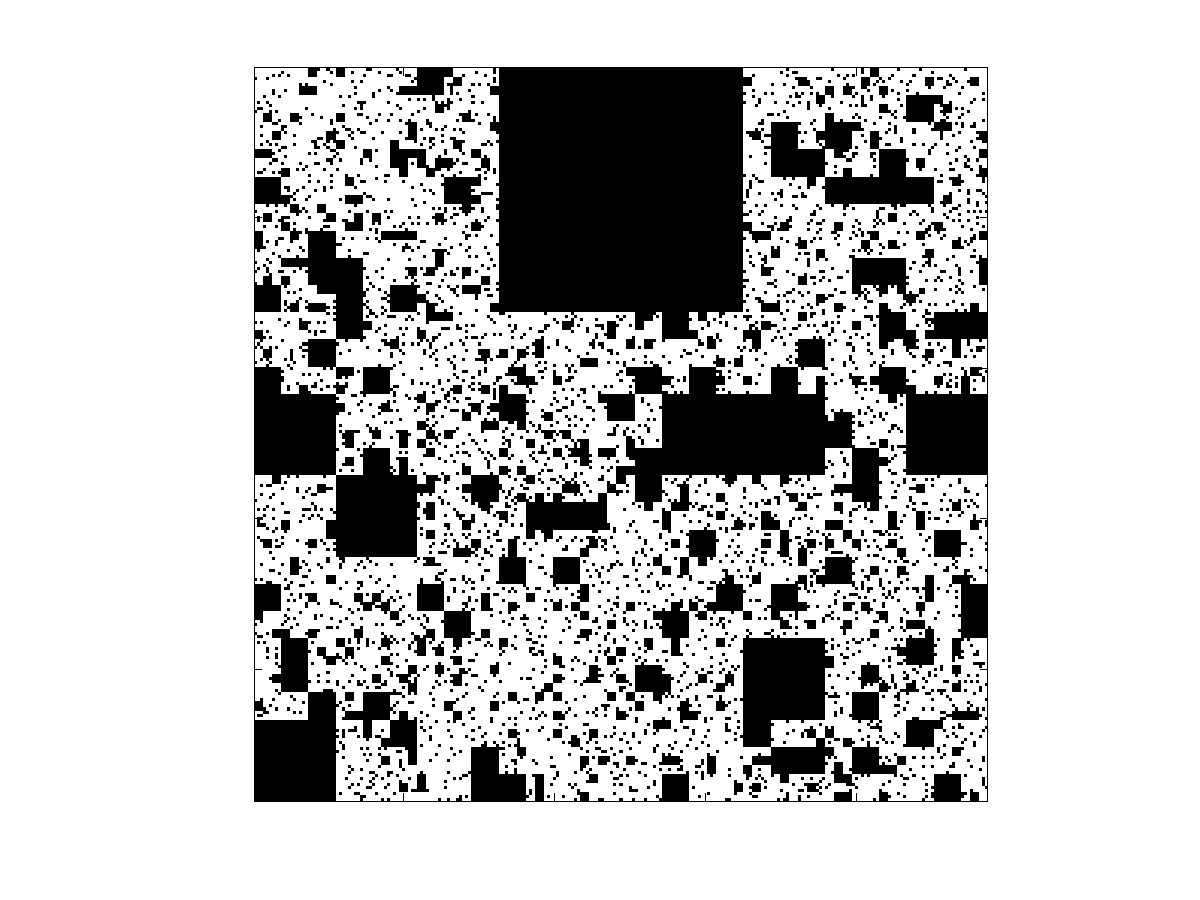}
\caption{Random fractal with the same Hausdorff dimension as the Sierpinski carpet.}
\label{fig:sierpinskiRand}
\end{figure}

The Sierpinski carpet is a deterministic fractal, since it looks identical each time it is generated. It is also possible to construct random fractals. For example, we might randomize the location from which the piece of fabric is cut at each iteration. Similarly, we might specify a probability that a given piece of fabric is removed at each iteration. The result of applying this second method is shown in Fig.~\ref{fig:sierpinskiRand}. Just like the deterministic fractal in the first example, there is an $8/9$ probability that a given sub-cell will be covered with fabric. The expected number of covered cells in the first iteration is thus $9(8/9)=8$. In the second iteration, a cell will be covered if it was covered at both the first iteration \emph{and} the next. Hence the expected number of cells covered after the second iteration is $81*(8/9)*(8/9)=8^2$. Continuing in this fashion, it is clear that $N=8$, and hence the Hausdorff dimension is unchanged from the deterministic Sierpinski carpet.

\begin{figure}[tbph]
\centering
\includegraphics[width=0.8\linewidth,natwidth=1200,natheight=900]{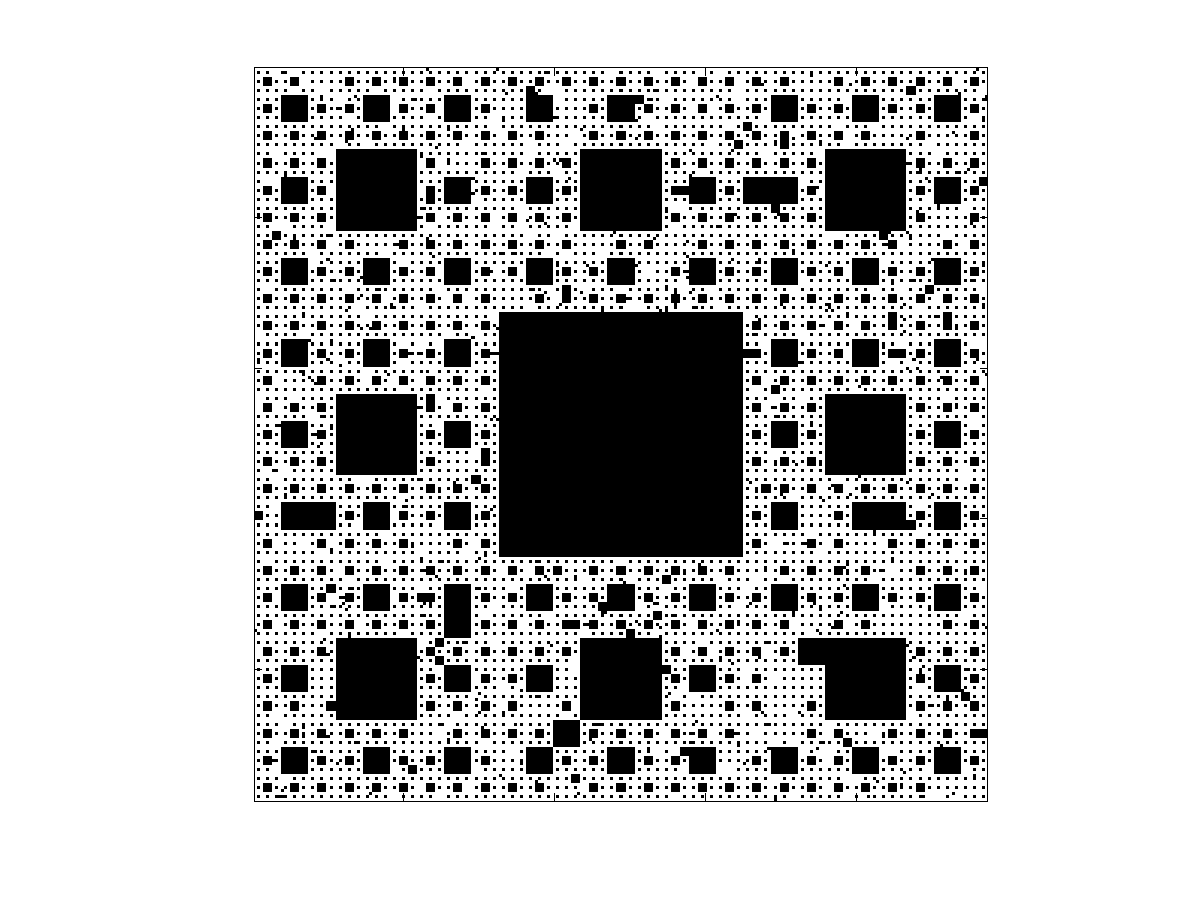}
\caption{“Dirty” Sierpinski carpet, for which there is a 92\% probability of removing the fabric from the center square, and 1\% probability of removing the fabric from each of the other squares.}
\label{fig:sierpinskiDirty}
\end{figure}

Figures~\ref{fig:sierpinski} and \ref{fig:sierpinskiRand} demonstrate that objects with very dissimilar appearances can have the same Hausdorff dimension. Another interesting example, which bridges the gap between the purely deterministic and purely random carpets, is shown in Fig.~\ref{fig:sierpinskiDirty}. This could be called a ``dirty'' Sierpinski carpet. It is like the normal, deterministic carpet, except that the chance of removing any one of the eight pieces of fabric surrounding the center is $0.01$, rather than 0. To maintain an overall probability keeping 8 pieces at each iteration, we set the probability of removing the center piece to $0.92$, instead of 1. The expected value of the number of pieces \emph{retained} in the first generation is thus $N_1=(8\cdot 0.99)+(1\cdot 0.08)=8$. At the next generation, we have $N_2=8\cdot 0.99[(8\cdot 0.99)+(1\cdot 0.08)]+0.08[(8\cdot 0.99)+(1\cdot 0.08)]=8^2$. Continuing in this fashion, we find $N=8$. Thus, the Hausdorff dimension of Fig.~\ref{fig:sierpinskiDirty} remains the same as Fig.~\ref{fig:sierpinski}.

\subsection{Cratering Processes}
\label{sec:crater}

\begin{figure}[tbh]
\centering
\includegraphics[width=0.8\linewidth,natwidth=1200,natheight=900]{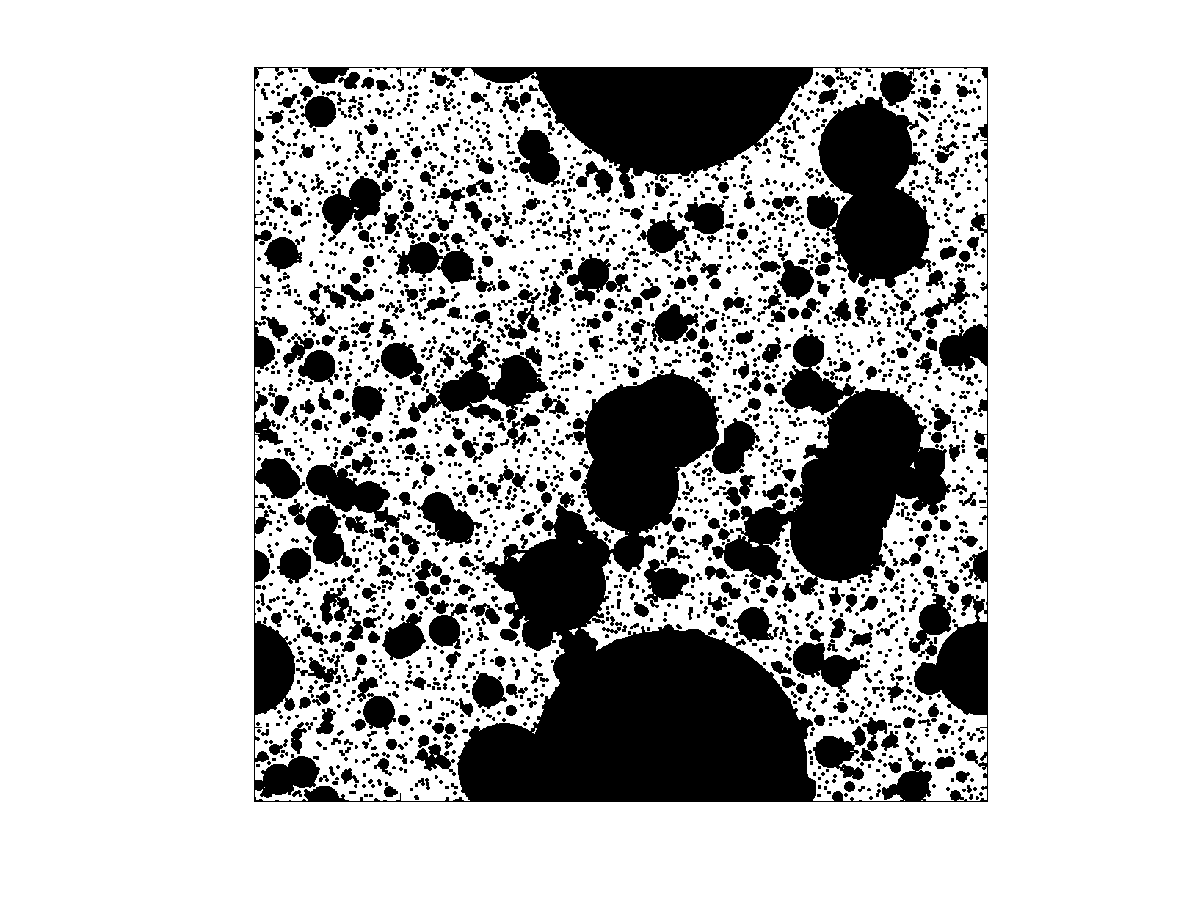}
\caption{Random fractal created by cratering (circular objects). Each size class (generation) has a radius of $1/3$ the previous class. The fractional area occupied by the craters of each size class, $\phi$, is $1/9$.}
\label{fig:craterFractal}
\end{figure}

The examples in the previous section were all based on removing and adding square shapes. Fractals can be constructed from other shapes also. An example with randomly placed, circular cut-outs is shown in Fig.~\ref{fig:craterFractal}. Mandelbrot (1977) compared the appearance of these circular cut-outs to craters on the moon and Swiss cheese. We thus refer to the process underlying the construction of Fig.~\ref{fig:craterFractal} as a \emph{cratering process}.

The randomly placed craters (circles) in Fig.~\ref{fig:craterFractal} have radii equal to $a_i=(1/3)^i/\sqrt{\pi}$ (where $i=1,2,\ldots$ indicates the \emph{generation} of the cratering process) times the side length of the square area in which they are placed. Hence their areas are $(1/3)^{2i}$, just like the sequence of square cut-outs in Fig.~\ref{fig:sierpinski}. At each generation, $9^{i-1}$ craters are added.\footnote{The reader may recall from the previous section that construction of the Sierpinski carpet involved $8^{i-1}$ new cut-outs at each iteration. Hence, for consistency, it might seem we should introduce $8^{i-1}$ new craters. The reason for this apparent discrepancy is that some of the craters overlap with the previous ones, and thus play essentially no role in the construction. (The construction of the Sierpinski carpet does not allow such overlap.) For example, since the first iteration produces a single crater covering $1/9$ of the total area, of the next $9$ randomly placed craters, one will on average overlap with the previous crater. This notion will be formulated more rigorously later in this section.} Let us define the \emph{packing fraction} as the total area of the craters for a given size class (that is, the number of craters, $N_i$, times the area of each, $A_i=\pi a_i^2$), divided by the area of the square ($A$):
\begin{equation}
\phi_i=\frac{N_i A_i}{A}.
\label{def:phii}
\end{equation}
In this example, $\phi_i=1/9$ for all $i$. Four generations are simulated ($I=4$).

\begin{figure}[tbh]
\centering
\includegraphics[width=0.8\linewidth,natwidth=1200,natheight=900]{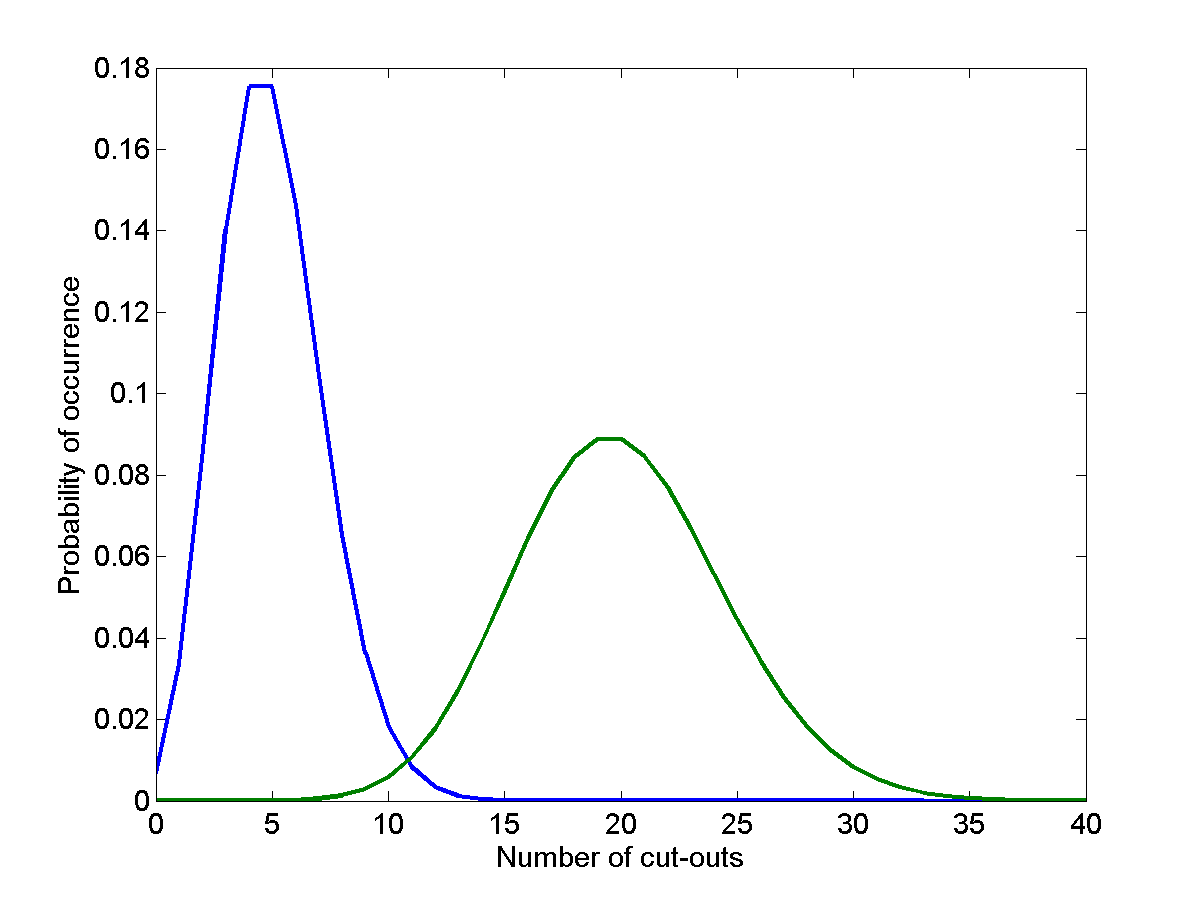}
\caption{Probability of occurrence for a given number of craters. Two cases are shown: one for the mean number of craters equal to 4 (blue line), the other for the mean number of craters equal to 20 (green line).}
\label{fig:Poisson}
\end{figure}

The method for randomly placing the craters in Fig.~\ref{fig:craterFractal} is based on a Poisson process. First, the overall area $A$ is partitioned into $R$ rows and $C$ columns of pixels. The area of each pixel is thus $A/RC$. The probability $p_i$ of the center of a circle in size class $i$ occurring within a particular pixel equals the expected number of circles of that size class divided by the number of possible locations, $N_i/RC=(\phi_i A)/(\pi a_i^2 RC)=(\phi_i\Delta x\Delta y)/(\pi a_i^2)$, where $\Delta x$ and $\Delta y$ are the grid spacing. Hence, the finer we make the grid, the smaller $p_i$ becomes. If $p_i$ is sufficiently small, the actual number of circles $n_i$ of size class $i$ present in realizations of the cratering process is a random variable with Poisson distribution with mean $N_i$, $P_{N_i}(n_i=\nu)$, where:\footnote{The applicability of the Poisson distribution to a situation such as this is described in many introductory texts in statistics, such as Lapin \citeyearpar{lapin1980statistics}.}
\begin{equation}
P_N(n=\nu)=\frac{N^\nu e^{-N}}{\nu!}.
\label{eq:Poisson}
\end{equation}
Figure~\ref{fig:Poisson} shows this probability distribution for $N_i=4$ and $N_i=20$. For example, when $N_i=4$, there is a probability of about $17.5\%$ that four cut-outs will actually be present in the realization, and $0.7\%$ chance that none will be present.

Clearly, as more craters are added, the amount of remaining, uncratered area (corresponding to the fabric in the Sierpinski carpet example) decreases. After the first generation, which on average has one crater of area $1/9$, we would have approximately $8/9$ of the uncratered area remaining. If the next set of craters, of which there are $9$ each with area $1/81$, do not overlap with any other craters, we would have $8/9-9(1/81)=8/9-1/9=7/9$ of the uncratered area left. Carrying on, we would have $6/9$, then $5/9$, etc. However, this sequence cannot continue indefinitely, because eventually the assumption that there is no overlap must fail. 

A more precise analysis would thus account for the overlapping of the craters. Suppose some fraction $\sigma$ of the area is uncratered after randomly placing a number of craters. This value equals the probability that a randomly selected point will be a part of the uncratered area. Next, suppose a crater from generation $i$, with fractional area $A_i/A$, is randomly added. The fractional area that is not a part of this crater is $1-A_i/A$. For a random point to fall within the uncratered area, that point must have been uncratered prior placement of the new crater, and must also be outside the new crater. The probability of both of these conditions being met is $\sigma(1-A_i/A)$. Carrying on in this manner, the probability of a given point remaining a part of the uncratered area, after the placement of $N_i$ craters of each generation (size class), is
\begin{equation}
\sigma=\sigma_0\prod_{i=1}^{I}(1-A_i/A)^{N_i}, 
\label{eq:siga}
\end{equation}
where $\sigma_0$ is the initial uncratered fraction (normally 1) and $I$ is the number of generations (sizes) of the circles. The fractional uncratered area remaining after the addition of generation $i$ is
\begin{equation}
\sigma_i=\sigma_0\prod_{j=1}^{i}(1-A_j/A)^{N_j}.
\label{def:sigi}
\end{equation}
If $\phi_i=N_i A_i/A\ll 1$ for all $i$, we have
\begin{equation}
\sigma_i\approx\sigma_0\prod_{j=1}^{i}(1-\phi_j). 
\label{eq:sig2a}
\end{equation}
This approximation amounts to neglecting the overlap between craters within each size class (but not between size classes). Finally, when $\phi_i$ is the same for all $i$, we have the simple result 
\begin{equation}
\sigma_i\approx\sigma_0(1-\phi)^i\approx \sigma_0\left({1-i\phi+\frac{i(i-1)}{2}\phi^2-\cdots}\right).
\label{eq:sigapp}
\end{equation}
Here, we use $\phi$ without a subscript to indicate the constant value for all size classes. The second approximation follows from the binomial expansion, assuming that $\phi\ll 1$. In iterative form, Eq.~\ref{eq:sigapp} implies 
\begin{equation}
\sigma_{i+1}=(1-\phi)\sigma_i.
\label{eq:sigiter}
\end{equation} 
The cratered area at each iteration $\delta_i$, equals $1-\sigma_i$. One then finds $\delta_{i+1}=(1-\phi)\delta_i+\phi$.

\begin{figure}[tbh]
\centering
\includegraphics[width=0.8\linewidth,natwidth=1200,natheight=900]{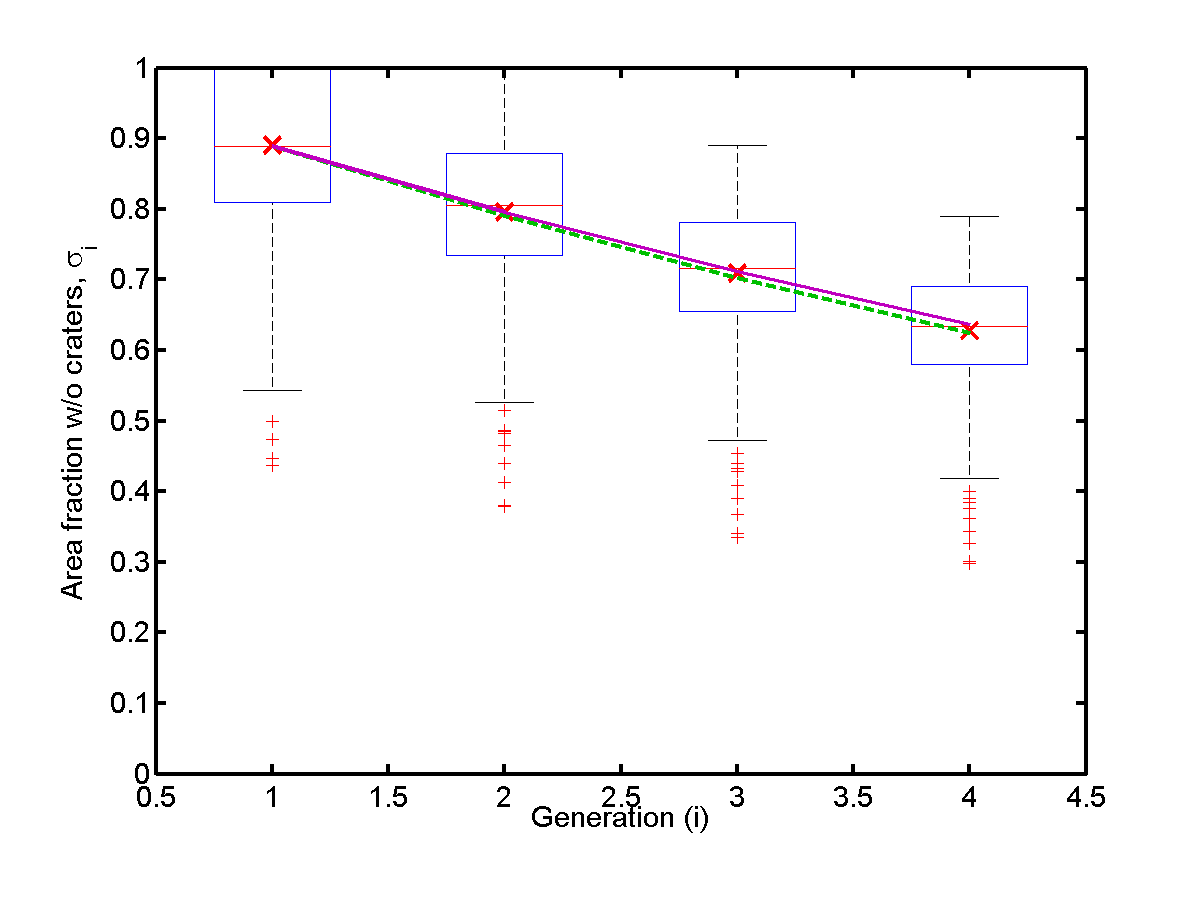}
\caption{Fractional area remaining uncratered after each generation ($\sigma_i$) for a cratering process with $\phi=1/9$. Each generation (size class) has a radius of $1/3$ the previous class. The \textsf{x}'s are the mean of 1024 random simulations. The boxes indicate the 25th, 50th (median), and 75th percentiles; the whiskers and $+$ signs indicate the extreme values. The solid line is the exact theoretical prediction, and the dashed line is an approximation assuming no overlap between craters in a particular size class.}
\label{fig:craterArea}
\end{figure}

Figure~\ref{fig:craterArea} compares the results of the cratering simulation with $\sigma_0=1$ and $\phi=1/9$ to Eq.~\ref{eq:siga} and the approximation, Eq.~\ref{eq:sigapp}. A total of random 1024 cratering experiments were performed. The \textsf{x}'s represent the mean of these trials, and the boxes indicate the 25th, 50th (median), and 75th percentiles. For this case, Eqs.~\ref{eq:siga} and \ref{eq:sigapp} both agree nearly exactly with the mean, thus indicating that the overlap between craters is statistically negligible. 

\begin{figure}[tbph]
\centering
\includegraphics[width=0.8\linewidth,natwidth=1200,natheight=900]{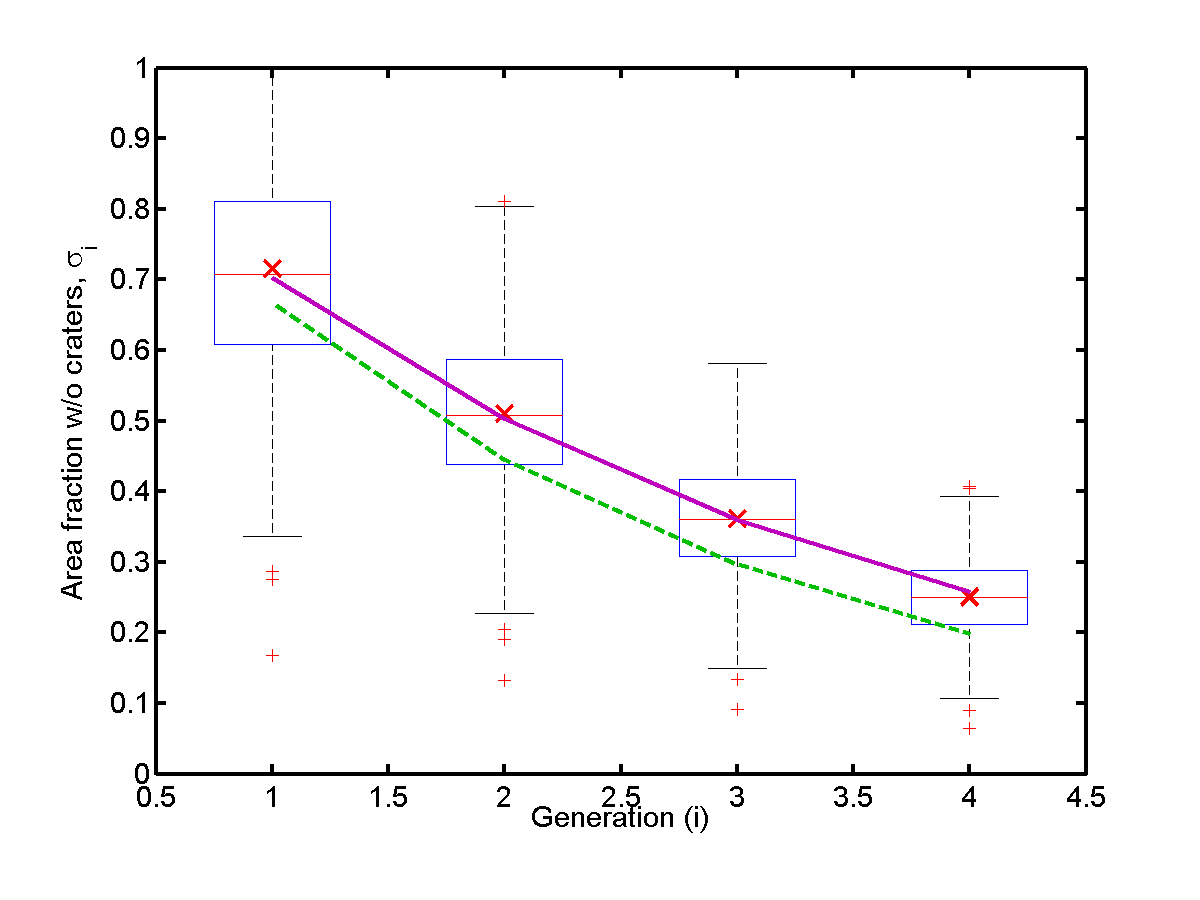}
\caption{Same as Fig.~\ref{fig:craterArea}, except that $\phi=1/3$.}
\label{fig:craterAreaMed}
\end{figure}

A similar simulation, but with $\phi$ increased to $1/3$ to create more overlap, is shown in Figure~\ref{fig:craterAreaMed}. In this case, Eq.~\ref{eq:siga} still agrees well with the mean, but the approximation, Eq.~\ref{eq:sigapp}, is no longer very accurate.

\begin{figure}[tbph]
\centering
\includegraphics[width=0.8\linewidth,natwidth=1200,natheight=900]{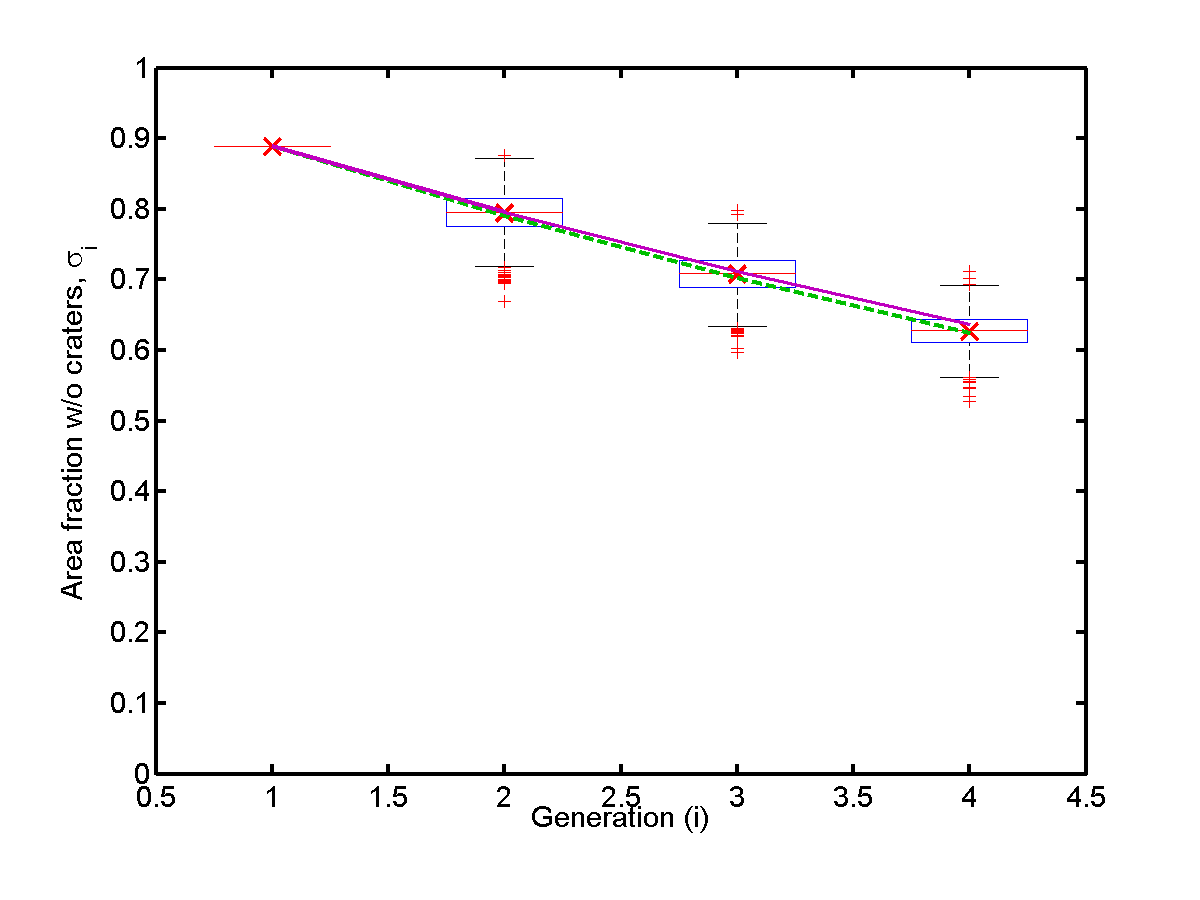}
\caption{Same as Fig.~\ref{fig:craterArea}, except that a single large crater is always added for the first generation.}
\label{fig:craterAreaFix}
\end{figure}

The quantiles shown in Figs.~\ref{fig:craterArea} and \ref{fig:craterAreaMed} indicate substantial variability in the uncratered fraction from trial to trial. Since the variability actually diminishes as more generations are added, it appears that the variability may be primarily caused by randomness in the number of the largest craters, as results from the Poisson process. Figure~\ref{fig:craterAreaFix} tests this hypothesis. This figure is the same as Fig.~\ref{fig:craterArea}, except that a single crater is \emph{always} added for the first generation. The variability in the uncratered fraction thus disappears for the first generation; however, the variability remains greatly diminished as additional generations are added, thus demonstrating that most of the variability is linked to the largest craters. 

\begin{figure}[tbph]
\centering
\includegraphics[width=0.8\linewidth,natwidth=960,natheight=720]{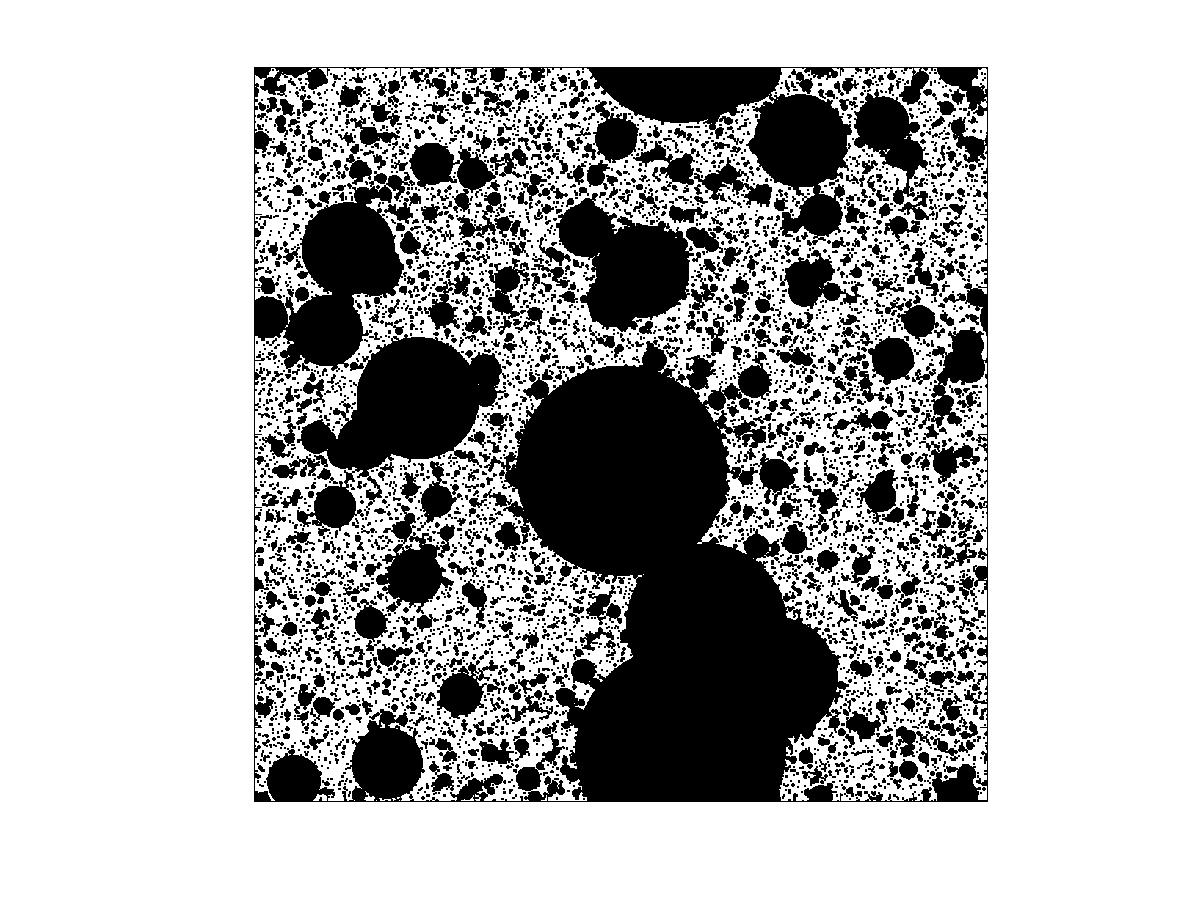}
\caption{Same as Fig.~\ref{fig:craterFractal}, except that each size class has a radius of $(1/3)^{1/4}$ the previous class.}
\label{fig:craterFractal2}
\end{figure}

Up to this point, the ratio of one size class to the next generation has always been an integer value. This is necessary for fractals generated by partitioning cells into exactly fitting sub-cells, such as in Sec.~\ref{sec:carpet}. The cells and sub-cells need not be square (or rectangular); for example, a fractal carpet built from triangles is described in Schroeder \citeyearpar{schroeder1991fractals}. Randomly placed shapes like the circles in Fig.~\ref{fig:craterFractal}, do not need to fit together, however, so the value $1/\ell$ can readily be set to a non-integer. Figure~\ref{fig:craterFractal2} shows craters with radii $a_i=(1/3)^{i/4}/\sqrt{\pi}$ (where $i=1,2,\ldots$), and hence $1/\ell=a_i/a_{i+1}=3^{1/4}$. Because there are four times as many size classes as before, the packing fraction has been decreased to $1/36$ to produce a similar coverage in area. To the eye, the distribution of sizes appears nearly continuous, like many natural phenomena. 

\subsection{Fractal Supports}
\label{sec:support}

In the previous sections, we described the black areas in Figs.~\ref{fig:sierpinski}--\ref{fig:sierpinskiDirty} as being ``cut-outs'' from the carpet fabric, whereas the white areas were the remaining fabric. The black areas in Figs.~\ref{fig:craterFractal} and \ref{fig:craterFractal2} were craters, whereas the white areas were uncratered. In terms of the construction process, the black areas are intended to represent area that is progressively removed from the 2D region, thus creating a fractal set. In this sense, $\sigma_i$ is viewed the \emph{occupied fraction}, whereas $\alpha_i=1-\sigma_i$ is the \emph{void fraction}. Still, if for some reason it is so preferred, the terminology and associations could be reversed. The white areas could be viewed as voids, and the black areas as objects. Then $\sigma_i$ would be the void fraction, and $\alpha_i$ the occupied fraction. 

A more abstract interpretation is that the black regions become in some sense ``inactive'' or ``disallowed'' regions. This viewpoint is closer to that espoused by \cite{mandelbrot1974intermittent} and \cite{frisch1978simple} when they discussed the fractal properties of turbulence. Specifically, the dissipation of turbulent kinetic energy (TKE) was viewed as taking place on a fractal set embedded in a 3D volume. The dissipation surface might be conceptualized as rotating and twisting sheets. A useful intuitive analogy might be made to the construction of houses on dry land. In this analogy, the inactive region, where no housing construction is possible, consists of bodies of water. As one examines maps at a finer and finer scale, more small bodies of water, and hence inactive construction regions, become evident. This does not mean that construction is equally vigorous at all dry land locations, but rather that construction is spatially confined to these active regions.

We thus might think of the white regions in the figures as comprising the spatial \emph{support} for the cascade process. Within these regions, the cascade process is fully active in some sense, whereas in the black regions it is inactive. This idea could be taken a step further, with ``shades of gray'' where the cascade process occurs in an intermediate state of activity. For consistency with the fractal model, the support should be created such that the active area decreases with decreasing object size. This can be done, for example, by using a cratering process to create the inactive regions cascade process. Objects of a particular size may not occur within an inactive region that is larger than the object. 

\begin{figure}[tbh]
\centering
\includegraphics[width=0.8\linewidth,natwidth=960,natheight=720]{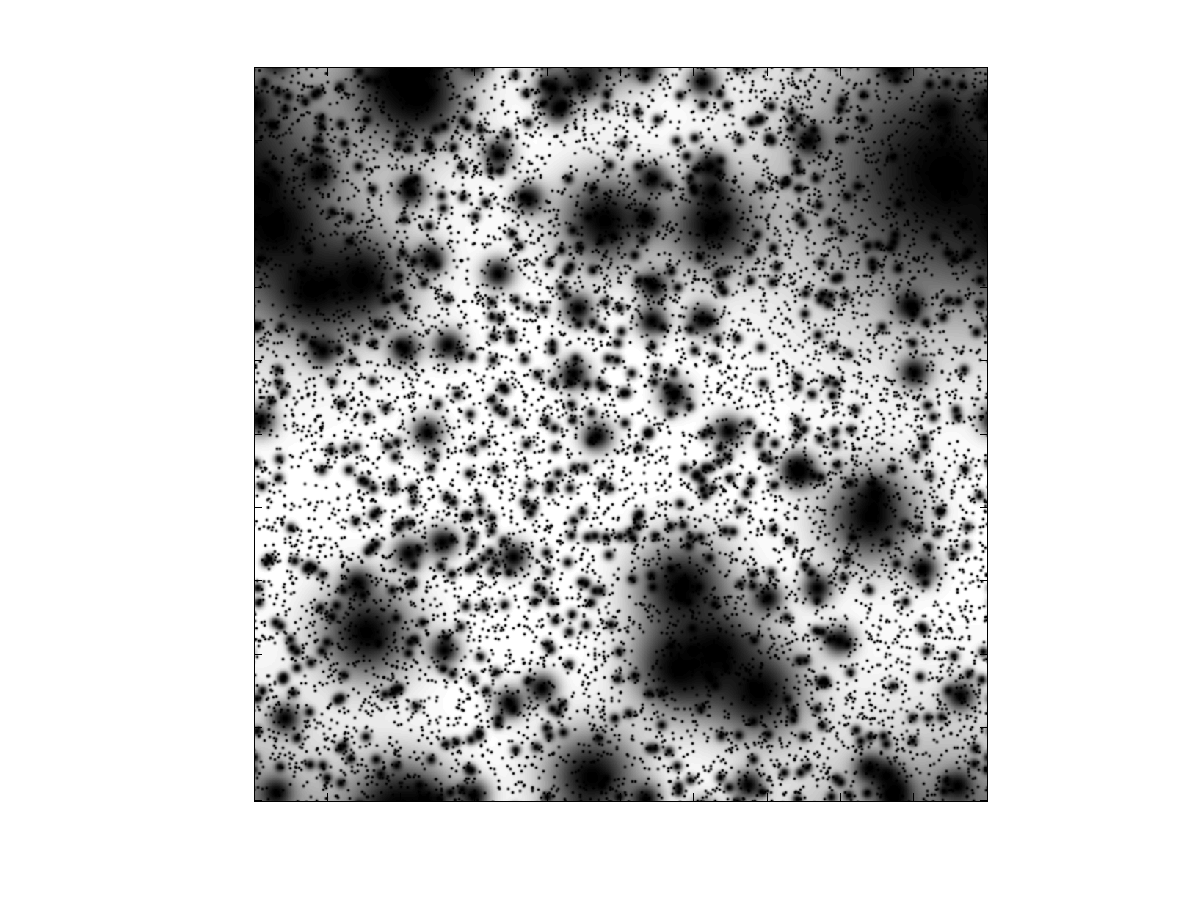}
\caption{Cratering process constructed for $\phi=1/9$ and by the same statistical method as in Fig.~\ref{fig:craterFractal}, except that the craters are now described by a Gaussian function. Plotted is the product of the crater functions for all five size classes. This plot is intended to represent the activity level (support) for a random process, where black represents inactive regions and white fully active regions.}
\label{fig:FractalSupport}
\end{figure}

As a concrete example, let us construct the support from Gaussian craters, namely
\begin{equation}
g_i(r)=\exp\left({-{r^2}/{a_i^2}}\right),
\label{eq:Gauss}
\end{equation}
where $r$ is the distance from the center of the crater and $a_i$ its effective radius. (It can be shown, by integrating Eq~\ref{eq:Gauss} over all area, that the two-dimensional support is $\pi a_i^2$, just like the circular crater.). A value of $1$ for $g_i(r)$ indicates an inactive region, whereas $0$ indicates a fully active region. The inactive region for a particular size class $i$ is determined by multiplying all craters of size $i$ and larger. The active region, or support for the process, equals $1$ minus the inactive region. An example realization of such a cratering process is shown in Fig.~\ref{fig:FractalSupport}. This process was constructed with the same statistical method and parameters as Fig.~\ref{fig:craterFractal}, except for the change to a Gaussian function (instead of perfect circles), and application of the just-described algebraic rules for constructing the support.

As indicated by Eq.~\ref{eq:sigapp} above, the active fractional area of the support is  $\sigma_i\approx(1-\phi^\prime)^{i-1}$ for size class $i$, where $\phi^\prime$ is the packing fraction of the cratering process used to create the support.\footnote{Here it is assumed that the active region for the first size includes the entire area, i.e., $\sigma_1=1$. Hence the exponent $i$ in Eq.~\ref{eq:sigapp} is replaced by $i-1$.} The packing fraction for each size class is approximately the full activity packing fraction for the objects, $\phi$, times the active fractional area. Hence, for size class $i$, $\phi_i\approx\phi(1-\phi^\prime)^i$. In the particular example shown in Figs.~\ref{fig:craterFractal} and \ref{fig:FractalSupport}, $\phi^\prime=\phi=1/9$, so $\phi_i\approx(1/9)(8/9)^i$.

\begin{figure}[tbh]
\centering
\includegraphics[width=0.8\linewidth,natwidth=960,natheight=720]{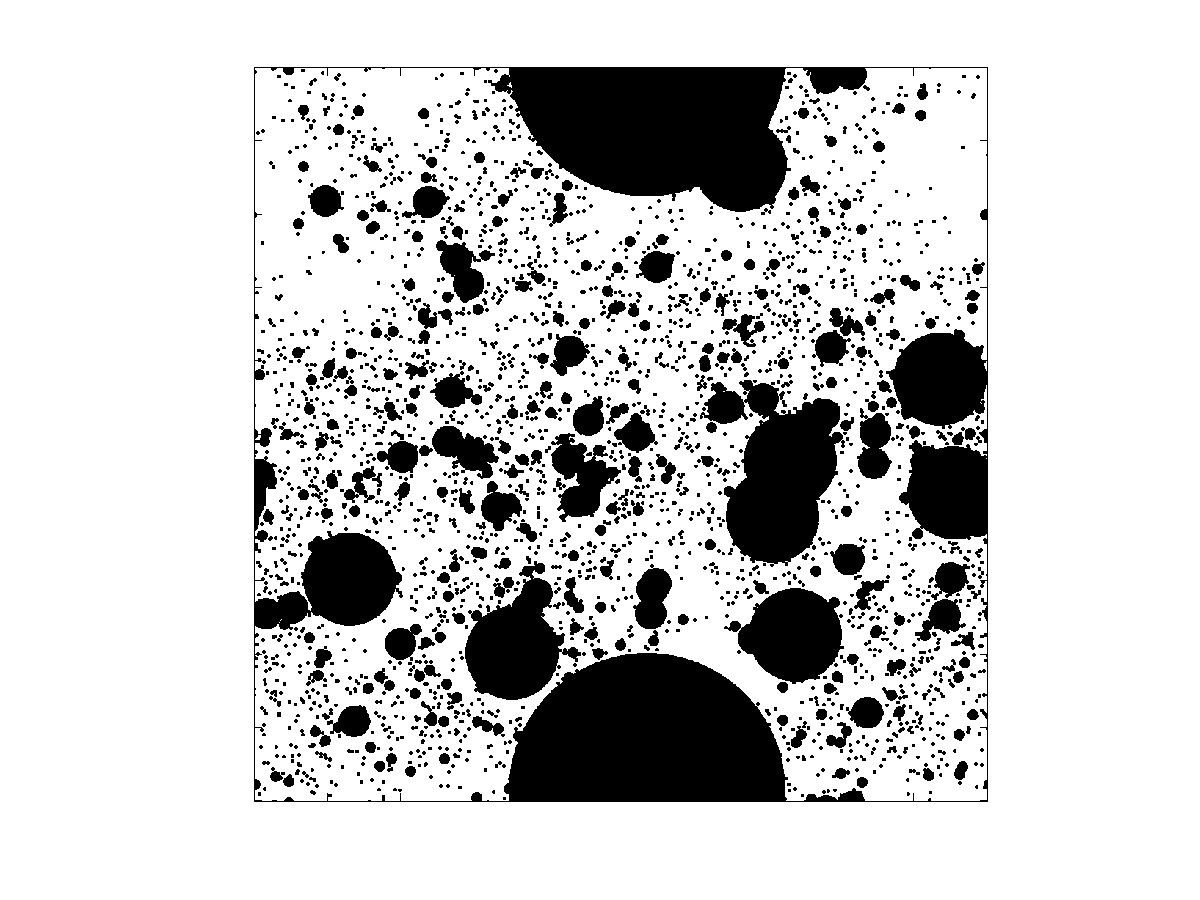}
\caption{Another cratering process constructed with $\phi=1/9$ and the same statistical method as in Fig.~\ref{fig:craterFractal}, except that the probability of a crater occurring has been multiplied by the random support function shown in Fig.~\ref{fig:FractalSupport}.}
\label{fig:CraterSupport}
\end{figure}

Having developed a model for the support of the process, we now use this model to modulate the activity level of a second random cratering process. The previous section described the cratering process using a Poisson model, in which there is a probability $p_i$ of the center of a circle of size class $i$ occurring within a particular pixel. In the current context, we may regard $p_i$ as the probability of a crater appearing when the process is fully developed. In an inactive region, the probability drops to zero. Thus, we multiply $p_i$ by the value of the support, and then implement the second cratering process. Figure~\ref{fig:CraterSupport} shows the result. Qualitative intermittency effects similar to real-world random media of interest, such as the Amboy site (Fig.~\ref{fig:amboypic2}) and ice floes (Fig.~\ref{fig:icefloe}), are now apparent.

\begin{figure}[tbph]
\centering
\includegraphics[width=0.8\linewidth,natwidth=960,natheight=720]{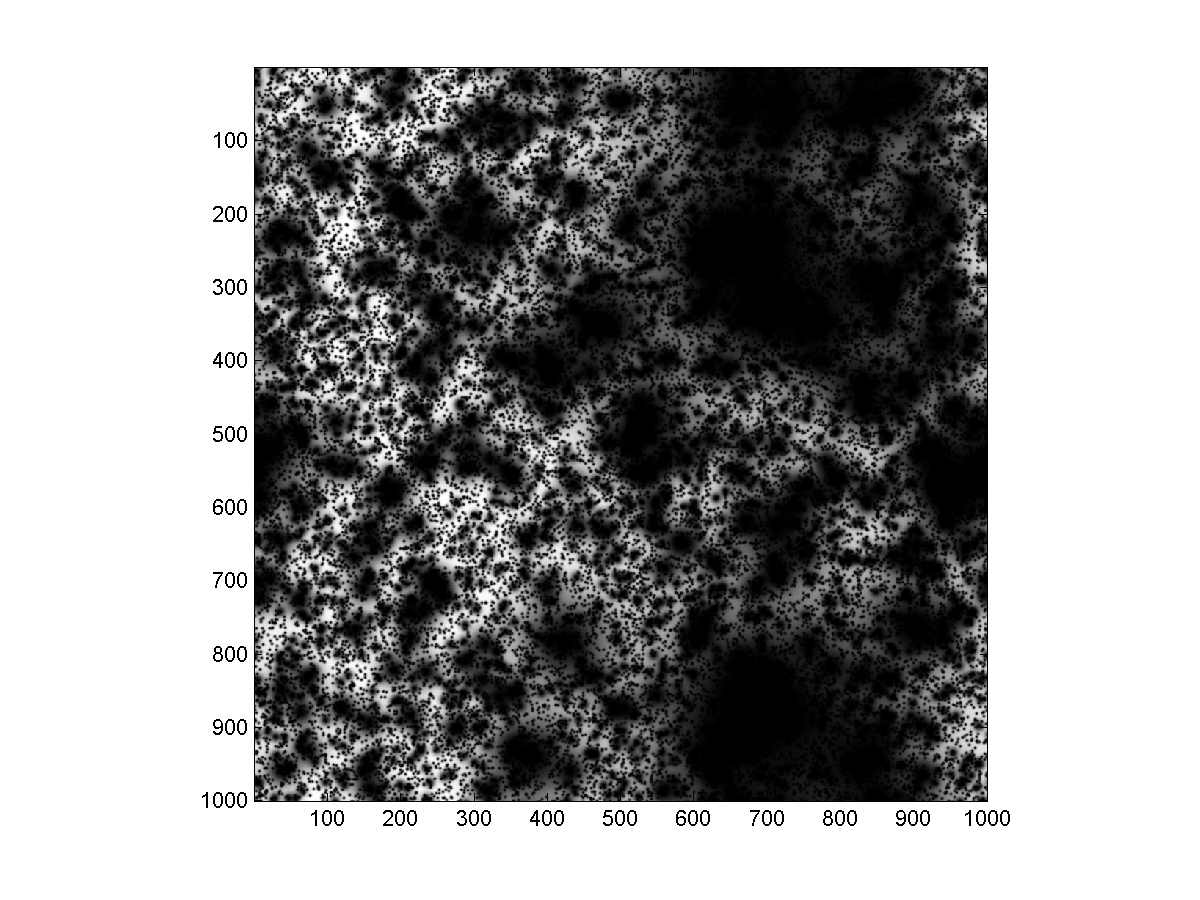}
\caption{Same as Fig.~\ref{fig:FractalSupport}, except that $\phi^\prime=1/3$.}
\label{fig:FractalSupport2}
\end{figure}

\begin{figure}[tbph]
\centering
\includegraphics[width=0.8\linewidth,natwidth=960,natheight=720]{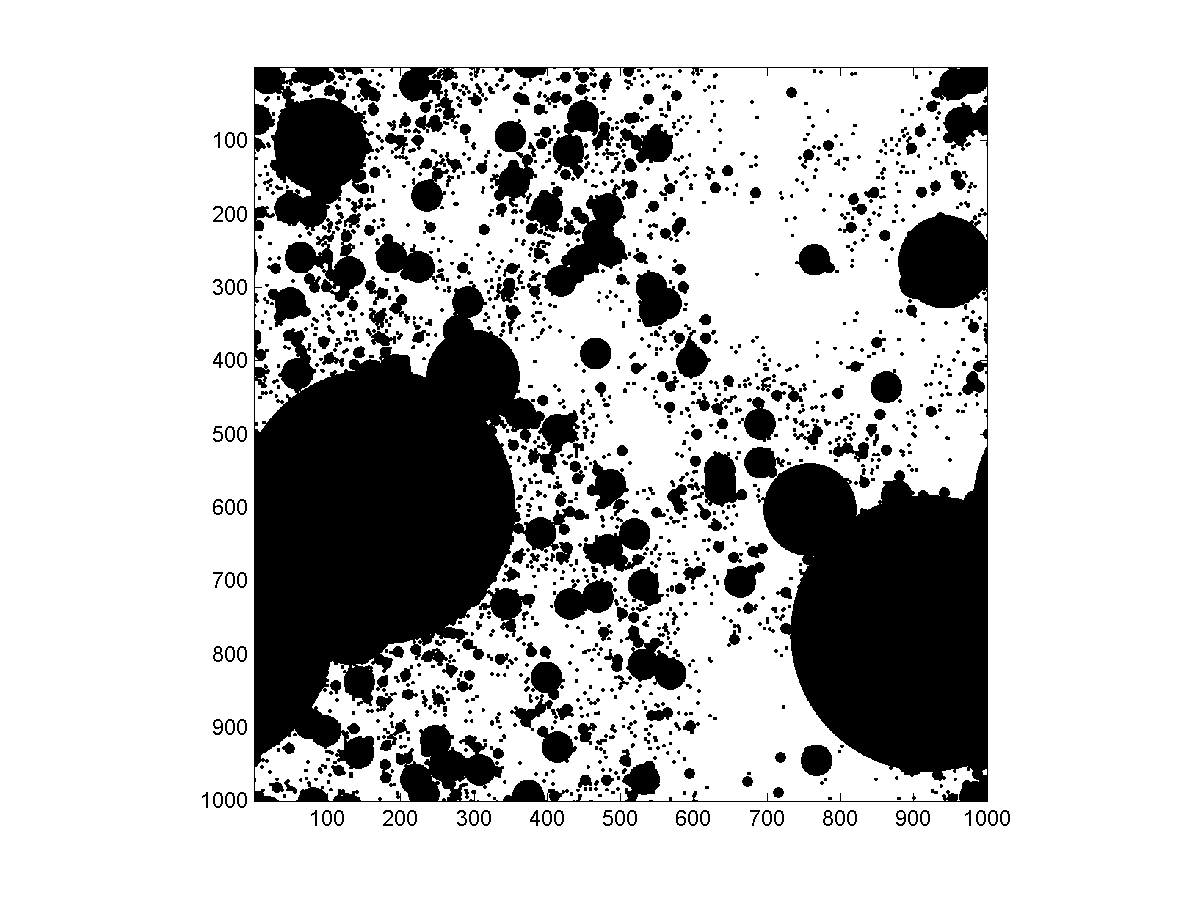}
\caption{Same as Fig.~\ref{fig:CraterSupport}, except that $\phi=1/3$, and the support shown in Fig.~\ref{fig:FractalSupport2} has been used.}
\label{fig:CraterSupport2}
\end{figure}

Figures~\ref{fig:FractalSupport2} and \ref{fig:CraterSupport2} are similar to \ref{fig:FractalSupport} and \ref{fig:CraterSupport}, respectively, except that $\phi^\prime$ and $\phi$ have been increased to $1/3$, so $\phi_i\approx(1/3)(2/3)^i$. The result is a medium with more voids, and for which the small craters are more concentrated into smaller regions of space.

\section{Basic Quasi-Wavelet Model}
\label{sec:QW}

The previous section introduced fractals constructed from squares and circles. In this section, a more flexible type of object, which we term a \emph{quasi-wavelet}, is introduced for the purpose of constructing random fields with a greater degree of realism. Wavelets \cite[e.g.,][]{torrence1998practical} make convenient analogs for physical objects because they are spatially localized. The prefix \emph{quasi} is added here to indicate that, strictly speaking, our objects are not always true wavelets. In particular, restrictions such as a zero mean may be relaxed (to represent a rock, for example), and the QWs may be spherically symmetric in multiple dimensions, which is not normal practice in wavelet analysis. Like customary wavelets, QWs are always based on translations and dilations of a parent function; however, their positions and orientations are randomized.

\subsection{Field Representation and Packing}
\label{sec:field}

We assume for now that the parent function $f$ is spherically symmetric. That is, $f$ depends only on the magnitude $\xi$ of the vector $\boldsymbol{\xi}\equiv\left(\mathbf{r} -\mathbf{b}^{ij}\right)/a_i$, where $\mathbf{r}$ is the spatial coordinate, $\mathbf{b}^{ij}$ is the location of the center of the QW, and $a_i$ is a length scale, which can be taken as the radius of the QW, or another convenient measure. The index $i$ ($i=1,\ldots,I$) indicates the \emph{size class} of the QW, with $i=1$ being the largest size (called the \emph{outer} scale), and $i=I$ the smallest size (called the \emph{inner} scale). The index $j$ indicates a particular QW belonging to the size class $i$. (The notational convention followed here is that a superscript $ij$ indexes a single QW, whereas a subscript $i$ indicates a property associated with all QWs of a particular size class.)

The parent function is normalized such that\footnote{This normalization was not used in previous papers on QWs, such as \cite{wilson2009quasi}, in which the integral of $f^2(\xi)$ over volume was flexible and designated as $I_D$. Thus, for this report, $I_D=1$. The normalization helps to standardize the size of the QW volume. For example, a QW in the shape of a cube, after normalization of the parent function, would have always have a volume of $V_i=a_i^D$ as expected.}  
\begin{equation}
\int_V d^{D}\xi\, f^{2}\left(\xi\right) =1. 
\label{eq:norm}%
\end{equation}
The following Gaussian parent function meets this normalization condition, and is convenient for analysis and rapid calculation on computers:%
\begin{equation}
f\left(\xi\right)=\exp\left({-\frac{\pi\xi^{2}}{2}}\right).
\label{eq:parent_Gauss}%
\end{equation}
Other choices for spherically symmetric parent functions are of course possible; \cite{goedecke2004quasi} and \cite{wilson2009quasi} discuss several alternatives and their utility.

The field perturbation associated with an individual QW is written as
\begin{equation}
Q^{ij}\left(\mathbf{r}\right)=h^{ij} q_i f\left(  \frac
{\left\vert \mathbf{r}-\mathbf{b}^{ij}\right\vert }{a_i}\right),
\label{eq:QWi}%
\end{equation}
where $h^{ij}$ is a random coefficient that may be positive or negative, and $q_i$ is an amplitude factor for the size class. (The amplitude will be associated with particular physical quantities later on.) The total scalar field $Q\left(\mathbf{r}\right)$ is then constructed by summing the contributions of each individual QW in the ensemble:\footnote{Here we deal entirely with synthesis of scalar fields. QW models for turbulent velocity fields are discussed in \cite{goedecke2004quasi}.}
\begin{equation}
Q\left(\mathbf{r}\right)=\sum_{i=1}^{I}\sum_{j=1}^{N_i}
Q^{ij}\left(\mathbf{r}\right), 
\label{eq:decomp}%
\end{equation}
where $N_i$ is the number of QWs in the size class $i$ that contribute to the field in the observation volume $V$. \emph{Volume} is used here in a generic sense to indicate a $D$-dimensional region. 

The quantity $V_i=N_i a_i^D$ corresponds to the total volume occupied by QWs of size $a_i$; hence $V_i/V$ is the effective fraction of $V$ occupied by QWs of size $a_i$. This motivates the definition of the \emph{packing fraction} as
\begin{equation}
\phi_i=\frac{N_i a_i^D}{V}=\mathcal{N}_i a_i^D,
\label{def:pack}
\end{equation}
where $\mathcal{N}_i=N_i/V$ is the number density (number per unit volume) for the size class. Note that the packing fraction is unaffected by overlap among individual QWs; it simply reflects the number of QWs in a size class per unit volume, scaled by the effective volume of QWs in that size class.

\subsection{Scales and Self-Similarity}
\label{sec:scales}

This section describes construction of a \emph{self-similar} ensemble of QWs. By this, we mean that the properties of the QWs, and the rules by which the ensemble are constructed, are independent of the size class $i$. (Exceptions may be made for the largest size class, $i=1$, and the smallest, $i=I$.) We also define in this section additional quantities associated with the QWs, such as amplitudes and energies, and relate these to the length scales and number densities.

Assuming self similarity, the ratio of the scales between adjacent classes, $a_{i+1}/a_i$, must be independent of $i$. Thus an invariant ratio $\ell$ can be defined as follows:\footnote{In previous formulations of QW models \citep[e.g.,][]{goedecke2006quasi}, the ratio $a_{i+1}/a_i$ was set to $e^{-\mu}$, where $\mu$ is an adjustable parameter equal to $-\ln\ell$. Adoption of $\ell$ in this paper is for convenience.}
\begin{equation}
\ell=\frac{a_{i+1}}{a_i}.
\label{def:ellK}
\end{equation}
We will later associate $\ell$ with the ratio of length scales between generations of the cascade process. In the construction of many fractals and the beta-model for turbulence \citep{frisch1978simple}, for example, at each iteration of the construction $a_{i+1}$ is set to $a_i/2$, which implies $\ell=1/2$. For the time being, however, the spacing between the size classes may be regarded as arbitrary. From Eq.~\ref{def:ellK}, it follows that
\begin{equation}
a_i=a_1 \ell^{i-1}.
\label{eq:ellseq}
\end{equation}

Ratios of other properties between adjacent size classes must also be invariant in order to preserve self similarity. The ratio of packing fractions between adjacent size classes, i.e., 
\begin{equation}
\overline{\phi}=\frac{\phi_{i+1}}{\phi_i}
\label{def:phi}
\end{equation}
is of particular interest. A decreasing packing fraction ($\overline{\phi}<1$) indicates that activity is concentrated in less volume as the size of the QWs decreases; this relates to the fractal dimension of the process.\footnote{When $\ell=1/2$, the ratio ${\phi_{i+1}}/{\phi_i}$ corresponds to the parameter $\beta$ in the beta model \citep{frisch1978simple}. Each generation in the beta model consists of halving the size of the eddies.} Presuming that $\phi_1$ and $\overline{\phi}$ are known, we can determine the packing fractions for the remaining size classes as 
\begin{equation}
\phi_i=\phi_1 \overline{\phi}\,^{i-1}.
\label{eq:phiseq}
\end{equation}

The packing fraction is related to the more commonly encountered \emph{volume fraction}. The volume fraction is analogous to the fractional area occupied by the craters, as introduced in Sec.~\ref{sec:support} and indicated by the symbol $\delta=1-\sigma$. The packing fraction tends to be more convenient for analysis of energetics, whereas the volume fraction is usually more convenient for comparisons with observations. For the case where the QWs are distributed uniformly and randomly over space, the relationship between the two can be derived analytically. In particular, Eqs.~\ref{eq:siga}--\ref{eq:sigiter} apply, although we replace $A_i/A$ with $V_i/V$. (Or, we could use $\phi_i/N_i=V_i/V=A_i/A$.)  

The value of $\overline{\phi}$ as defined by Eq.~\ref{def:phi} is dependent upon the choice of $\ell$. To better understand this linkage, we observe that self similarity implies that quantities such as the packing fraction must have a \emph{power-law dependence} on the length scale $a$. Defining $\phi(a)$ as the packing fraction at $a$ (that is, $\phi(a_i)=\phi_i$), we have $\phi(a)\propto a^\beta$, where $\beta$ is an invariant parameter. Then 
\begin{equation}
\frac{\phi_{i+1}}{\phi_i}=\left( \frac{a_{i+1}}{a_i}\right)^\beta=\ell^\beta.
\end{equation}
Comparing to Eq.~\ref{def:phi}, we see
\begin{equation}
\overline{\phi}=\ell^\beta.
\label{def:beta}
\end{equation}

The ratio of number densities between adjacent classes, $\overline{N}={N_{i+1}}/{N_i}$, follows from Eqs.~\ref{def:pack} and \ref{def:ellK} as:
\begin{equation}
\overline{N}=\overline{\phi}\ell^{-D}.
\label{eq:N}
\end{equation}

As indicated in Sec.~\ref{sec:field}, each QW of the size class $i$ has a characteristic amplitude $q_i$. For self-similarity, the ratio $\overline{q}={q_{i+1}}/{q_i}$ must be invariant. Analogously to the packing fraction, we define a power-law exponent $\lambda$ such that $q(a)\propto a^\lambda$. It then follows
\begin{equation}
\overline{q}=\ell^\lambda.
\label{def:lambda}
\end{equation}

\begin{figure}[tbph]
\centering
\includegraphics[width=\linewidth,natwidth=1231,natheight=459]{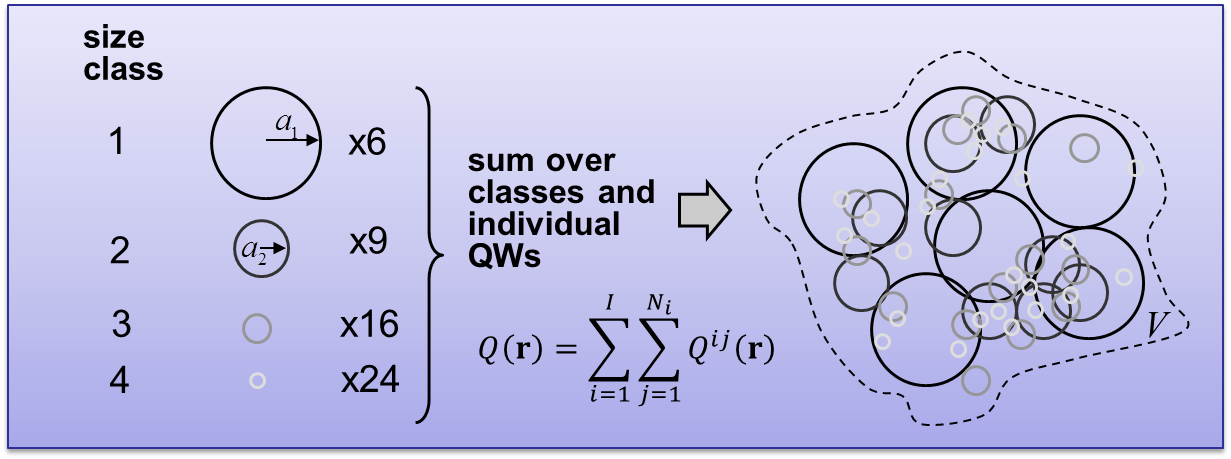}
\caption{Illustration of a QW field constructed from four size classes.}
\label{fig:QWfield}
\end{figure}

Figure~\ref{fig:QWfield} conceptually illustrates the construction of a self-similar QW field with four different size classes. 

\subsection{Definition of a Generation}
Section~\ref{sec:scales} introduced four distinct ratios between adjacent QW size classes: $\ell$, $\overline{\phi}$, $\overline{N}$, and $\overline{q}$. Eq.~\ref{eq:N} provides a constraint on the values of the first three of these ratios. Furthermore, one of the ratios must be regarded as a \emph{reference} for defining the size classes, which otherwise would be completely arbitrary. Hence only two of the ratios may be considered truly independent.  

In this paper, we follow the precedent \cite{frisch1978simple} (and other references on construction of fractals) by setting $\ell=1/2$ as the reference ratio. This means that the QWs of each class are half the size of the previous. We refer to each size class constructed in this manner as a \emph{generation} of the model. Other approaches could of course be contemplated for defining the reference ratio. For example, the energy or amplitude between adjacent size classes might be set to a fixed value. But, fixing the value of $\ell$ seems like the most natural choice when constructing a spatial process. 

We next must choose the two adjustable parameters in the model. In this paper, they are taken to be $\beta$ and $\lambda$, since these power-law exponents are \emph{independent} of the method for defining a generation. The remaining model parameters can now all be calculated from $\ell$, $\beta$, and $\lambda$. From Eqs.~\ref{def:beta}, \ref{eq:N}, and \ref{def:lambda}, we have, for $\ell=1/2$,
\begin{equation}
\overline{\phi}=2^{-\beta},
\label{eq:phi2}
\end{equation}
\begin{equation}
\overline{N}=2^{D-\beta},
\label{eq:N2}
\end{equation}
and
\begin{equation}
\overline{q}=2^{-\lambda}.
\label{eq:q2}
\end{equation}

Hence we have reduced the description of the scaling between size classes to two free parameters, $\beta$ and $\lambda$ (or, alternatively, $\overline{\phi}$ and $\overline{q}$). There are still a number of other free parameters in the model representing the largest size class, namely $a_1$, $\phi_1$, and $q_1$, as well as $a_I$, representing the smallest size class. (The number of size classes $I$ can be determined from $a_1$ and $a_I$.)

\subsection{Scale Densification and Fractional Size Classes}
\label{sec:densify}

In earlier presentations of the QW model, the spacing between size classes (effectively, the parameter $\ell$) was arbitrary. This allowed models to be constructed with fine gradations between the size classes. By associating $\ell$ with a generation of the cascade process and thus fixing its value to $1/2$, however, the scales of the size classes are fixed to values that may be undesirably coarse. To address this problem, we now introduce \emph{fractional} generations as an approach to densifying the spacing between modeled scales.

The basic idea behind scale densification is simply to allow the generation index $i$ in Eq.~\ref{eq:ellseq} and similar equations to assume non-integer values, although still with constant spacing, so that the length scales still occur in a geometric series. Specifically, we allow $i=1+(j-1)/K$, where $K$ is called the \emph{scale densification} factor and $j=1,2,\ldots,IK$.  There are a factor of $K$ more size classes present than would occur for $K=1$. For example, a cascade process with three generations and no scale densification ($K=1$) would have the three length scales $a_1$, $a_2=\ell a_1$, and $a_3=\ell^2 a_1$. When this process is densified to $K=2$, additional size classes would be introduced at $a_{3/2}=\ell^{1/2} a_1$, $a_{5/2}=\ell^{3/2} a_1$, and $a_{7/2}=\ell^{5/2} a_1$. Note that these additional size classes form a new generational sequence, i.e., $a_{5/2}=\ell a_{3/2}$, and $a_{7/2}=\ell^2 a_{3/2}$. 

Normally, we would desire to perform the scale densification so that it has no (or little) net effect on the number of QWs. This motivates the generalized definition of the packing fraction for a densified model as
\begin{equation}
\phi_i=\frac{K N_i a_i^D}{V}=K \mathcal{N}_i a_i^D.
\label{def:pack2}
\end{equation}
The densification factor $K$ was not present in Eq.~\ref{def:pack} or in the definition of the packing fraction for earlier formulations of the QW model. When $K=1$, the various definitions coincide. When the number of size classes are increased by a factor $K$, however, $N_i$ must be diminished by a factor of $1/K$ to preserve the packing fraction. Then a scale densification would not alter the packing fractions. Inclusion of $K$ in Eq.~\ref{def:pack2} does not affect $\overline{\phi}$, because the factor of $K$ cancels out in Eq.~\ref{def:phi}.

\subsection{Level Cuts}
\label{sec:cut}

Often we wish to model RHM with two or more distinct phases. Examples from the Introduction are Fig.~\ref{fig:amboypic2}, for which the phases may be regarded as volcanic rock and sand, and Fig.~\ref{fig:melt_ponds}, for which the phases may be regarded as water and ice. In such situations, we may use \emph{level cuts} through the random field $Q\left( \mathbf{r}\right)$ to partition a continuous random medium into discrete phases \citep{grigoriu2003random,field2012level}. A single-sided level cut would consist of assigning all values of $Q\left( \mathbf{r}\right)>=c$ (where $c$ is a constant) to one phase, and values $Q\left( \mathbf{r}\right)<c$ to the other phase. A double-sided level cut assigns $|Q\left( \mathbf{r}\right)|>=c$ (where $c$ is a positive constant) to one phase, and $|Q\left( \mathbf{r}\right)|<c$ to the other. Since $Q$ is proportional to $q_1$, the level cut does not, in principle, introduce a new model parameter; it depends on the value of the ratio $c/q_1$.

Qualitatively, when a double-sided level cut is used, we would expect contributions from size classes $i$ such that $q_i\gtrsim c$ to dominate. (This relationship is only approximate because, at any given location in space, QWs from multiple size classes can overlap and cause the superposition to fluctuate about $c$.) Since $q_i=q_1 \overline{q}^{i-1}$, conversely we would conclude that spatial scales $a_i$ such that
\begin{equation}
\overline{q}^{(i-1)}\lesssim \frac{c}{q_1}
\end{equation}
tend to be filtered out of the representation. This is roughly analogous to setting the inner scale to $a_i$, although the transition is more gradual. 

To avoid the distinctly different behaviors imposed by the level cut above and below the cut-off length scale $a_i$, one could set $\overline{q}=1$ ($\lambda=0$). Then, the distribution of object sizes will be solely controlled by the parameter $\overline{\phi}$ ($\beta$) everywhere, instead of being affected by the level cut at some length scales but not others. Examples of level-cut model will be given in the following sections.

\section{Static QW Model}
\label{sec:staticmodel}
A static QW model does not \emph{explicitly} incorporate any evolution in time; such models apply to a single ``snapshot,'' or image, of a process. A static ensemble should be statistically consistent with the expected packing fractions for each size class, $\phi_i$, and thus the QW counts for each size class, $N_i$. If we interpret $N_i$ as the expected value for the volume $V$, any spatial random process yielding a mean value of $N_i$ could conceivably be used. Once random positions for the QWs in each size class $i$ are generated from such a process, they are each assigned a corresponding amplitude $q_i$, and the random field associated with the resulting ensemble can be calculated from Eq.~\ref{eq:decomp}.

\subsection{Basic Poisson Model}

Perhaps the simplest implementation of this idea would be to simulate the spatial placements of QWs using a spatially homogeneous Poisson process with spatial rate $N_i/V=\mathcal{N}_i$. Realizations for the number of QWs in $V$, $n_i$, then would follow a Poisson distribution with mean $N_i$ (Eq.~\ref{eq:Poisson}, with $N=N_i$ and $n=n_i$). Numerically, we could synthesize the positions of QWs for the Poisson process by a couple approaches. First, more directly, we could partition $V$ into a large number of subvolumes with size $\delta V$, such that $\delta V \mathcal{N}_i \ll 1$. (This condition is necessary so that the probability of more than 1 QW occuring in the subvolume is negligible.) Then, within each subvolume, a random number $\epsilon$ is drawn with uniform distribution between 0 and 1. If $\epsilon<\delta V \mathcal{N}_i$, a QW is placed in the subvolume. This process must be repeated for each subvolume. Second, but more efficiently, we can draw a random QW count $n_i$ from the Poisson distribution Eq.~\ref{eq:Poisson}. The $n_i$ QWs are then placed at random locations in $V$ using a uniform random number generator. This second approach avoids partitioning and random number generation for a large number of small subvolumes. QW fields generated by either of these two approaches are termed \emph{disorganized}, since the positions of the QWs are mutually independent uniformly random over space.

\begin{figure}[tbh]
\centering
\includegraphics[width=\linewidth,natwidth=1200,natheight=900]{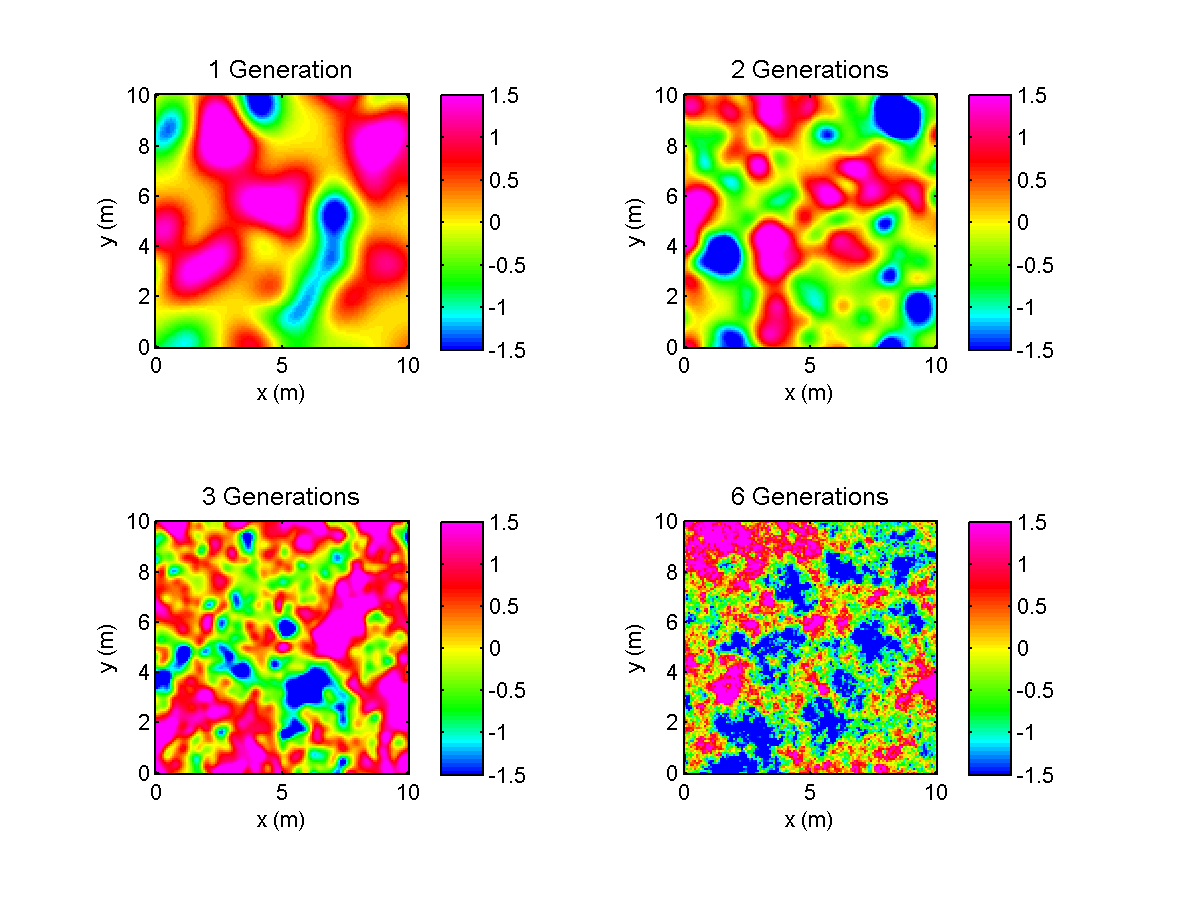}
\caption{Realizations of static QW fields with one, two, three, and six generations.}
\label{fig:StaticExamp}
\end{figure}

Figure~\ref{fig:StaticExamp} shows a sequence of realizations of static, disorganized QW fields. The sequence incorporates progressively more generations (1, 2, 3, and lastly 6). These realizations were calculated with $\ell=1/2$, $\beta=0$, $\lambda=1/3$, $a_1=1$ m, $\phi_1=0.3$, and $q_1=1$. A Gaussian parent function was used. The realization for a single generation (top left), since it contains only one relatively large QW size, has a well defined length scale and smooth appearance. When six generations are present (bottom right), the realization includes considerably more fine structure and appears to be self-similar.  

\begin{figure}[tbh]
\centering
\includegraphics[width=\linewidth,natwidth=1200,natheight=900]{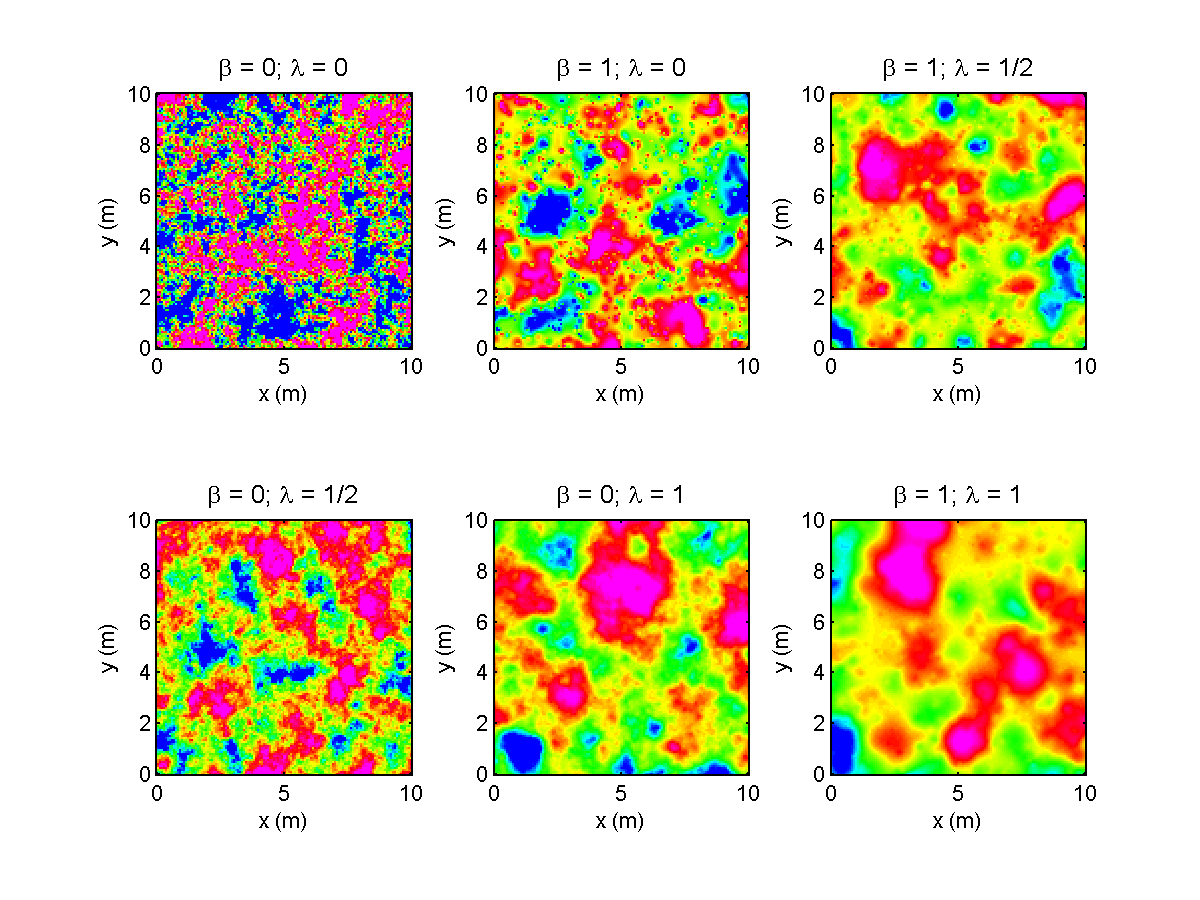}
\caption{Same as Fig.~\ref{fig:BetaLambda}, except that static realizations (after six generations) for various values of $\beta$ (power-law exponent for the packing fraction) and $\lambda$ (power-law exponent for the amplitude). Limits on the color axis are $\pm 1.5$.}
\label{fig:BetaLambda}
\end{figure}

Figure~\ref{fig:BetaLambda} shows a six-generation process with the same parameters Fig.~\ref{fig:StaticExamp}, except that $\beta$ and $\lambda$ are varied. The realizations with $\lambda=0$ have QW amplitudes that are independent of size. The large amplitude fluctuations at small scales do not look like any of the examples in the Introduction, except perhaps for the urban terrain (Fig.~\ref{fig:rosslyn}). The realization for $\beta=1$ and $\lambda=0$ has fewer small QWs, but this leads to a rather peculiar appearance of rapid small variations occurring within larger objects. The realizations for $\lambda=1/2$ and $\lambda=1$ seem more realistic. The latter value attenuates the amplitude at small scales very noticeably. The case $\beta=1$ and $\lambda=1$ has a very smooth appearance, not unlike the realizations with 1 or 2 generations shown in Fig.~\ref{fig:StaticExamp}.

\begin{figure}[tbh]
\centering
\includegraphics[width=\linewidth,natwidth=1200,natheight=900]{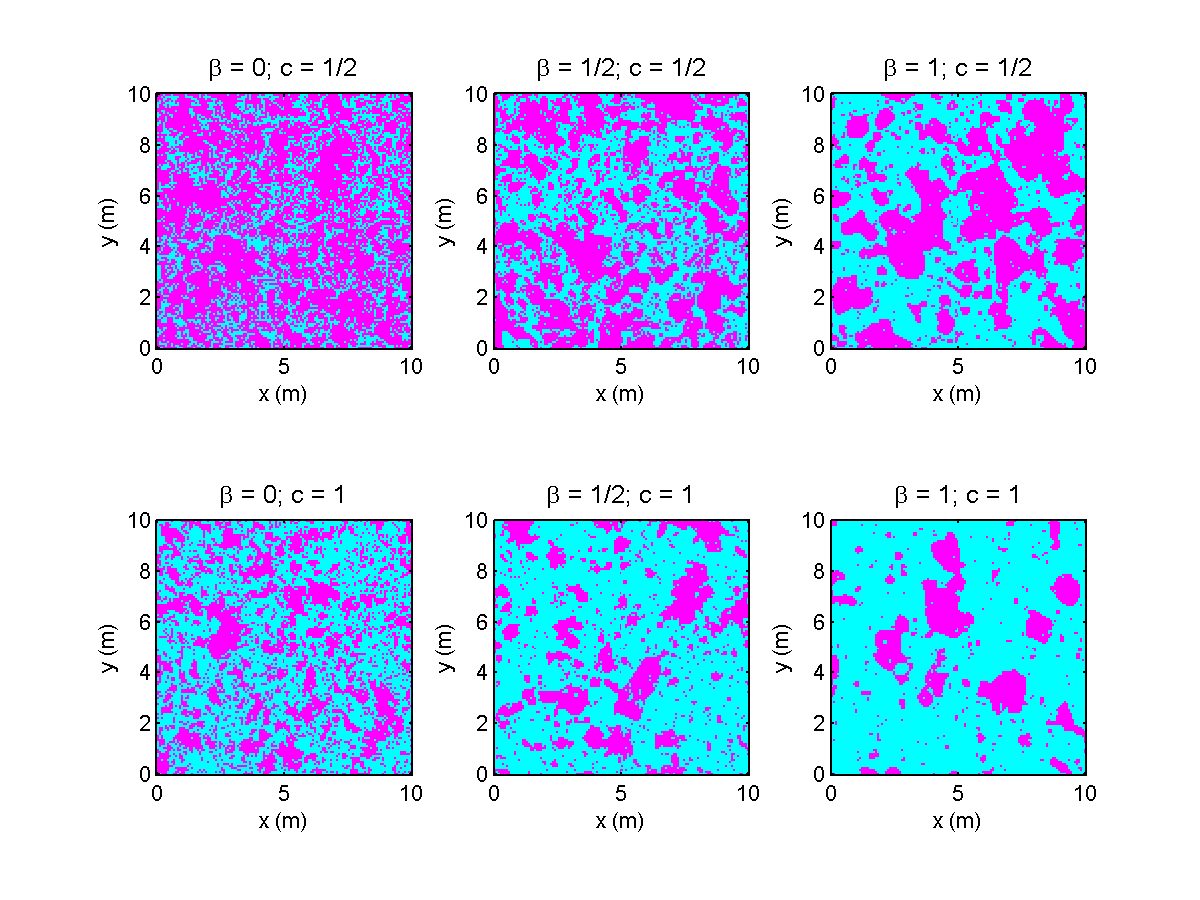}
\caption{Static realizations of two-phase media for various values of $\beta$ (power-law exponent for the packing fraction) and the level-cut parameter $c$, for a double-sided level cut. Cyan represents the background phase (regions where $Q<|c|$), whereas magenta represents the occupied phase (regions where $Q\geq|c|$).}
\label{fig:CutExample}
\end{figure}

Figure~\ref{fig:CutExample} shows example realizations of two-phase media, as calculated by the double-sided level-cut method described in Sec.~\ref{sec:cut}. As the parameter $\beta$ is increased, the number of small objects decreases. For a given value of $\beta$, larger values of the level cut increase the void fraction.

\subsection{Spatially Organized Model}
\label{sec:org}

As mentioned earlier, the preceding modeling approach leads to mutually independent QW positions; that is, realizations with no spatial organization. This is analogous to the basic cratering process described in Sec.~\ref{sec:crater}. However, the fractal support techniques described in Sec.~\ref{sec:support} could also be used to modulate the Poisson process for QW placements, and thus organize the QWs into active regions. 

In Sec.~\ref{sec:support}, it was shown that a cratering process with a packing fraction of $\phi^\prime$ for the support leads to a packing fraction for size class $i$ of $\phi_i\approx\phi(1-\phi^\prime)^{i-1}$, where $\phi$ is the ``full activity'' level for the packing fraction, which applies to the initial size class. Comparing to Eq.~\ref{eq:phiseq}, we see that the desired packing fractions of the individual classes can be obtained by setting $\overline{\phi}=1-\phi^\prime$ and $\phi_1=\phi$. This simple prescription allows a spatially organized QW model with decreasing packing fraction ($\overline{\phi}<1$, or equivalently $\beta>0$) to be simulated from a cratering process for the fractal support, with a constant packing fraction $\phi^\prime=1-\overline{\phi}$, which modulates a QW model, with a constant packing fraction $\phi=\phi_1$.

\begin{figure}[tbh]
\centering
\includegraphics[width=\linewidth,natwidth=1200,natheight=900,trim = 0in 1.2in 0in 1.2in, clip]{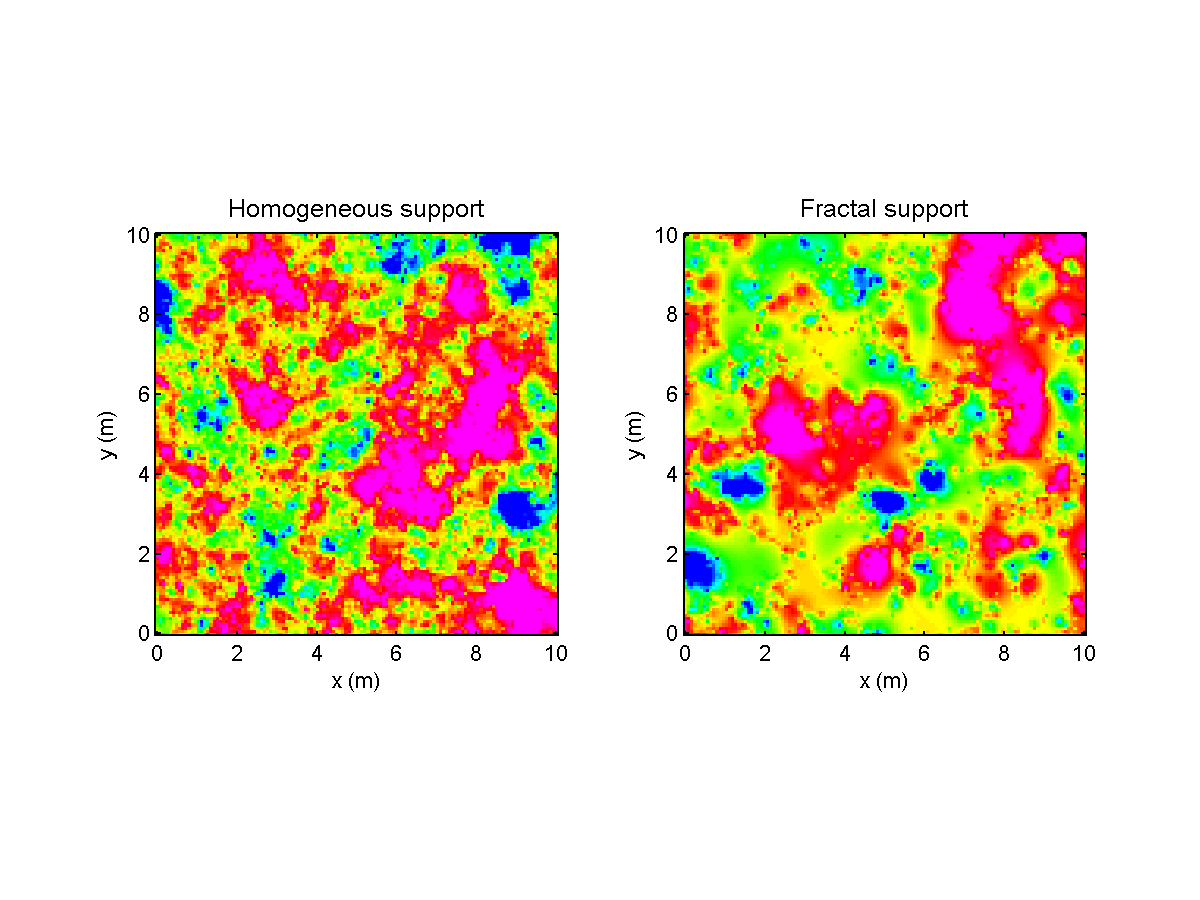}
\caption{Realizations of QW fields without spatial organization (a homogeneous support, left) and with spatial organization (a fractal support, right). The realizations are both for $\beta=\lambda=1/4$, with all other parameters being the same as Fig.~\ref{fig:BetaLambda}.}
\label{fig:FractalSupp}
\end{figure}

Example realizations of processes without and with spatial organization (i.e., homogeneous and fractal supports, respectively) are shown in Fig.~\ref{fig:FractalSupp}. Both realizations are for $\beta=\lambda=1/4$. (These values of $\beta$ and $\lambda$ satisfy the constraint for turbulence, to be derived in Sec.~\ref{sec:turb}.) The fractal support causes the smaller QWs to be concentrated in certain regions; other regions have a relatively smooth, inactive appearance. This behavior is reminiscent of many of the example RHM shown in the Introduction, such as the simulations of turbulence (Fig.~\ref{fig:dns}).

\subsection{Turbulence}
\label{sec:turb}

Let us consider constraints between the scaling parameters pertinent to the modeling of turbulence. For turbulent velocity fluctuations, the amplitudes $q_i$ may be identified as the turbulent velocities associated with the QWs, $\upsilon_i$. According to Kolmogorov's (1941) hypothesis, the transfer rate of specific turbulent kinetic energy (TKE) from one scale to the next is invariant, and equal to the dissipation at the molecular scale, $\epsilon$. Let us define $\Delta\Lambda_i$ as the net transfer rate of energy (the \emph{flux}) from scale $i$ to $i+1$. This quantity has dimensions of (length)$^2$ (time)$^{-3}$. By dimensional analysis, one has
\begin{equation}
\Delta\Lambda_i=\epsilon\propto\phi_i\frac{\upsilon_i^3}{a_i}.
\end{equation}
The factor $\phi_i$ adjusts for the volume occupied by the QWs, so that $\Delta\Lambda_i$ represents the transfer specific to the active volume. Setting $\Delta\Lambda_i=\Delta\Lambda_{i+1}$, we find 
\begin{equation}
\overline{q}=\ell^{1/3}\overline{\phi}^{(-1/3)}=\ell^{(1-\beta)/3}.
\label{eq:lambdaTurb}
\end{equation}
Comparing to Eq.~\ref{def:lambda}, we have the following constraint for turbulence:
\begin{equation}
\lambda=(1-\beta)/3.
\end{equation}
Later, in Sec.~\ref{sec:stat}, we will see that this constraint is consistent with Kolmogorov's well known ``$-11/3$ law'' for the spectrum.

\section{Steady-State Cascade QW Model}
\label{sec:steadycascade}

In Sec.~\ref{sec:org}, an approach was described for imposing organization on the positions of QWs. This section describes a conceptually different approach to organizing the QWs, namely one based on an iterative cascade process. As discussed earlier, self-similar random media may result from a cascade process, in which progressively smaller objects emerge from larger ones. The turbulence cascade, where large eddies break down into smaller ones that are eventually dissipated through viscous forces, is an example. Fracturing and weathering of rocks is another. Some cascade processes may actually operate in the reverse of this pattern. As a forest matures, it tends to support fewer but larger trees. Progressively larger structures are built in a growing city. Unlike the steady-state description, the cascade process introduces the element of time into the modeling.

In the following sections, we initially consider steady-state processes, which means that the creation of QWs of a particular size class is balanced by destruction. This balance is assumed to hold only in an average, statistical sense. This approach keeps with a general philosophy that the model is not a deterministic one describing the precise physics of the cascade process; rather, it is a stochastic model intended to realistically describe spatial statistics of a random field. 

The cascade model generically specifies that $M_i$ parent QWs of size $a_i$ decay, on average, into $M_{i+1}$ offspring of size $a_{i+1}$. The decay occurs at a time $\tau_{i}$ after the parent QWs were initially created. Alternatively, we could consider an aggregation process, in which $M_{i+1}$ QWs of size $a_{i+1}$ coalesce into $M_i$ QWs of size $a_i$ after a time $\tau_{i}$. In either case, self similarity of the cascade reaction implies that ratios of $M_i$ and $\tau_i$ between adjacent size classes are invariant. Thus, in keeping with our earlier notation, we define the two new parameters
\begin{equation}
\overline{M}=\frac{M_{i+1}}{M_i},
\label{eq:Mcasc}
\end{equation}
which represents the number offspring produced by one generation relative to the previous, and 
\begin{equation}
\overline{\tau}=\frac{\tau_{i+1}}{\tau_i},
\label{eq:tau}
\end{equation}
which represents the time scale of one generation relative to the previous.

\subsection{One-Way (Decay) Model}
\label{sec:oneway}

\subsubsection{QW Production and Counts}
\label{sec:general}

We begin by considering a process in which QWs are steadily created at the largest size class, and then uniformly decay into smaller QWs. In a steady-state model, creation of the largest QWs must take place at a steady rate. The number of QWs of each size class, $N_i$, must subsequently remain steady. 

Let us first consider initiation of the cascade process when there is no scale densification ($K=1$), i.e., there are only integer size classes. Supposing there are $N_1$ QWs at $t=0$, they will all decay by a time $t=\tau_1$. To maintain a steady state, $N_1$ new QWs of the largest size must be created during this same time interval. The QWs could be generated, for example, by a Poisson process with rate parameter $R_1=N_1/\tau_1$. On a per-volume basis, the corresponding rate would be $\mathcal{R}_1=\mathcal{N}_1 /\tau_1$, which will maintain the packing fraction at an average value of $\phi_1=\mathcal{N}_1 a_1^D$.

Each QW created for the size class $i=1$ eventually decays into an average of $M_i=\overline{M}\,^{i-1}$ offspring of the size class $i$. Since these offspring have a lifetime of $\tau_i=\tau_1\overline{\tau}\,^{i-1}$, there are an average of $N_i=N_1(\overline{M}\overline{\tau})^{i-1}$ such offspring at any particular time. This observation implies
\begin{equation}
\frac{N_{i+1}}{N_i}=\overline{N}=\overline{M}\overline{\tau},
\label{eq:Ncasc}
\end{equation}
Thus only one of the two new parameters, $\overline{M}$ and $\overline{\tau}$, may be regarded as independent.

The distinction between $\overline{N}$ and $\overline{M}$ should be kept clearly in mind. $\overline{N}$ is the ratio of the expected QW count from one generation to the preceding one, as present in a particular realization. $\overline{M}$ is the ratio of the expected number of offspring in one generation to the preceding one. The density of QWs in a particular realization depends on both the number of offspring per generation and the rate at which the decay reaction occurs. The ratios $\overline{N}$ and $\overline{M}$ are the same only if the generations all decay at the same rate, i.e., $\overline{\tau}=1$. Single realizations of the static and a steady-state models with the same values of $\overline{N}$ differ only with regard to the dynamic approach of the latter for \emph{placing} the offspring relative to the parents. If a steady-state model were constructed by uniformly dispersing the offspring across the volume, it would be indistinguishable from a disorganized static model. In fact, statistics that are not sensitive to placement of the QWs cannot reveal differences between the disorganized and cascade-based models. In particular, second-order statistics such as variances, correlation functions, and spectra, are not sensitive to these differences. High-order statistics, such as the kurtosis, are generally needed to reveal information on the clustering (intermittency) of the QWs.

The packing fraction can also be related to $\overline{M}$ and $\overline{\tau}$. From Eqs.~\ref{eq:N} and \ref{eq:Ncasc}, we have
\begin{equation}
\overline{\phi}=\overline{M}\overline{\tau}\ell^D,
\label{def:phiCasc}
\end{equation}
The effect of increases in $\overline{M}$ on the packing fraction can thus be compensated by decreases in $\overline{\tau}$, or vice versa. Considering the three-dimensional case, for example, we would have the same sequence of packing fractions when $\overline{\tau}=1$ and there are 8 children per parent, as we would when $\overline{\tau}=1/2$ and there are 16 children per parent. Given a single snapshot of the random medium, \emph{we cannot distinguish between the compensating effects of the decay time and the number of children per parent}. Presumably, a sequence of correlated snapshots would be needed to distinguish these effects.

\subsubsection{Mass Conservation}

Let us next consider application of conservation principles to the one-way cascade model. First, we consider a process which conserves mass from one generation to the next. This might be appropriate, for example, for rocks that break down into smaller ones. Assuming all of the QWs have the same mass density, where the density $\rho$ is defined as mass per unit volume (where the volume is $D$-dimensional), it follows that the $m_i\propto\rho a_i^D$, where $m_i$, the mass of QWs in size class $i$, is a conserved quantity. The assumption that the density does not change from one generation to the next eliminates one free parameter: $q$ must equal one (indicating that $q_i$ is independent of $i$), or, equivalently, $\lambda$ must be zero. Defining $q_i=\sqrt{\rho}$, we have $m_i\propto q_i^2 a_i^D$.\footnote{Setting $q_i$ to $\sqrt{\rho}$, instead of, say, $1$ and $\rho$, has no impact on the subsequent derivation, but results in a convenient parallel to an energy-conserving cascade, as will be discussed shortly.} Hence ${m_{i+1}}/{m_i}=\ell^D$. Since $M_i$ parents of mass $m_i$ produce $M_{i+1}$ offspring of mass $m_{i+1}$, and conservation of mass implies $M_i m_i=M_{i+1} m_{i+1}$, we must also have ${m_{i+1}}/{m_i}=\overline{M}^{-1}$. Thus
\begin{equation}
\overline{M}=\ell^{-D}.
\label{eq:Mgenp}
\end{equation}
Substitution into Eq.~\ref{def:phiCasc} implies
\begin{equation}
\overline{\phi}=\overline{\tau}.
\end{equation}

If the cascade is to be steady, it must also be the case that the \emph{rate} of mass transferred into size class $i$ equals the rate of mass transferred out of size class $i$. From the definitions in Sec.~\ref{sec:general}, the rate of mass transfer out of $i$ is $N_i m_i/\tau_i$. The rate of transfer into $i$ must equal the rate of transfer out of $i-1$, which is $N_{i-1} m_{i-1}/\tau_{i-1}$. Setting $N_{i-1} m_{i-1}/\tau_{i-1}=N_i m_i/\tau_i$, we have
\begin{equation}
\overline{N}=\frac{m_{i-1}}{m_i}\frac{\tau_i}{\tau_{i-1}}=\ell^{-D}\overline{\tau}=\overline{M}\overline{\tau}.
\end{equation}
This result agrees with Eq.~\ref{eq:Ncasc}, which is to be expected.

\subsubsection{Energy Conservation}

We next consider an energy-conserving process. For many phenomena of interest, the square of the amplitude is proportional to energy, or to another quantity which must be conserved by the cascade process. For example, if the amplitude represents velocity fluctuations in a turbulent flow, the specific kinetic energy would be proportional to velocity squared. The total energy associated with a QW of size class $i$, $E_i$, must be proportional to the amplitude squared times its volume: 
\begin{equation}
E_i\propto a_i^D q_i^2. 
\label{def:e}
\end{equation}
Defining $\overline{E}={E_{i+1}}/{E_i}$ as the ratio of the energy for individual QWs of one generation to the previous, we thus have
\begin{equation}
\overline{E}=\ell^D\overline{q}^2=\ell^{D+2\lambda}.
\label{eq:e}
\end{equation}
The spatial energy density (energy per unit volume) in a size class, $\mathcal{E}_i$, equals the number density times the energy associated with an individual QW, i.e., $\mathcal{E}_i=\mathcal{N}_i E_i$. Defining $\overline{\mathcal{E}}$ as the ratio of the energy per unit volume from one generation to the previous, we have, from Eqs.~\ref{eq:N} and \ref{eq:e},
\begin{equation}
\overline{\mathcal{E}}=\overline{N}\overline{E}=\overline{\phi}\overline{q}^2=\ell^{\beta+2\lambda}.
\label{eq:E}
\end{equation}
For $\ell=1/2$, we have simply
\begin{equation}
\overline{E}=2^{-D-2\lambda}.
\label{eq:e2}
\end{equation}
and
\begin{equation}
\overline{\mathcal{E}}=2^{-\beta-2\lambda}.
\label{eq:E2}
\end{equation}

The value of any conserved quantity for a set of parent QWs must equal the total for the offspring. Since $M_i$ QWs of the generation $i$, each with energy $E_i$, produce $M_{i+1}=\overline{M}M_i$ offspring of the generation $i+1$, each with energy $E_{i+1}$, we must have  
\begin{equation}
\overline{E}=\overline{M}^{-1}.
\label{eq:ecasc}
\end{equation}
From Eq.~\ref{eq:e}, we then find
\begin{equation}
\overline{M}=\ell^{-D}\overline{q}^{-2}=\ell^{-D-2\lambda}.
\label{eq:Mgen}
\end{equation}
This result is a generalization of Eq.~\ref{eq:Mgenp}, which was specifically for the case $\overline{q}=1$. Now, by substituting Eq.~\ref{eq:ecasc} into Eq.~\ref{eq:E}, we find 
\begin{equation}
\overline{\mathcal{E}}=\overline{N}/\overline{M}.
\end{equation} 
By comparison with Eq.~\ref{eq:Ncasc}, we have the simple result
\begin{equation}
\overline{\tau}=\overline{E}=\overline{\phi}\overline{q}^2.
\label{def:phiGen}
\end{equation}
Interestingly, by assuming that $q_i^2$ is proportional to energy, and that energy is conserved, the ratios $\overline{M}$ and $\overline{\tau}$ turn out to have a fixed dependence on the basic cascade ratios $\ell$, $\overline{\phi}$, and $\overline{q}$. Thus, the energy-conserving steady-state model has no more free parameters than the static model.

Another constraint on the energy-conserving steady-state model is that $E_i$ should remain \emph{steady} for all size classes. As the size class $i$ decays into class $i+1$, energy is transferred at a net rate (per unit volume) of $E_i/\tau_{i}$. But the relationship $\overline{E}=\overline{\tau}$ (Eq.~\ref{def:phiGen}) implies that $E_i/\tau_{i}$ is independent of $i$. Thus, when energy is conserved by individual reactions in the cascade, energy also remains steady within each size class. 

Let us return to the example of a turbulent velocity field from Sec.~\ref{sec:turb}. By dimensional arguments, the velocity fluctuation $q\to \upsilon$ must be proportional to $a_i/\tau_i$. Hence $\upsilon_{i+1}/\upsilon_i=(a_{i+1}/a_i)/(\tau_{i+1}/\tau_i)$, and we have 
\begin{equation}
\overline{\upsilon}=\frac{\ell}{\overline{\tau}}.
\label{def:qCasc}
\end{equation}
Substituting for $\overline{\tau}$ with Eq.~\ref{def:phiGen}, we have $\overline{q}^3=\ell/\overline{\phi}$, in agreement with Eq.~\ref{eq:lambdaTurb} from the static model. Furthermore, from Eq.~\ref{eq:Mgen}, we now have a relationship between $\overline{M}$ and $\overline{\tau}$:
\begin{equation}
\overline{M}=\ell^{-D-2}\overline{\tau}^2.
\label{eq:Meta}
\end{equation}

For a conserved scalar in turbulence, Eqs.~\ref{def:e}--\ref{def:phiGen} all apply, with $q\to c$ now representing the amplitude of the scalar. In particular, Eq.~\ref{eq:Mgen} implies that $\overline{M}=\ell^{-D}\overline{c}^{-2}$. Since $\overline{M}$ is constrained by Eq.~\ref{eq:Meta}, in order to conserve TKE, we must have $\ell^{-D}\overline{c}^{-2}=\ell^{-D-2}\overline{\tau}^2$ and hence $\overline{c}=\ell/\overline{\tau}=\overline{\upsilon}$.

\subsubsection{Number of Offspring}
\label{sec:steadydense}

For the model described in Sec.~\ref{sec:staticmodel}, the number of QWs in each size class followed a Poisson distribution with mean $N_i$ (i.e., Eq.~\ref{eq:Poisson}). We would like to formulate the steady-state cascade model such that the distributions are unchanged from the static model. 

Suppose, as a starting point, that $n_i$ QWs in the size class $i$ are present. Next, suppose that each of the $n_i$ parent QWs decays into a number of offspring. The simplest approach that might be considered is to assume that the number of offspring is fixed. A total of $n_{i+1}=n_i \overline{M}$ offspring must be produced, meaning that each reaction must produce exactly $\overline{M}$ offspring. However, this decay process is realizable only if $\overline{M}$ is an integer, which should not be assumed, in general. Another, perhaps more fundamental, issue with this formulation is that, even if $n_i$ has a Poisson distribution, it does not follow that $n_{i+1}$, has a Poisson distribution. This can be readily deduced from the fact that $n_{i+1}$ can only equal multiples of $\overline{M}$, rather than any integer value.

An alternative is to assume that the number of offspring from each parent satisfies a Poisson distribution with mean $\overline{M}$. The total number of offspring is thus given by the sum of $n_i$ Poisson-distributed random variables, each with mean $\overline{M}$. Let us approximate this sum as itself being a Poisson-distributed random variable, with mean $n_i \overline{M}$. The expected value of the number of offspring is $\langle n_{i+1}\rangle=\langle n_i\rangle \overline{M}=N_i\overline{M}$. Hence, if we initialize the cascade process with QW counts drawn from a Poisson distribution, subsequent generations will also have size counts satisfying Poisson distributions, with the desired mean values. 

Note that the procedure described in the preceding paragraph does \emph{not} enforce conservation of energy for each individual decay of a parent into offspring. Conservation is enforced only in an average sense. Apparently, it is not possible to enforce conservation on individual decay events unless $\overline{M}$ happens to be an integer. The viewpoint taken here is that the model is formulated only to be a self-similar, stochastic process. No claim is made that the construction method captures the actual physics underlying the cascade process. For example, in many real cascade processes, multiple parents might interact to produce the offspring. This is true in turbulence, for example, where multiple interacting eddies produce localized regions of high shear, which in turn produce new  eddies. Thus there is no one-to-one correspondence between a parent and offspring. The goal of the modeling here is, rather, to mimic the random \emph{geometry} produced by such cascade processes, while not violating any applicable conservation laws in a larger sense. Other modeling choices than those made here are, of course, possible, and might be appropriate depending on the goals of the modeling.

As described in Sec.~\ref{sec:densify}, we may wish to create a more continuous range of QW sizes than dictated by the choice of $\ell$, the ratio of length scales between generations in the cascade. For this purpose, the number of size classes is increased by a factor of $K$ by introducing non-integer size classes. While the ratio of length scales between adjacent size classes then becomes $\ell^{1/K}$, generations are still separated by whole integers of the size class index and have length scale ratios $\ell$. If we seed the cascade process for the largest size class at a rate $\mathcal{R}_1$ with QWs of size $a_1$, the next generation will consist of QWs with size $a_2=\ell a_1$ produced at a rate $\mathcal{R}_1\overline{M}$, followed by QWs with size $a_3=\ell^2 a_1$ produced at a rate  $\mathcal{R}_1\overline{M}^2$, etc. Note that, in order to leave the packing fraction unchanged relative to the case $K=1$, the rate $\mathcal{R}_1$ must be a factor of $1/K$ less. To populate the non-integer size classes, we must introduce QWs into the non-integer classes $i=1+1/K,1+2/K,\ldots 1+(K-1)/K$ in the correct proportions. Specifically, the size class $1+1/K$ must be seeded at a rate $\mathcal{R}_1\overline{M}^{1/K}$, the size class $1+2/K$ at a rate of $\mathcal{R}_1\overline{M}^{2/K}$, etc., up to size class $1+(K-1)/K$ at a rate of $\mathcal{R}_1\overline{M}^{(K-1)/K}$. These classes will in turn decay into the size classes $i=2+1/K,2+2/K,\ldots 2+(K-1)/K$, and so forth. This process ensures that all scales in the cascade, including the non-integer classes, are represented in the correct proportions. Hereafter we will refer to the size classes $i=1,1+1/K,1+2/K,\ldots 1+(K-1)/K$ as the \emph{production classes}, since these are the ones that are used to seed the cascade process.

\subsubsection{Placement of Offspring}
\label{sec:placement}

Having devised a procedure for determining the desired number of offspring from a parent, we now discuss the spatial placement of the offspring relative to the parent. Assuming the process is self similar, we can define a probability density function $g(\xi)$ which describes the probability of a QW of size $a_{i+1}$ being placed at a normalized distance $\xi=|\mathbf{r}-\mathbf{b}^{in}|/a_i$ from the parent of size $a_i$ at location $\mathbf{b}^{in}$. This formulation assumes that the placement of each child QW is independent of the others, which could be a significant idealization of the cascade processes.

One might assume, for example, that $g(\xi)$ is a normal distribution:
\begin{equation}
g(\xi)=\sqrt{\frac{2}{\pi\sigma^2}}\exp\left(-\frac{\xi^2}{2\sigma^2} \right),
\label{eq:GaussPlace}
\end{equation}
where $\sigma$ controls the dispersion of the offspring relative to the parent. (Note that $\xi$ is defined to be positive, so the usual normal distribution is multiplied by 2.) This selection for $g(\xi)$ assumes that there is a higher probability that the offspring will be placed near to the center of the parent. Alternatively, it might be assumed that there is a higher probability that the offspring will be placed around the edges of the parent QW, where the derivative of the envelope has the highest magnitude. In this case, we could choose a chi-square model, 
\begin{equation}
g(\xi)=\frac{(\xi/s^2)^{\nu/2-1}}{2^{\nu/2}s^2\Gamma(\nu/2)}\exp\left(-\frac{\xi}{2s^2} \right),
\label{eq:Chi2Place}
\end{equation}
where $s$ is a dispersion parameter and $\nu$ is the degrees of freedom. In this report, we will always set $\nu=4$, which yields a density with a broad peak around $\xi=2$.

\begin{figure}[tbh]
\centering
\includegraphics[width=\linewidth,natwidth=1200,natheight=900]{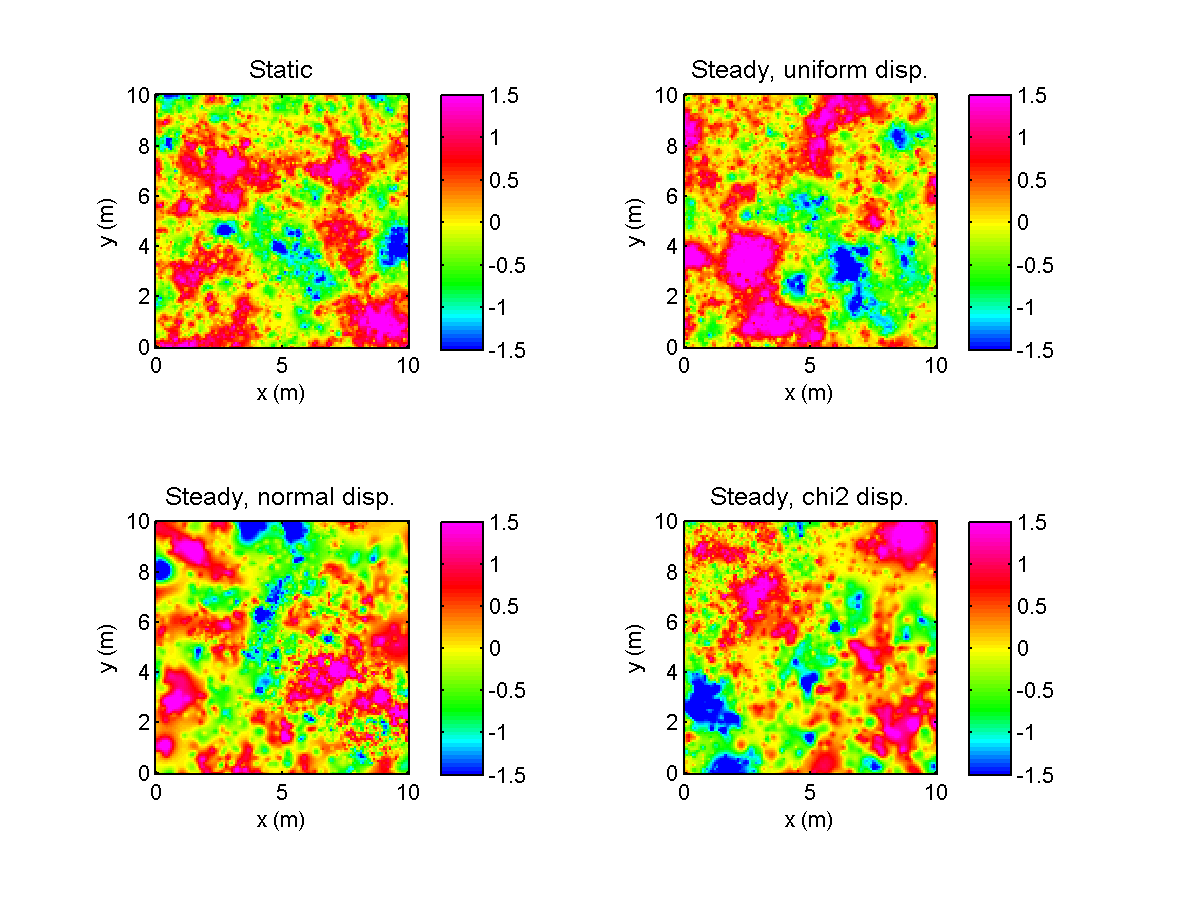}
\caption{Upper left is a realizations of a static QW field for an energy-conserving process. The remaining three figures are steady-state cascade QW fields created from the same parameters with uniform dispersion (upper right), a normal dispersion model (lower left), and a $\chi^2$-dispersion model (lower right).}
\label{fig:SteadyState}
\end{figure}

\begin{figure}[tbh]
\centering
\includegraphics[width=\linewidth,natwidth=1200,natheight=900]{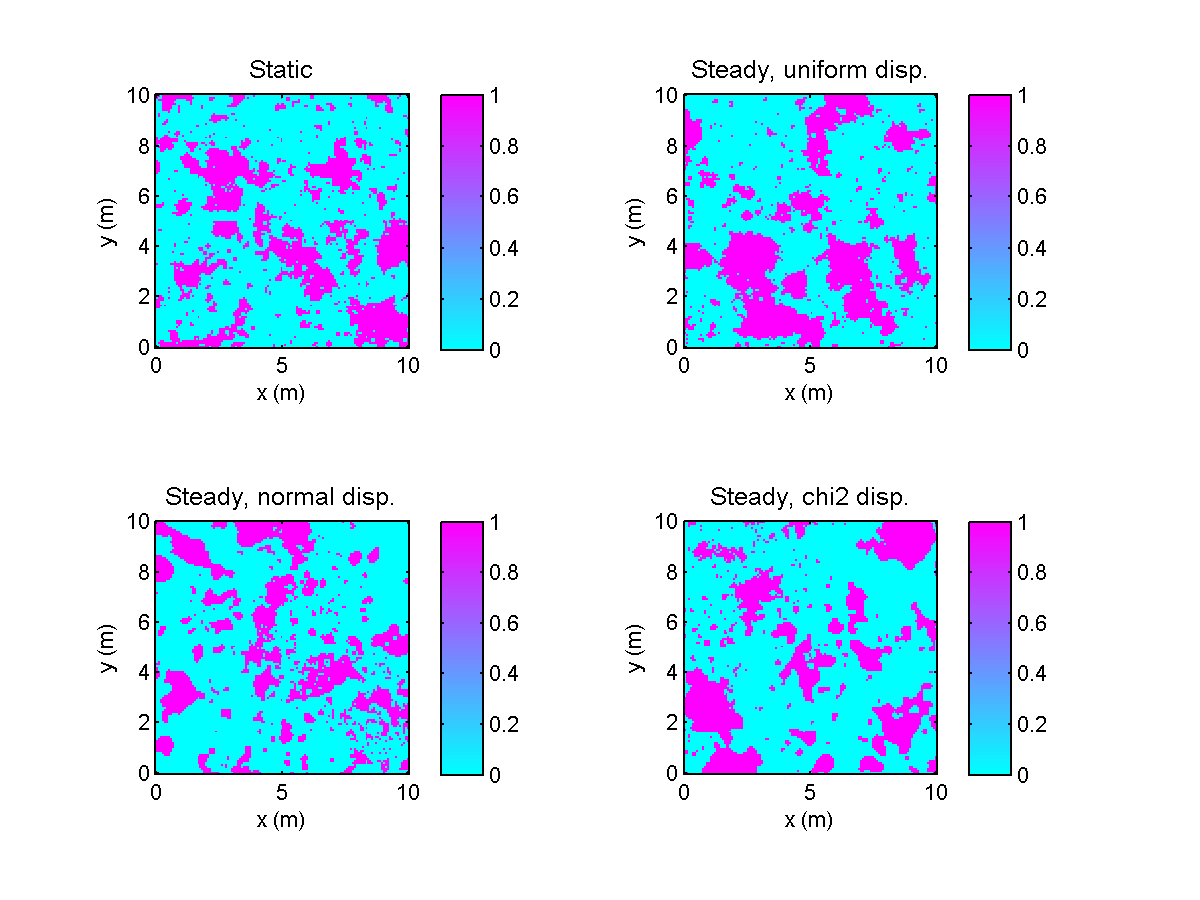}
\caption{Same as Fig.~\ref{fig:SteadyState}, except that level cuts through the fields at $\pm 0.7$ are shown.}
\label{fig:SteadyStateCut}
\end{figure}

Figures~\ref{fig:SteadyState} and \ref{fig:SteadyStateCut} illustrate the construction of steady-state cascade QW fields with varying dispersion models: uniform dispersion over the entire realization volume (i.e., no clustering relative to the parents), normal dispersion (clustering near the center of the parent), as given by Eq.~\ref{eq:GaussPlace} with $\sigma=4$, and chi-square dispersion (clustering around the edge of the parent), as given by Eq.~\ref{eq:Chi2Place} with $\nu=4$ and $s=2$. Also shown for reference is a static QW realization for the same parameters, which in principle is statistically indistinguishable from the steady-state cascade with uniform dispersion. All realizations were calculated with $\ell=1/2$, $\beta=0$, $\lambda=1/3$, $a_1=1$ m, $\phi_1=0.1$, $q_1=1$, and $\tau_1=1$. The cascades are calculated through five generations. The main observation to be made from these figures is how the normal and chi-square dispersion models lead to realistic intrinsic intermittency and sorting by size, as observed in Fig.~\ref{fig:amboypic2} and other images in the Introduction.
 
A potentially interesting issue is how the area or volume occupied by the cascade expands (or contracts) over time. The centroid of the cascade will also experience a random walk. We define these quantities based on the conservative quantity in the cascade (mass or energy). The centroid at the $i$th generation of the $j$th cascade is thus
\begin{equation}
\mathbf{x}(i,j) = 
\frac{\int_V \mathbf{r} Q^2_{(i,j)}\left(  \mathbf{r}\right)\,dV}
{\int_V Q^2_{(i,j)}\left(  \mathbf{r}\right)\,dV},
\end{equation}
whereas the dispersion relative to the centroid is
\begin{equation}
\mathbf{s}^2(i,j) = 
\frac{\int_V |\mathbf{r}-\mathbf{x}(i,j)|^2 Q^2_{(i,j)}\left(  \mathbf{r}\right)\,dV}
{\int_V Q^2_{(i,j)}\left(  \mathbf{r}\right)\,dV},
\end{equation}
where
\begin{equation}
Q^2_{(i,j)}\left(  \mathbf{r}\right)  =\sum_{n\in j}
q_i^2 f^2\left(  \frac
{\left\vert \mathbf{r}-\mathbf{b}^{i n}\right\vert }{a_{i}}\right)
\end{equation}

Presumably, statistics for the centroid and dispersion could be derived for a given $g(\xi)$, although that is not attempted here. The drift of the centroid and the dispersion would be characterized on a \emph{per generation} basis.

\subsection{Two-Way Energy-Conserving Model}

\begin{figure}[tbph]
\centering
\includegraphics[width=0.4\linewidth,natwidth=455,natheight=556]{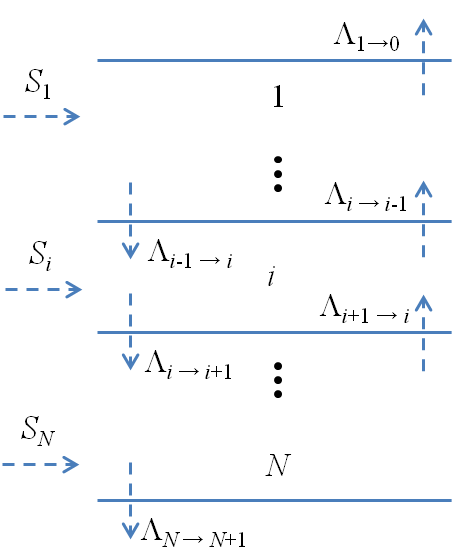}
\caption{Energy transfer between QW size classes.}
\label{fig:energyflow}
\end{figure}

Consider now a cascade process which transfers energy from large QWs to smaller ones, as well as vice versa. The former process is termed \emph{decay} or \emph{forward scatter}, whereas the latter process is termed \emph{backscatter} \citep{kim2008wavelet}. Let us define $\Lambda_{i\rightarrow j}$ as the energy flux from generation (size class) $i$ to $j$. (In this section, it will be assumed that there is no scale densification, i.e., the size classes are integers.) A steady-state model must balance the energy being transferred into the size class with that going out. Assuming that transfers occur only between adjacent size classes, the following equation describes the flux balance for class $i$:
\begin{equation}
\Lambda_{i\rightarrow i-1}+\Lambda_{i\rightarrow i+1}=\Lambda_{i-1\rightarrow i}+\Lambda_{i+1\rightarrow i}+S_i.
\label{eq:balance}
\end{equation}
The terms on the left side, $\Lambda_{i\rightarrow}\equiv\Lambda_{i\rightarrow i-1}+\Lambda_{i\rightarrow i+1}$, represent transfer \emph{from} class $i$ to larger and smaller scales (back and forward scatter), respectively. The terms on the right, $\Lambda_{\rightarrow i}\equiv\Lambda_{i-1\rightarrow i}+\Lambda_{i+1\rightarrow i}$, represent transfer \emph{to} class $i$ from larger and smaller scales (forward and back scatter), respectively. Also present is a source term $S_i$, which represents the creation of QWs of size $a_i$ by an external process. This situation is depicted graphically in Fig.~\ref{fig:energyflow}. The previous one-way cascade model corresponds to a per-volume energy production rate at the largest scale of $S_1=\mathcal{N}_1 E_1/\tau_1=\mathcal{R}_1 E_1$, whereas $S_i=0$ for $i\ne 1$. The backscatter terms, $\Lambda_{i\rightarrow i-1}$, are all zero.

Of particular interest is the \emph{net} energy transfer between two adjacent size classes. The net transfer from class $i$ to $i+1$ is
\begin{equation}
\Delta\Lambda_{i}=\Lambda_{i\rightarrow i+1}-\Lambda_{i+1\rightarrow i}.
\end{equation}
From Eq.~\ref{eq:balance}, we have
\begin{equation}
\Delta\Lambda_{i}=\Delta\Lambda_{i-1}+S_i.
\label{eq:balance2}
\end{equation}
Since it is assumed there is no transfer from unresolved scales, we define $\Delta\Lambda_0=-\Lambda_{1\rightarrow 0}$ and $\Delta\Lambda_N=\Lambda_{N\rightarrow N+1}$. In the absence of a source, the net transfers to and from each size class must balance. Consider the situation when there is the only non-zero source term, for $i=j$. Then $\Delta\Lambda_{j}=\Delta\Lambda_{j-1}+S_j$, $\Delta\Lambda_0=\Delta\Lambda_1=\cdots=\Delta\Lambda_{j-1}$, and $\Delta\Lambda_N=\Delta\Lambda_{N-1}=\cdots=\Delta\Lambda_{j}$. The case $j=1$ corresponds to Kolmogorov's hypothesis; that is, all of the $\Delta\Lambda_{i}$ must equal the source term $S_1$.

Let us now assume that a certain fraction $f$ of the total energy transferred into the size class is always subsequently transferred to smaller scales, i.e., forward scattered. Such an assumption implies that the energy cascade is scale invariant. Notationally,
\begin{equation}
\Lambda_{i\rightarrow i+1}=f\left(\Lambda_{\rightarrow i}+S_i\right),\ i=1,\cdots,N.
\label{eq:Lambda1}
\end{equation}
The remaining fraction, $1-f$, is transferred to larger scales (the backscatter). 
\begin{equation}
\Lambda_{i\rightarrow i-1}=(1-f)\left(\Lambda_{\rightarrow i}+S_i\right),\ i=1,\cdots,N.
\label{eq:Lambda2}
\end{equation}
Note that Eq.~\ref{eq:balance} is recovered when Eq.~\ref{eq:Lambda1} is added to Eq.~\ref{eq:Lambda2}. Eqs.~\ref{eq:Lambda1} and~\ref{eq:Lambda2} together provide a system of $2N$ linear equations that can be solved by standard matrix techniques. (The fluxes $\Lambda_{0\rightarrow 1}$ and $\Lambda_{N+1\rightarrow N}$ are set to 0.)

\begin{figure}[tbph]
\centering
\includegraphics[width=0.95\linewidth,natwidth=1200,natheight=900]{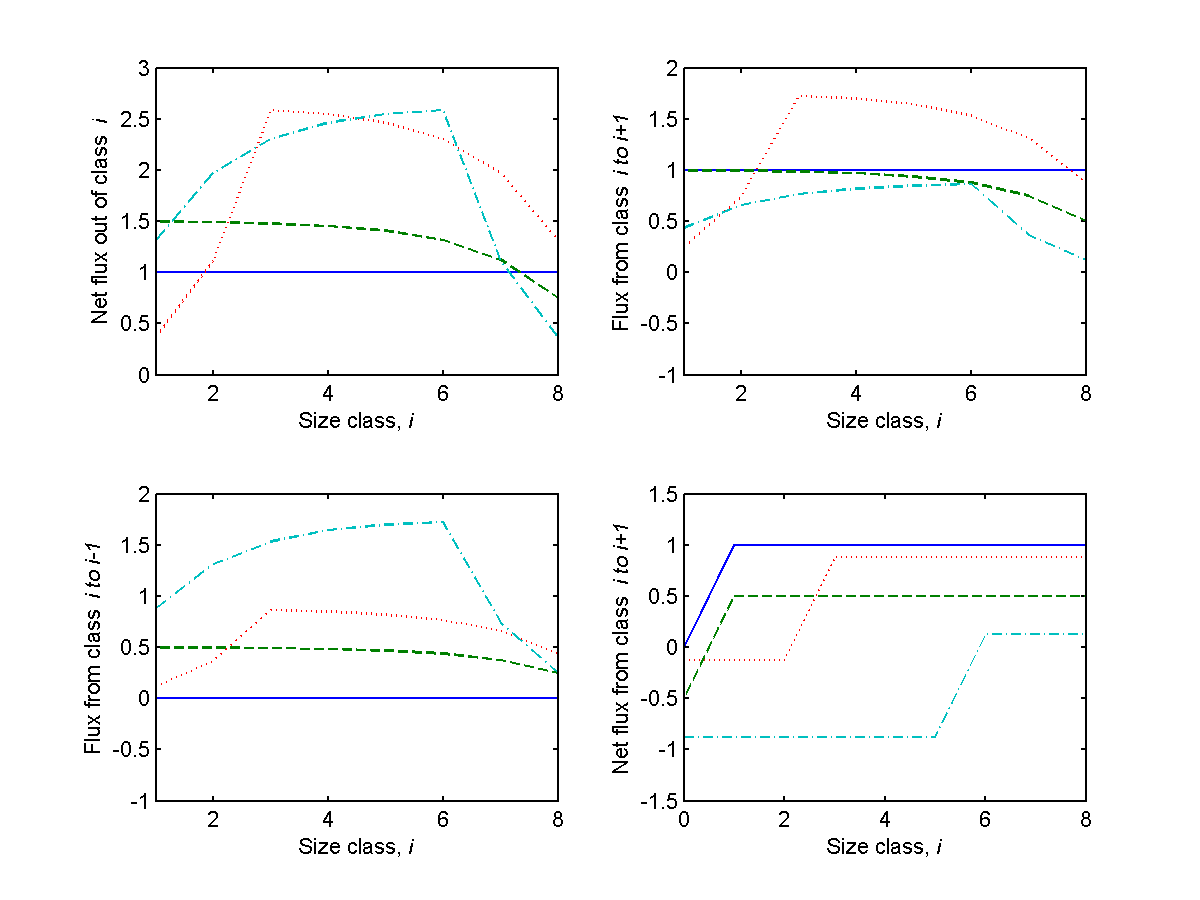}
\caption{Energy flux calculations for various 8-generation cascade processes. Solid line is a cascade with production in the largest size class, and only forward scatter ($f=1$). Dashed line is a cascade with production in the largest size class, and predominantly forward scatter ($f=2/3$). Dotted line is a cascade with production in the third largest size class, and predominantly forward scatter ($f=2/3$). Dash-dotted line is a cascade with production in the third smallest size class, and predominantly back scatter ($f=1/3$). The figures show $\Lambda_{i\rightarrow}$ (upper left), $\Lambda_{i\rightarrow i+1}$ (upper right), $\Lambda_{i\rightarrow i-1}$ (lower left), $\Delta\Lambda_{i}$ (lower right).}
\label{fig:cascades}
\end{figure}

Figure~\ref{fig:cascades} illustrates the energy flux for various cascade processes with eight generations ($N=8$) and $S_j=1$ for various values of $j$ and $f$. Two of the processes have production in the largest size class ($j=1$). One of these has only forward scatter ($f=1$); the other has mostly forward scatter ($f=2/3$). The other two processes are mirror images of each other. One has production at the third-largest size class ($j=3$) and predominantly forward scatter ($f=2/3$), while the other has production at the third-smallest size class ($j=6$) and predominantly back scatter ($f=1/3$). 

Calculation of the number densities of QWs is actually quite straight forward once solutions of Eqs.~\ref{eq:Lambda1} and~\ref{eq:Lambda2}, such as those shown in Fig.~\ref{fig:cascades}, are available. The energy loss rate for size class $i$, $\Lambda_{i\rightarrow}$, equals $\mathcal{N}_i E_i/\tau_i$, where $\mathcal{N}_i$ is the total number of QWs (per unit volume) that are present in size class $i$. We thus have $\mathcal{N}_i=\Lambda_{i\rightarrow}\tau_i/E_i$. Since all of the $\Lambda$'s are proportional to $S_j$ (when there is only one non-zero value of $S_j$), if desired, we could solve the equations for $S_j=1$, and then rescale the $\Lambda$'s by the actual value of $S_j$. Note that $S_j=\mathcal{R}_j E_j$. It should be kept in mind that $\mathcal{N}_j$ does not necessarily equal $\mathcal{R}_j \tau_j$, though; this would be the case only when $\Lambda_{\rightarrow j}=0$, so that the source term is exactly balanced by the transfer out of the size class.

\section{Second-Order Statistics for the Static and Steady-State Models}
\label{sec:stat}

As described in Sec.~\ref{sec:general}, second-order statistics, such as correlation functions and spectra, do not provide information on the clustering of objects. Single realizations of the static and steady-state models are indistinguishable on the basis of these statistics. Nonetheless, second-order statistics can provide very useful information on the spatial scales and distribution of object sizes in an RHM. Several previous papers \citep{goedecke2004quasi,goedecke2006quasi,wilson2009quasi,wilson2008sound} have derived correlation functions and spectra for the QW model. Here, we summarize some of these previous results. The appearance of the equations is somewhat different from the original references, due to changes in notation and normalization conventions.

\subsection{General Results for Spectra and Correlations}

The spectrum may be defined as \citep{bendat2011random}:
\begin{equation}
\Phi\left(\mathbf{k}\right)  =\lim_{V\rightarrow\infty}\frac{\left(
2\pi\right)^{D}}{V}\left\langle \left\vert \widetilde{Q}\left(
\mathbf{k}\right)\right\vert ^{2}\right\rangle , \label{def:specT}%
\end{equation}
where $\mathbf{k}$ is the vector wavenumber and $\widetilde{Q}\left(\mathbf{k}\right)$ is the Fourier transform of the field, defined as%
\begin{equation}
\widetilde{Q}\left(\mathbf{k}\right)=\frac{1}{\left(2\pi\right)^{D}%
}\int d^{D}r\,Q\left(\mathbf{r}\right) e^{-i\mathbf{k}\cdot\mathbf{r}}.
\label{eq:Tft}%
\end{equation}
Eq.~\ref{eq:decomp} implies that the spectra of the individual QWs add linearly to produce the total spectrum:
\begin{equation}
\widetilde{Q}\left(\mathbf{r}\right)=\sum_{i=1}^{I}\sum_{j=1}^{N_i}
\widetilde{Q}^{ij}\left(\mathbf{k}\right). 
\label{eq:decomp2}%
\end{equation}
Transformation of Eq.~(\ref{eq:QWi}) yields the spectrum for an individual QW:
\begin{equation}
\widetilde{Q}^{ij}\left(\mathbf{k}\right)  =a_i^{D}q_i\exp\left(  -i\mathbf{k}\cdot\mathbf{b}^{ij}\right)  F\left(ka_i\right), 
\label{eq:Tft2}%
\end{equation}
where%
\begin{equation}
F\left(y\right)=\frac{1}{\left(2\pi\right)^{D}}\int d^{D}\xi\,f\left(\xi\right)  e^{-i\mathbf{y}\cdot\boldsymbol{\xi}} 
\label{eq:Fft}%
\end{equation}
is the spectrum of the parent function and $\mathbf{y}=\mathbf{k}a$ is the normalized wavenumber. For the Gaussian parent function, Eq.~\ref{eq:parent_Gauss}, we find
\begin{equation}
F\left(y\right)=\left(\sqrt{2}\pi\right)^{-D}\exp\left(-y^{2}/2\pi\right)
\label{eq:parent_Gauss_xfrm}
\end{equation}
By Parseval's theorem, the normalization condition, Eq.~\ref{eq:norm}, requires
\begin{equation}
(2\pi)^{D}\int d^{D}y\,{F^2\left(y\right)}=1. 
\label{eq:norm2}%
\end{equation}

Substituting Eq.~(\ref{eq:decomp2}) into (\ref{def:specT}), and assuming statistical independence of the individual QWs, yields
\begin{equation}
\Phi\left(\mathbf{k}\right)=\sum_{i=1}^{N}\Phi_{i}\left(\mathbf{k}\right), 
\label{eq:QWspec}%
\end{equation}
where%
\begin{equation}
\Phi_{i}\left(k\right)=\left(2\pi\right)^{D}\frac{N_i}{V}q_{i}^{2}a_{i}^{2D}F^{2}\left(ka_{i}\right)=\frac{\left(2\pi\right)^{D}}{K}\phi_i q_{i}^{2}a_{i}^{D}F^{2}\left(ka_{i}\right) . 
\label{eq:specalpha}
\end{equation}
is the contribution to the spectrum from size class $i$. By integrating over wavenumber, we find the contribution to the variance from size class $i$:
\begin{equation}
\sigma_{i}^{2}=\int d^{D}k\,\Phi_{i}\left(k\right)=\frac{N_i}{V}q_{i}^{2}a_{i}^{D}=\frac{1}{K}\phi_i q_{i}^{2}.
\end{equation}

The correlation function of the random field is also of interest. In a homogeneous, isotropic medium, as considered here, it can be defined as
\begin{equation}
B\left(\mathbf{r}\right)=\left\langle Q\left(\mathbf{r}_0\right)Q\left(\mathbf{r}_0+\mathbf{r}\right)\right\rangle , 
\label{def:corrT}
\end{equation}
where $\mathbf{r}_0$ is a reference observation point. The correlation function can be shown to be the Fourier transform of the spectrum, namely
\begin{equation}
B\left(\mathbf{r}\right)=\int d^{D}k\,\Phi\left(\mathbf{k}\right)  e^{i\mathbf{k}\cdot\mathbf{r}}.
\end{equation}
Since the Fourier transform is a linear operator, we can decompose $B\left(r\right)$ by size class as in Eq.~\ref{eq:QWspec}, with result
\begin{equation}
B\left(\mathbf{r}\right)=\sum_{i=1}^{N}B_{i}\left(\mathbf{r}\right), 
\label{eq:QWcorr}
\end{equation}
where, from Eq.~\ref{eq:specalpha},
\begin{equation}
B_{i}\left(r\right)=\frac{\phi_i q_{i}^{2}}{K} f_2\left(\frac{r}{a_i}\right), 
\label{eq:corrlpha}
\end{equation}
in which we have \emph{defined}
\begin{equation}
f_2(\xi)=\left(2\pi \right)^{D}\int d^{D}y\,F^{2}\left(y\right)  e^{i\mathbf{y}\cdot\boldsymbol{\xi}}.
\end{equation}
The normalization condition, Eq.~\ref{eq:norm2}, implies that $f_2(0)=1$. Hence $B_i(0)=\phi_i q_{i}^{2}/K$, which equals $\sigma_i^2$, as it should. For the Gaussian parent function, substitution with Eq.~\ref{eq:parent_Gauss_xfrm} and the assistance of a standard table of Fourier transforms yields
\begin{equation}
f_2(\xi)=\exp\left({-\frac{\pi\xi^2}{4}}\right).
\label{eq:parent_Gauss_f2}
\end{equation}

The overall spectrum, correlation function, and variance are found by summing over the size classes. In the limit of a highly densified representation ($K\to\infty$), the summation over classes transforms to an integration. Such an integral representation, with the QW size as the integration variable, can be derived as follows. Since $a_i=a_1\ell^{i-1}$, we have $a_{i+\Delta i}=a_1 \ell^{i+\Delta i-1}$, where $i+\Delta i=i+1/K$ is the size class following $i$. We then find 
\begin{equation}
\frac{a_{i+\Delta i}}{a_i}=\ell^{1/K}.
\end{equation} 
Defining $a_{i+1}=a_i+\Delta a_i$, it follows that $1+\Delta a_i/a_i=\ell^{1/K}$. Taking the logarithm of both sides and assuming $\Delta a_i\ll a_i$, we finally have 
\begin{equation}
\Delta a_i\simeq\frac{a_i}{K}\ln\ell. 
\end{equation}
We then find, through substitution of Eq.~\ref{def:pack},
\begin{equation}
\Phi\left(k\right)=\sum_{\mathrm{classes}}\Phi_{i}\left(k\right)=
\frac{\left(2\pi\right)^{D}}{\ln\ell}\sum_{\mathrm{classes}}\phi_i q_{i}^{2}
F^{2}\left(ka_{i}\right)a_{i}^{D-1}\Delta a_i,
\label{eq:Phiclass}
\end{equation}
\begin{equation}
B\left(r\right)=\sum_{\mathrm{classes}}B_{i}\left(r\right)=
\frac{1}{\ln\ell}\sum_{\mathrm{classes}}\phi_i q_{i}^{2}
f_2\left(\frac{r}{a_{i}}\right)\frac{\Delta a_i}{a_{i}},
\label{eq:Bclass}
\end{equation}
and
\begin{equation}
\sigma^2=\sum_{\mathrm{classes}}\sigma_{i}^{2}=
\frac{1}{\ln\ell}\sum_{\mathrm{classes}}\phi_i q_{i}^{2}\frac{\Delta a_i}{a_{i}}.
\end{equation}
In these equations, $\phi_i$ and $q_i$ are implicitly functions of $a_i$. The desired relationships follow from Eqs.~\ref{def:beta} and \ref{def:lambda}:
\begin{equation}
\phi_i=\phi_1\left(\frac{a_i}{a_1}\right)^{\beta},
\label{eq:phii}
\end{equation}
and
\begin{equation}
q_i=q_1\left(\frac{a_i}{a_1}\right)^{\lambda}.
\label{eq:qi}
\end{equation}
Hence we have for the spectrum, correlation function, and variance, respectively,
\begin{equation}
\Phi\left(k\right)=
\frac{\left(2\pi\right)^{D}\phi_1 q_1^2}{a_1^{\beta+2\lambda}\ln\ell} 
\sum_{\mathrm{classes}} F^{2}\left(ka_{i}\right)a_{i}^{\beta+2\lambda+D-1}\Delta a_i,
\label{eq:tempSpec}
\end{equation}
\begin{equation}
B\left(r\right)=
\frac{\phi_1 q_1^2}{a_1^{\beta+2\lambda}\ln\ell} 
\sum_{\mathrm{classes}} f_2\left(\frac{r}{a_{i}}\right) a_{i}^{\beta+2\lambda-1}\Delta a_i,
\label{eq:tempCorr}
\end{equation}
and
\begin{equation}
\sigma^2=
\frac{\phi_1 q_1^2}{ a_1^{\beta+2\lambda}\ln\ell}
\sum_{\mathrm{classes}} a_{i}^{\beta+2\lambda-1}\Delta a_i.
\label{eq:tempVar}
\end{equation}
Setting $y=k a_i$ in Eqs.~\ref{eq:tempSpec} and \ref{eq:tempVar}, and taking the limit $\Delta y\rightarrow 0$, results in
\begin{equation}
\Phi\left(k\right)=
\frac{\left(2\pi a_1\right)^{D}\phi_1 q_1^2}{\left(ka_1\right)^{D+\beta+2\lambda}\ln\ell} 
\int_{ka_1}^{ka_N} F^{2}\left(y\right) y^{\beta+2\lambda+D-1}\,dy
\label{eq:specint}
\end{equation}
and
\begin{equation}
\sigma^2=
\frac{\phi_1 q_1^2}{\left( ka_1\right)^{\beta+2\lambda}\ln\ell} 
\int_{ka_1}^{ka_N} y^{\beta+2\lambda-1}\,dy.
\end{equation}
Similarly, setting $\xi=r/a_i$ in Eq.~\ref{eq:tempCorr}, and taking the limit $\Delta\xi\rightarrow 0$, results in
\begin{equation}
B\left(r\right)=
-\frac{\phi_1 q_1^2\left(r/a_1\right)^{\beta+2\lambda}}{\ln\ell} 
\int_{r/a_1}^{r/a_N} f_2\left(\xi\right) \xi^{-\beta-2\lambda-1}\,d\xi.
\label{eq:corrint}
\end{equation}

The integration for the variance is easily performed, and we find
\begin{equation}
\sigma^2=-\frac{\phi_1 q_1^2}{(\beta+2\lambda)\ln\ell}
\left[1-\left(\frac{a_N}{a_1}\right)^{\beta+2\lambda} \right].
\label{eq:sig2}
\end{equation}
Since $\ln\ell$ is negative and $a_N<a_1$, the variance is always positive. The second term in square brackets is negligible when $a_N\ll a_1$, as is normally the situation of interest.\footnote{This result is equivalent, for example, to Eq.~(33) in \cite{wilson2008sound}, after making the replacement $-\ln\ell=K\mu$ and accounting for the incorporation of $K$ into the packing fraction here. Furthermore, $2/3+\nu$ in that paper is equivalent to $\beta+2\lambda$ here, and a change in the definition of the parent function leads to a constant factor of $8\pi^3$.} Note also that Eq.~\ref{eq:sig2} could have been derived by setting $r=0$ in Eq.~\ref{eq:corrint}.

Results for the spectra and correlation function depend on the choice of parent function. For the Gaussian parent function, substitution of Eq.~\ref{eq:parent_Gauss_xfrm} into \ref{eq:specint} yields

\begin{equation}
\Phi\left(k\right)=-\frac{\phi_{1}q_{1}^{2}a_{1}^{D}}{2\pi^D\ln\ell}\left(
\frac{ka_{1}}{\sqrt{\pi}}\right)^{-D-\beta-2\lambda} 
\left\{\gamma\left[\frac{D+\beta+2\lambda}{2},\frac{
k^2 a_{1}^2}{\pi}\right]- \gamma\left[\frac{D+\beta+2\lambda}{2},\frac{k^2 a_{N}^2}{\pi}\right]\right\}, 
\label{eq:gammaspec}
\end{equation}
where 
\begin{equation}
\gamma(s,x)=\int_0^x t^{s-1} e^{-t}\,dt
\label{def:incgamma}
\end{equation}
is the (lower) incomplete gamma function. Substitution of Eq.~\ref{eq:parent_Gauss_f2} into \ref{eq:corrint} similarly yields
\begin{equation}
B\left(r\right)=-\frac{\phi_{1}q_{1}^{2}}{2\ln\ell}\left(
\frac{\sqrt{\pi}r}{2a_1}\right)^{\beta+2\lambda}
\left\{\Gamma\left[-\frac{\beta+2\lambda}{2},\frac{\pi r^2}{4a_1^2}
\right]-
\Gamma\left[-\frac{\beta+2\lambda}{2},\frac{\pi r^2}{4a_N^2}
\right]\right\}, 
\label{eq:gammacorr}
\end{equation}
where $\Gamma(s,x)=\Gamma(s)-\gamma(s,x)$ is the upper incomplete gamma function.\footnote{The primary reason for switching from the lower to the upper incomplete gamma function, when writing an equation for the correlation function, is that the argument $s=-(\beta+2\lambda)/2$ is negative. The upper function, unlike the lower, allows $s<0$. Strictly speaking, the equation only holds for $r\ne 0$.} Eq.~\ref{eq:gammacorr} also enables determination of the second-order structure function, which plays an important role in turbulence theory. It is defined as \begin{equation}
D\left(r\right)=\left\langle{[ Q\left(\mathbf{r}_0+\mathbf{r}\right)-Q\left(\mathbf{r}_0\right)]^2}\right\rangle.
\label{def:SF}
\end{equation}
It can be readily shown that
\begin{equation}
D\left(r\right)=2\left[\sigma^2-B(r)\right].
\label{eq:SF}
\end{equation}

The correlation function and spectrum generally possess three regions with distinctive dependence on spatial separation or wavenumber: $r\gg a_1$ ($k^{-1}\gg a_1$), $a_1\gg r\gg a_N$ ($a_1\gg k^{-1}\gg a_N$), and $a_N\gg r$ ($a_N\gg k^{-1}$). In the context of turbulence, these are called the \emph{energy-containing}, \emph{inertial}, and \emph{dissipation} subranges, respectively. We can use the approximations $\gamma(s,x)\to x^s/s$ for $x\to 0$, and $\gamma(s,x)\to \Gamma(s)$ for $x\to\infty$, to derive the behavior of the spectrum, Eq.~\ref{eq:gammaspec}, in these three regions. For the energy-containing subrange, the small-argument approximation ($x\to 0$) applies to both terms, and the term involving $a_N$ becomes negligible in comparison to the one involving $a_1$, thus yielding $\Phi\left(k\right)\simeq-({\phi_{1}q_{1}^{2}a_{1}^{D}})/({b\pi^D\ln\ell})$. For the inertial subrange, the large-argument approximation now applies to the term involving $a_1$, which yields
\begin{equation}
\Phi\left(k\right)\simeq-\frac{\phi_{1}q_{1}^{2}a_{1}^{D}\Gamma[(D+\beta+2\lambda)/2]}{2\pi^D\ln\ell}
\left(\frac{ka_{1}}{\sqrt{\pi}}\right)^{-D-\beta-2\lambda}, 
\end{equation}
The spectrum is thus proportional to $k^{-D-\beta-2\lambda}$ in the inertial subrange, which represents the self-similar part of the spectrum. For the dissipation subrange, the large-argument approximation applies to both terms, and they approximately cancel.

Considering next the correlation and structure functions, in the energy-containing subrange, the large-argument approximations apply, and we find that $B(r)\to 0$ and $D(r)\to \sigma^2$. In the inertial subrange, the small-argument approximation applies to the $a_1$ term and the large-argument approximation to the $a_N$ term, with result
\begin{equation}
B\left(r\right)\simeq-\frac{\phi_1 q_1^2}{(\beta+2\lambda)\ln\ell}\left[1-\Gamma\left(1-\frac{\beta+2\lambda}{2}\right)\left(\frac{\sqrt{\pi}r}{2a_1}\right)^{\beta+2\lambda}\right].
\end{equation}
Thus, for $a_1\gg a_N$, Eqs.~\ref{eq:sig2} and \ref{eq:SF} lead to
\begin{equation}
D\left(r\right)\simeq 2\sigma^2\Gamma\left(1-\frac{\beta+2\lambda}{2}\right)\left(\frac{\sqrt{\pi}r}{2a_1}\right)^{\beta+2\lambda}.
\end{equation}
Thus the structure function is proportional to $r^{\beta+2\lambda}$ in the inertial subrange. In the dissipation subrange, we find that $B(r)\to\sigma^2$ and $D(r)\to 0$.

\subsection{Turbulence}
Recalling that $\lambda=(1-\beta)/3$ for turbulence (Eq.~\ref{eq:lambdaTurb}), in the inertial subrange (with $D=3$) the spectrum is proportional to $k^{-11/3-\beta/3}$. When $\beta=0$ (i.e., there is no intrinsic intermittency), the power spectrum thus decays as $-11/3$, which is consistent with Kolmogorov's \citeyearpar{kolmogorov1941local} well known second hypothesis for turbulence. The $-\beta/3$ term is an adjustment representing intrinsic intermittency. For turbulence, the structure function is proportional to $r^{2/3+\beta/3}$. In the absence of intrinsic intermittency, the structure function thus increases as $r^{2/3}$.

\begin{figure}[tbp]
\centering
\includegraphics[width=0.7\linewidth,natwidth=1200,natheight=900]{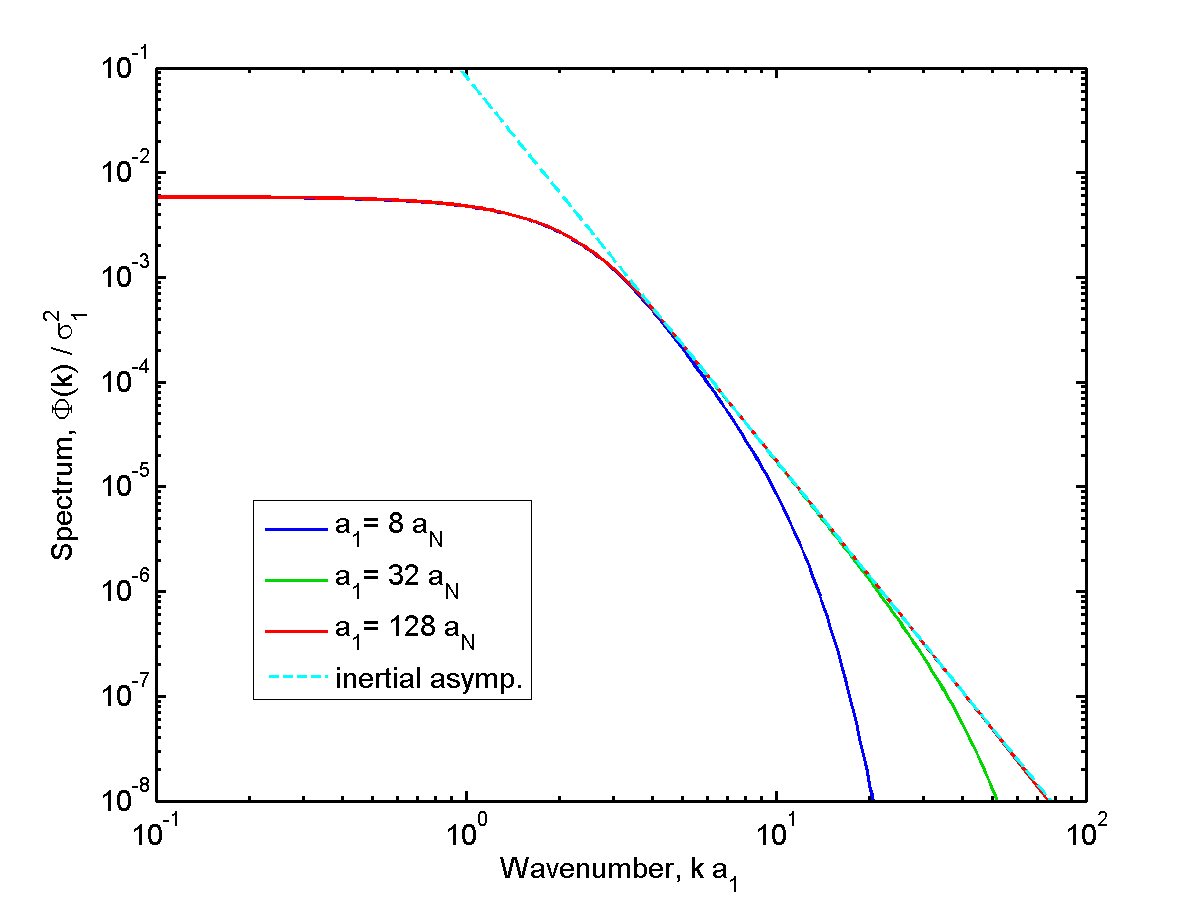}
\caption{Spectra of QW models for turbulence ($\lambda=1/3$, $\beta=0$, $\ell=1/2$) as based upon a Gaussian parent function. Shown are curves for various values of the ratio $a_1/a_N$, along with the asymptotic result for the inertial subrange.}
\label{fig:QWspectra}
\end{figure}

\begin{figure}[tbph]
\centering
\includegraphics[width=0.7\linewidth,natwidth=1200,natheight=900]{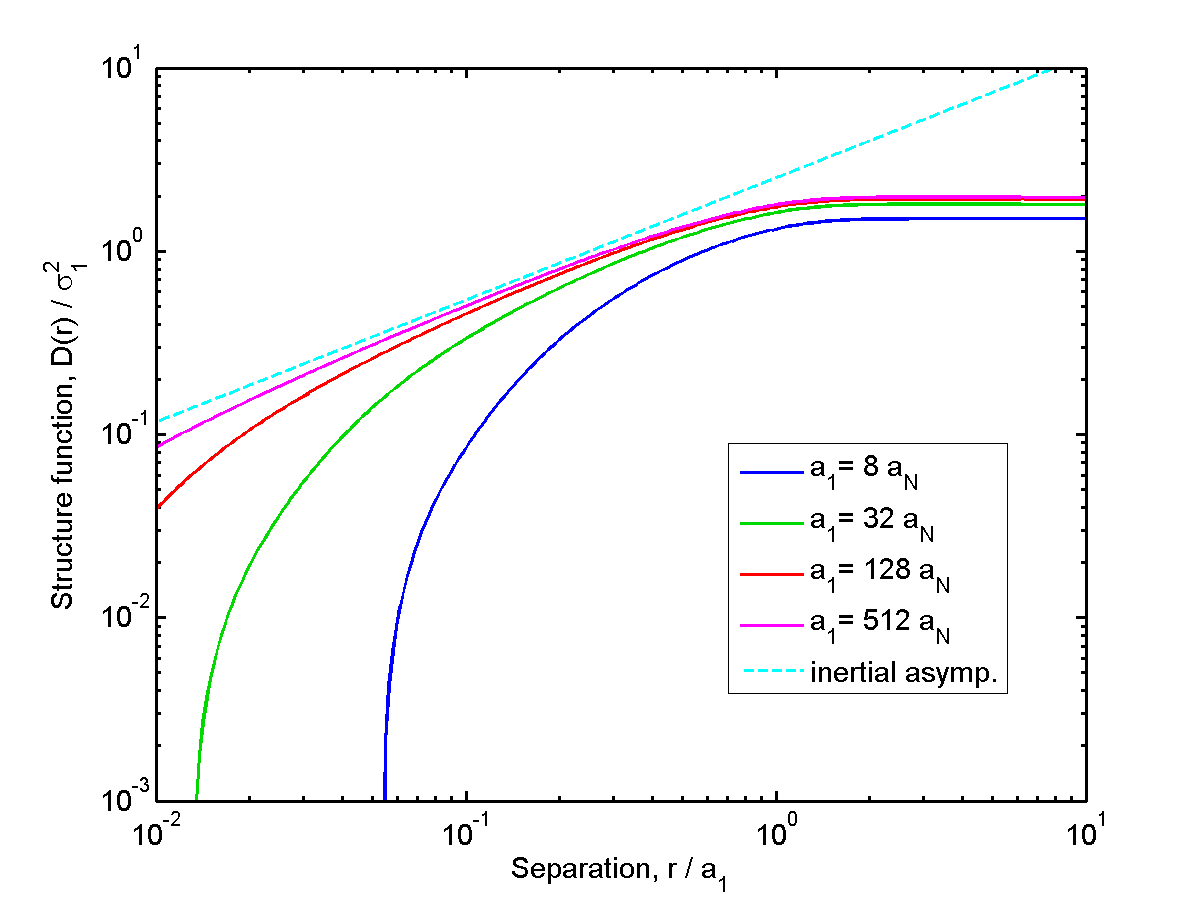}
\caption{Structure functions of QW models for turbulence ($\lambda=1/3$, $\beta=0$, $\ell=1/2$) as based upon a Gaussian parent function. Shown are curves for various values of the ratio $a_1/a_N$, along with the asymptotic result for the inertial subrange.}
\label{fig:QWstructure}
\end{figure}

Spectra and structure functions for turbulence are shown in Figs.~\ref{fig:QWspectra} and \ref{fig:QWstructure}, respectively. The figures show calculations based on Eqs.~\ref{eq:gammaspec} and \ref{eq:gammacorr}, with $\lambda=1/3$ and $\beta=0$. Curves for various values of $a_1/a_N$ are shown. The spectra and structure functions have been normalized by the variance for $a_1\gg a_N$, namely $\sigma_1^2=-(\phi_1 q_1^2)/[(\beta+2\lambda)\ln\ell]$. The figures demonstrate that the inertial subrange is evident in the spectrum when $a_1\gtrsim 32 a_N$, although it is not evident in the structure function until $a_1\gtrsim 128 a_N$.

\begin{figure}[tbph]
\centering
\includegraphics[width=0.7\linewidth,natwidth=1200,natheight=900]{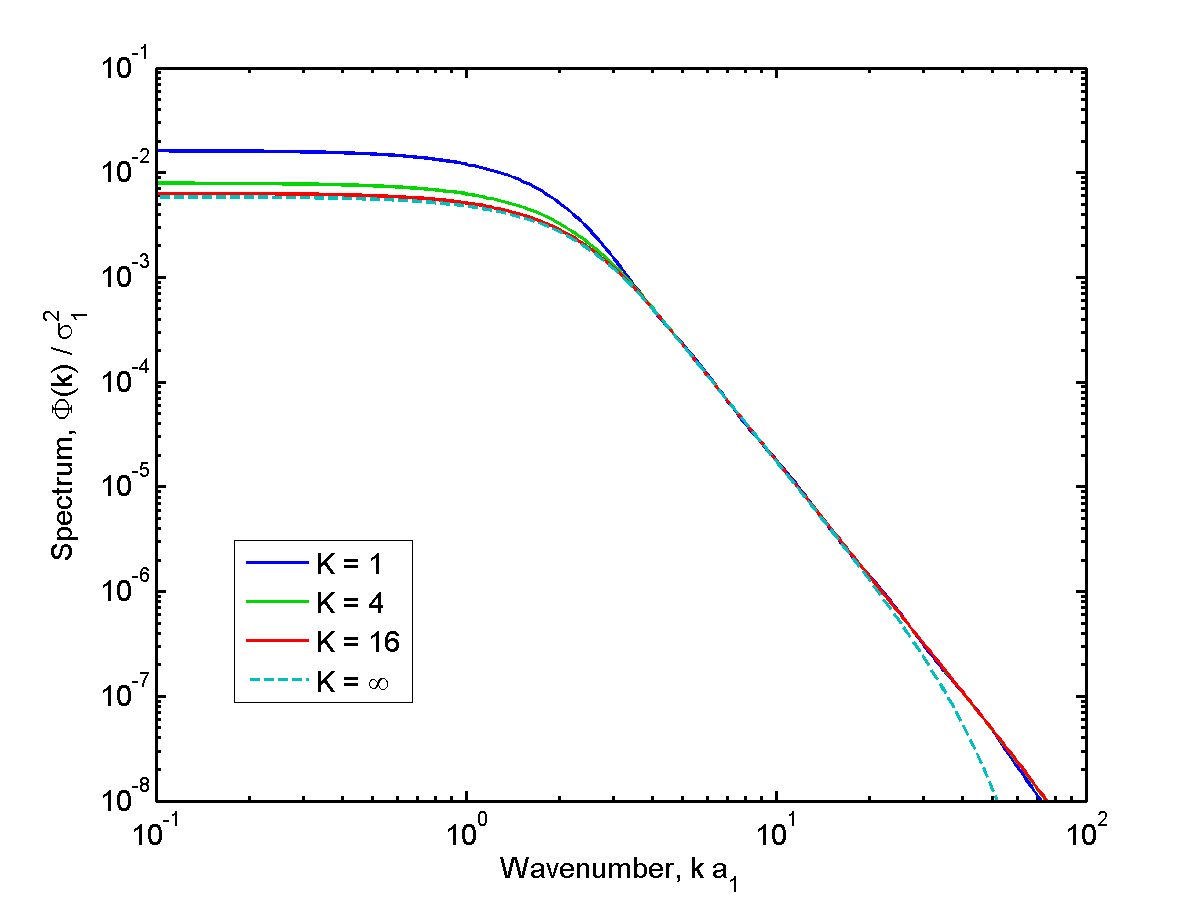}
\caption{Spectra of QW models for turbulence ($\lambda=1/3$, $\beta=0$, $\ell=1/2$) as based upon a Gaussian parent function. Shown are curves for $a_1/a_N=32$, along with calculations for discrete numbers of size classes with various densification factors ($K=1$, $4$, and $16$). Also shown is the calculation based on the integral equation ($K=\infty$).}
\label{fig:QWspecclass}
\end{figure}

\begin{figure}[tbph]
\centering
\includegraphics[width=0.7\linewidth,natwidth=1200,natheight=900]{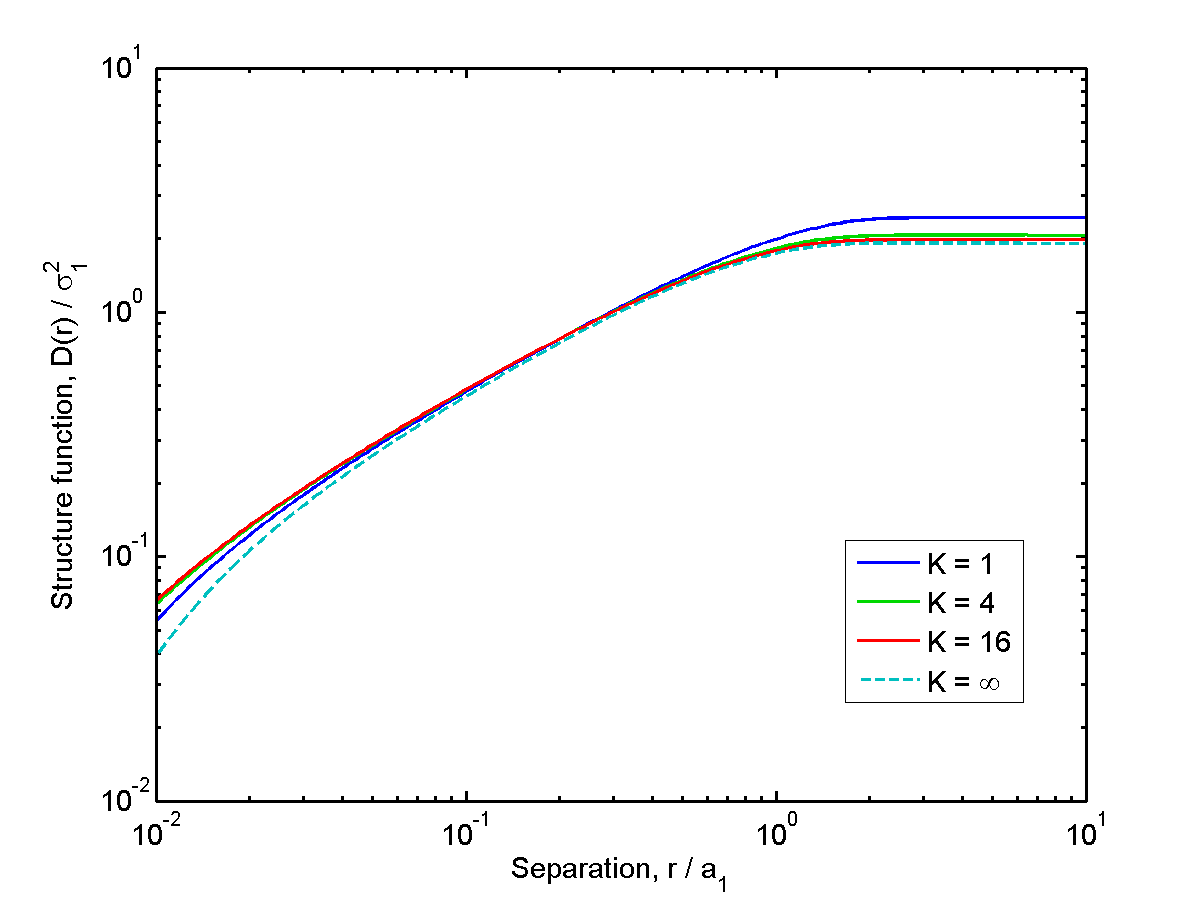}
\caption{Structure functions of QW models for turbulence ($\lambda=1/3$, $\beta=0$, $\ell=1/2$) as based upon a Gaussian parent function. Shown are curves for $a_1/a_N=128$, along with calculations for discrete numbers of size classes with various densification factors ($K=1$, $4$, and $16$). Also shown is the calculation based on the integral equation ($K=\infty$).}
\label{fig:QWstructclass}
\end{figure}

Figure~\ref{fig:QWspecclass} compares spectral calculations for discrete numbers of size classes (Eqs.~\ref{eq:QWspec} and \ref{eq:specalpha}) to the continuous calculation based on the integral equation (Eq.~\ref{eq:gammaspec}), for the case $a_1=32 a_N$. Figure~\ref{fig:QWstructclass} is an analogous comparison for the structure function (based on Eqs.~\ref{eq:QWcorr}, \ref{eq:corrlpha}, and \ref{eq:gammacorr}), for the case $a_1=128 a_N$. For both Figs.~\ref{fig:QWspecclass} and \ref{fig:QWstructclass}, the densification factor $K$ is stepped through the values $1$, $4$, and $16$. For $K=4$, the discrete calculations are quite close to the continuous calculation; for $K=16$, the discrete calculation is essentially equivalent to the continuous one. 

\subsection{Inference of Model Parameters}
Let us consider how the results in this section might be used to infer unknown QW model parameters. Fitting of the middle (inertial) subrange of the spectrum or structure function enables determination of $\beta+2\lambda$. A measurement of the variance then enables determination of $\phi_1 q_1^2/(\beta+2\lambda)$, which then provides $\phi_1 q_1^2$. The inner and outer length scales, $a_1$ and $a_N$, can also be determined from the spectrum, by examining the locations of the transitions between the subranges.\footnote{The behavior of the spectrum around the transitions does depend on the function $f(r)$ used for the individual QWs. Hence the fit will not be exact. Also, in many situations, we may have little interest in determining $a_N$, or the data may not be high enough resolution to resolve this part of the spectrum.}

Thus, on the basis of just the second-order statistics, we can determine $a_1$, $a_N$, $\phi_1 q_1^2$, and $\beta+2\lambda$. With scaling arguments, such as those for turbulence, constraints may furthermore be derived to distinguish the values of $\beta$ and $\lambda$. However, in general we cannot separate $\phi_1 q_1^2$ or $\beta+2\lambda$ using second-order statistics. Such a separation would require consideration of higher-order statistics. In \cite{wilson2008sound}, a relationship was found between the kurtosis and the packing fraction, for the case of turbulence. A more general equation could be derived and conceivably be used to separate $\phi_1$ from $q_1^2$. However, the equation depends on a uniformly random distribution of the QWs and other idealizations that are questionable in practice, which could have substantial impacts on any deductions that are made.

\section{Non-Steady Cascade QW Model}
\label{sec:nonsteady}

This section describes a non-steady, one-way (decay) cascade model. It is a generalization of the steady-state, one-way model considered in Sec.~\ref{sec:oneway}.

\subsection{Appearance Intervals}
We first consider a non-densified representation ($K=1$). Let us define $t_{i}$ as the time at which objects of size $a_i$ begin to appear. These objects subsequently begin to disappear when the next generation of objects, size $a_{i+1}$, begin to appear at $t_{i+1}$.  The interval between the onset of the appearance of class $i$, and the onset of the appearance of class $i+1$, is $\Delta t_i=t_{i+1}-t_{i}$. For the first interval, between the appearance of $a_1$ and $a_2$, $\Delta t_1=\tau_1$. The next interval, between the appearance of $a_2$ and $a_3$, is $\Delta t_2=\tau_2=\tau_1\overline{\tau}$. In general, $\Delta t_i=\tau_1\overline{\tau}^{i-1}$, and thus
\begin{equation}
\frac{\Delta t_i}{\Delta t_{i-1}}=\frac{t_{i+1}-t_{i}}{t_{i}-t_{i-1}}=\overline{\tau}
\label{eq:genint}
\end{equation}
for all $i$. For the appearance time relative to $t_1$, we thus have
\begin{equation}
t_{i+1}-t_1=\sum_{n=1}^i\Delta t_n=\tau_1\sum_{n=1}^i \overline{\tau}^{n-1},
\label{eq:series}
\end{equation} 
Utilizing a well known equation for the sum of a geometric series, we find
\begin{equation}
t_{i+1}-t_1=\tau_1\frac{1-\overline{\tau}^i }{1-\overline{\tau}}.
\label{def:ti}
\end{equation}

Let $\mathcal{R}_1(t)$ be the rate, per unit volume, at which the largest objects ($i=1$) are seeded, which is in general time dependent. Assume that $\mathcal{R}_1(t)=\mathcal{R}_1$ (where $\mathcal{R}_1$ is a fixed rate) during the time interval from $t=t_1$ to $t=t_1+T$, and zero otherwise. Then for the size class $i$, production will begin at time $t=t_i$ and persist until $t=t_i+T$. According to Eq.~\ref{eq:Ncasc}, during this interval the process must produce $a_i$-size objects at a rate $\mathcal{R}_1 (\overline{M}\overline{\tau})^{i-1}$. 

Note that when this process is observed at some time in the future, $t>t_i+T$, it becomes impossible (based on the number of QWs) to distinguish between the original process, seeded with $a_1$-size objects at a rate $\mathcal{R}_1$ during the interval $[t_1,t_1+T)$, and a process seeded with $a_i$-size objects at a rate $\mathcal{R}_1 (\overline{M}\overline{\tau})^{i-1}$ during the interval $[t_i,t_i+T)$. While a single snapshot thus does not reveal the timing and size of the objects initiating the cascade, it does provide information on the interval $T$. When $T$ is large compared to the decay times for the size classes, there will be a broad range of QW sizes. Specifically, from Eq.~\ref{def:ti},
\begin{equation}
T=t_j-t_i=\frac{\tau_1\overline{\tau}^{i}\left( 1-\overline{\tau}^{j-i}\right) }{1-\overline{\tau}},
\label{eq:T}
\end{equation}
where $t_i$ is the time at which the largest observed objects (class $i$) appeared, and $t_j$ is the time at which the smallest observed objects (class $j$) appeared. Hence, $T$ equals the sum of the QW lifetimes for the size classes between $i$ and $j$. In principle, given $\tau_1$ and $\tau$, we could solve Eq.~\ref{eq:T} for $T$.\footnote{Of course, in a discrete model, the relationship $t_i-t_j$ need not exactly equal $T$. We can solve the relationship more precisely by increasing the scale densification factor $K$.}

\begin{figure}[tbh]
\centering
\includegraphics[width=\linewidth,natwidth=1200,natheight=900]{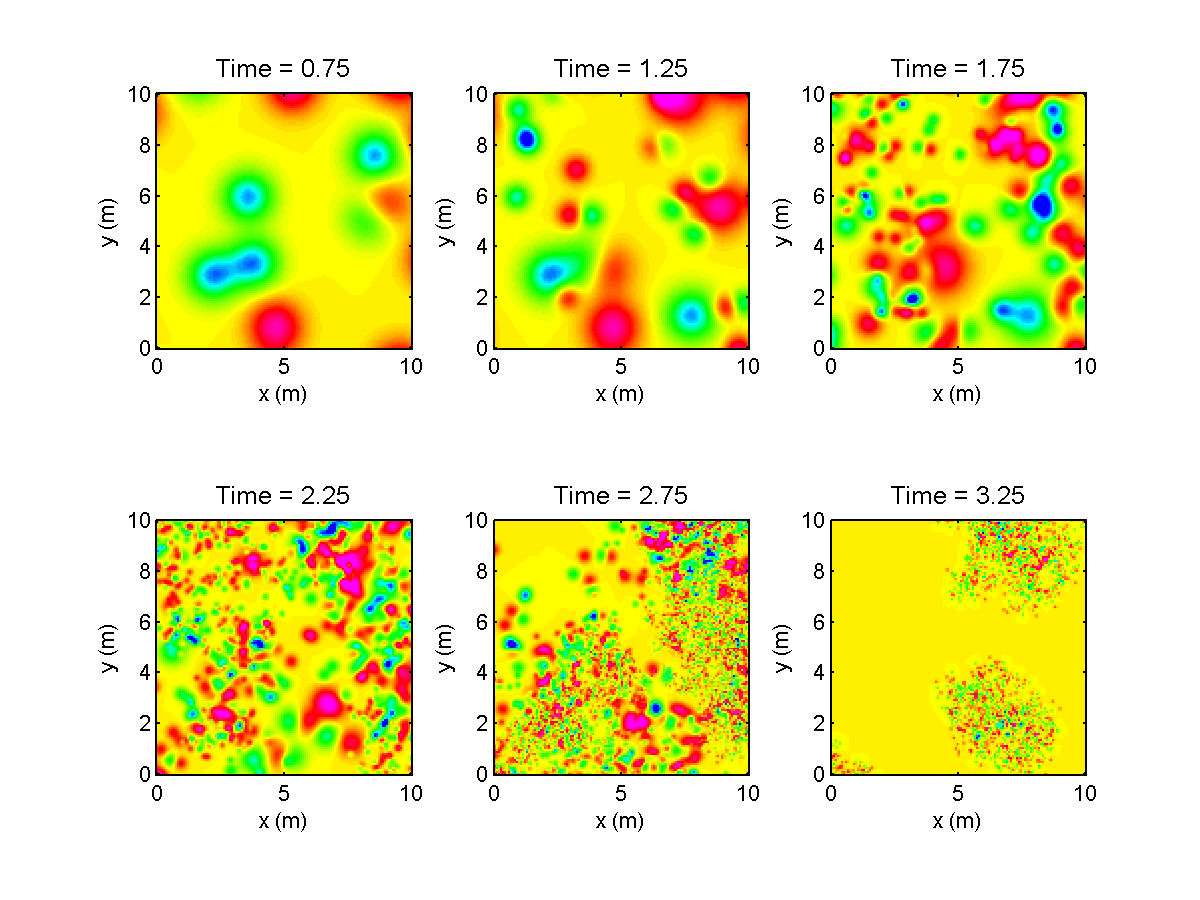}
\caption{Snapshots at six time steps of a non-steady cascade process. The cascade was initialized with QWs of size $a_1=1$ generated over an interval $[0,\tau_1=1]$. Limits on the color axis are $\pm 1.2$.}
\label{fig:NonSteadyState}
\end{figure}

\begin{figure}[tbh]
\centering
\includegraphics[width=\linewidth,natwidth=1200,natheight=900]{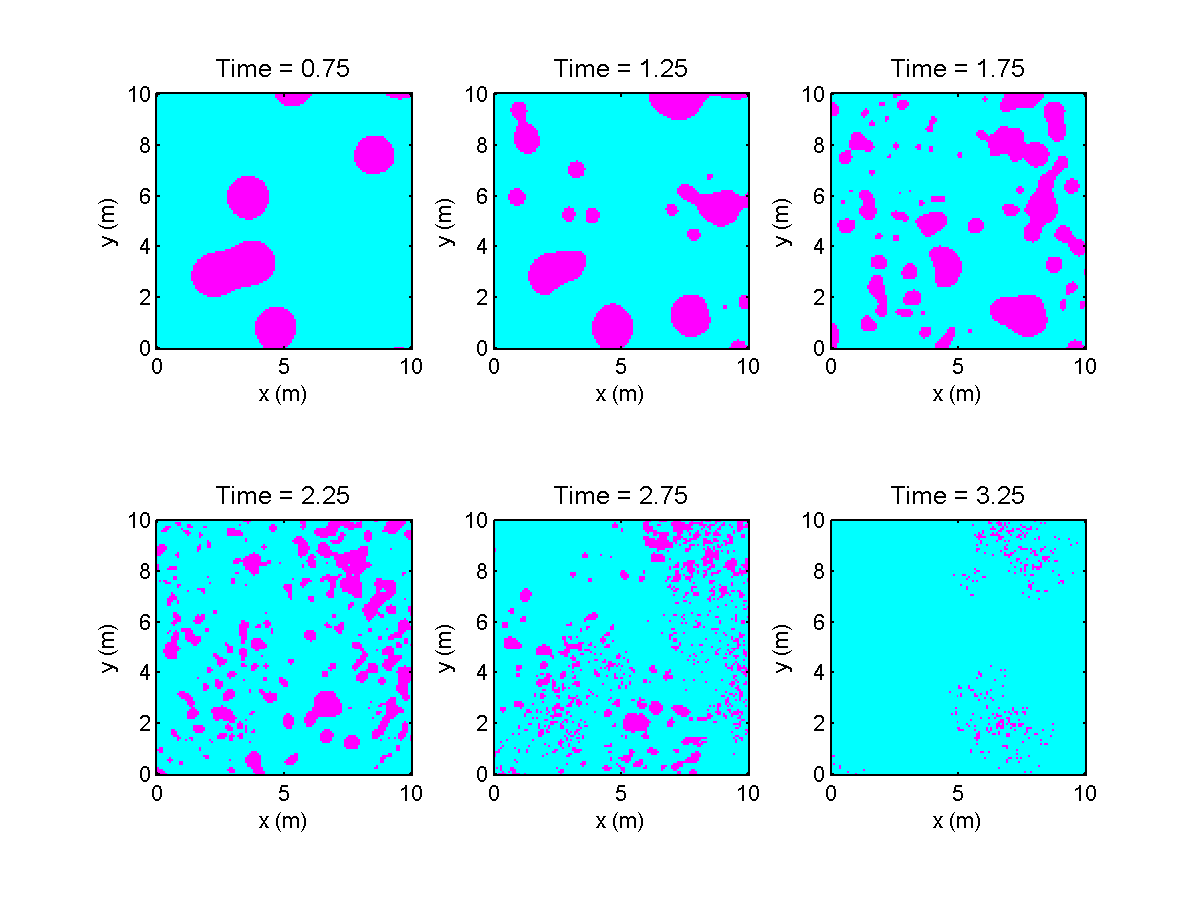}
\caption{Same as Fig.~\ref{fig:NonSteadyState}, except that level cuts through the fields at $\pm 0.7$ are shown.}
\label{fig:NonSteadyStateCut}
\end{figure}

\begin{figure}[tbph]
\centering
\includegraphics[width=0.8\linewidth,natwidth=1200,natheight=900]{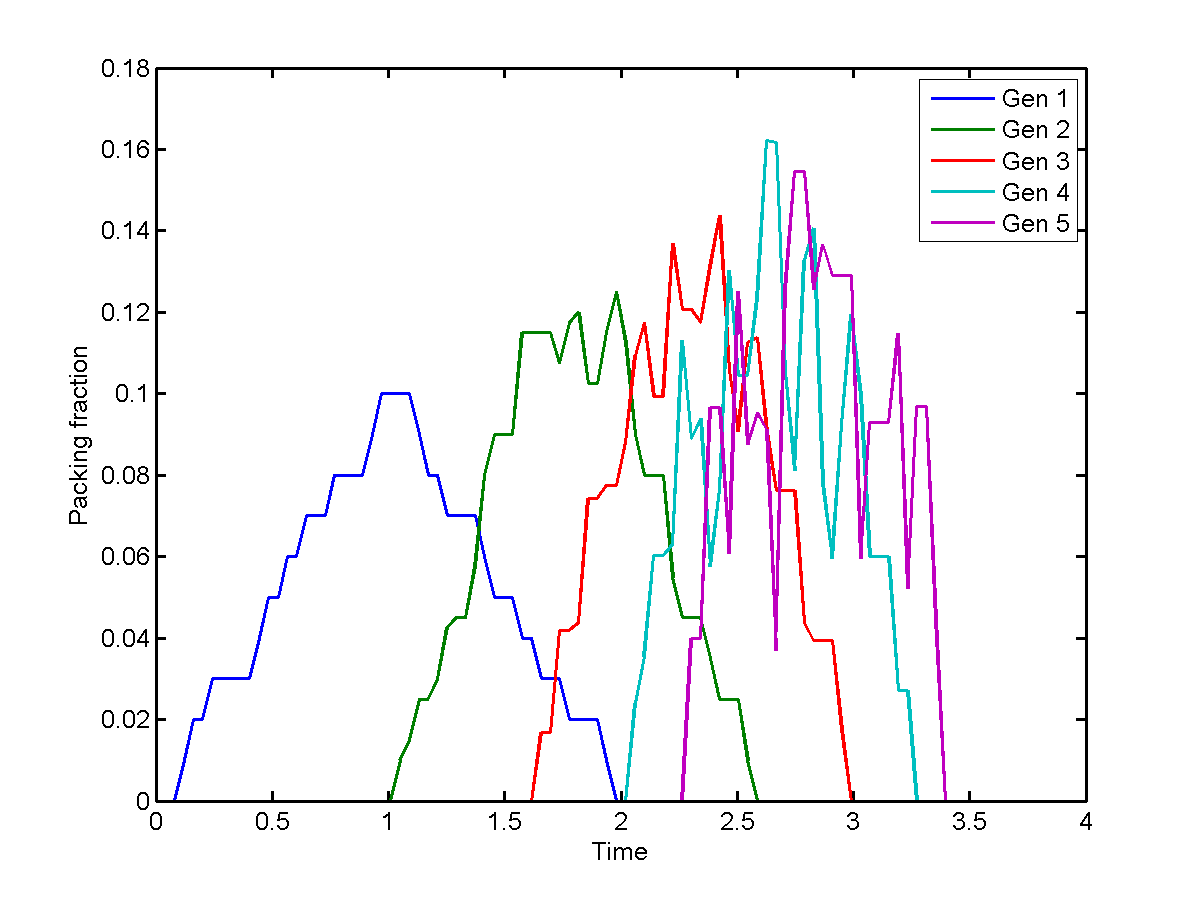}
\caption{Evolution of the empirical packing fractions for each of the five generations of the non-steady cascade process shown in Figs.~\ref{fig:NonSteadyState} and \ref{fig:NonSteadyStateCut}.}
\label{fig:NonSteadyPack}
\end{figure}

Figures~\ref{fig:NonSteadyState} and \ref{fig:NonSteadyStateCut} show an example non-steady cascade. QWs of size $a_1=1$ were seeded over an interval $[0,1]$, where the lifetime is $\tau_1=1$. The process is then observed at six times in the interval $[1,3.5]$. Five generations are simulated. Other parameters $\ell=1/2$, $\beta=0$, $\lambda=1/3$, $\phi_1=0.1$, and $q_1=1$. The accelerating breakdown of the large QWs into smaller ones, and eventual disappearance of the cascade at the end of the interval, are evident. Sorting by size is also observed, particularly in the later stages of the cascade. Shown in Fig.~\ref{fig:NonSteadyPack} is the temporal evolution of the empirical packing fractions for each of the five generations. The first generation peaks around $t=T=1$, when the seeding is turned off. This generation eventually disappears at $t=T+\tau_1=2$. The following generations appear and subsequently decay. As the decay times shorten, the time interval over which the generation persists approaches $T$.

\subsection{Scale Densification}
\label{nonsteadydense}

As described in Secs.~\ref{sec:densify} and \ref{sec:steadydense}, in the densified representation we allow fractional size class indices, $i=1+(j-1)/K$, where $K$ is the scale densification factor and $j=1,2,\ldots,IK$. The production size classes $i=1+1/K,1+2/K,\ldots 1+(K-1)/K$ are seeded directly and subsequently decay into the size classes $i=2+1/K,2+2/K,\ldots 2+(K-1)/K$, and so forth. In this section, we derive the appearance times $t_i$ for the fractional size classes.

The decay times for the fractional classes follow the same self-similarity relationships as the integer size class. In particular, the $i$th size class has a decay time of $\tau_i=\tau_1\overline{\tau}^{i-1}$. Since the decay of size class $i$ corresponds to the appearance of $i+1$, the interval between appearance times of \emph{consecutive generations} remains $t_{i+1}-t_{i}=\tau_i$. The ratio of appearance time intervals between consecutive generations $(t_{i+1}-t_i)/(t_{i}-t_{i-1})$, thus remains at $\overline{\tau}$ as indicated by Eq.~\ref{eq:genint}, for all $i$.

\begin{figure}[tbh]
\centering
\includegraphics[width=0.8\linewidth,natwidth=1043,natheight=225]{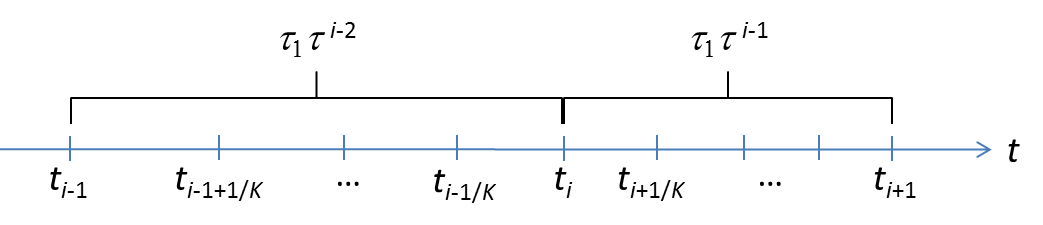}
\caption[Time line for appearance of QWs in a densified representation.]
{Time line for appearance of QWs in a densified representation. This example corresponds to a densification factor of $K=4$.}
\label{fig:timeline}
\end{figure}

In the densified representation, we wish to shorten the appearance times between the size classes while preserving the self-similarity of the representation. The situation is depicted in Fig.~\ref{fig:timeline}. For the densified representation, let us define $\Delta t_i$ now as the interval between appearances of adjacent size classes, i.e., $\Delta t_i=t_{i+1/K}-t_i$. We begin by setting
\begin{equation}
\frac{\Delta t_i}{\Delta t_{i-1/K}}=\frac{t_{i+1/K}-t_i}{t_i-t_{i-1/K}}=\overline{\tau}^{1/K}.
\label{eq:genintd}
\end{equation}
Applying this formula recursively, we find that $t_{i+1/K}-t_{i}=\overline{\tau}(t_{i-1+1/K}-t_{i-1})$. Thus
\begin{align}
t_{i+1}-t_{i}
&=(t_{i+1}-t_{i+1-1/K})+(t_{i+1-1/K}-t_{i+1-2/K})+\cdots+(t_{i+1/K}-t_{i}) \nonumber \\
&=\overline{\tau}\left[(t_{i}-t_{i-1/K})+(t_{i-1/K}-t_{i-2/K})+\cdots+(t_{i-1+1/K}-t_{i-1})\right] \nonumber \\
&=\overline{\tau}(t_{i}-t_{i-1}),\nonumber 
\end{align}
in agreement with Eq.~\ref{eq:genint}. This result also demonstrates that the appearance-time intervals have the same dependence on $\overline{\tau}$ as the decay times, i.e., \begin{equation}
\Delta t_i=\Delta t_1\overline{\tau}^{i-1},
\label{eq:Deltat}
\end{equation}
It is important to recognize that the scale densification does \emph{not} shorten the appearance time between generations; rather, it introduces $K$ more fractional time steps between the generations.

To find the appearance times relative to $t_1$, the summation in Eq.~\ref{eq:series} must be adjusted to include the fractional size classes: 
\begin{equation}
t_{i+1}-t_1=\Delta t_1\sum_{n=1}^{Ki} \overline{\tau}^{(n-1)/K}=\Delta t_1\frac{1-\overline{\tau}^i}{1-\overline{\tau}^{1/K}}.
\label{eq:delTemp}
\end{equation} 
Since $t_2-t_1=\tau_1$, the preceding result implies that $\tau_1=\Delta t_1(1-\overline{\tau})/(1-\overline{\tau}^{1/K})$. Solving for $\Delta t_1$, and substituting back into Eq.~\ref{eq:delTemp}, we have
\begin{equation}
t_{i+1}-t_1=\tau_1\frac{1-\overline{\tau}^{i}}{1-\overline{\tau}},
\label{def:ti2}
\end{equation}
which is the same as Eq.~\ref{def:ti}, except we have shown the result also holds for the fractional size classes. Recognizing from Eqs.~\ref{def:phiGen}, \ref{eq:phii}, and \ref{eq:qi} that
\begin{equation}
\tau_i=\tau_1\left( \frac{a_i}{a_1}\right)^{\beta+2\lambda},
\label{eq:taui}
\end{equation}
this result can be recast in a useful form in which the dependence on $\tau_i$ is replaced with $a_i$:  
\begin{equation}
t_i-t_1=\tau_1\frac{1-(a_i/a_1)^{\beta+2\lambda}}{1-\overline{\tau}},
\label{def:ti3}
\end{equation}
or, solving for $a_i$,
\begin{equation}
a_i=a_1\left[1-(t_i-t_1)\frac{1-\overline{\tau}}{\tau_1}\right]^{1/(\beta+2\lambda)}.
\label{eq:aiemerge}
\end{equation}

The discussion in Sec.~\ref{sec:steadydense}, regarding seeding of a steady-state one-way cascade model, mostly still applies to the non-steady model. The production classes, $i=1,1+1/K,\ldots 1+(K-1)/K$, should be seeded at a rate $\mathcal{R}_1\overline{M}^{i-1}$. The new aspect is that the seeding must begin at the times $t_i$ indicated by Eq.~\ref{def:ti2}. This process will ensure that the regular distribution of appearance times, as shown in Fig.~\ref{fig:timeline}, is maintained.

\begin{figure}[tbph]
\centering
\includegraphics[width=0.8\linewidth,natwidth=1200,natheight=900]{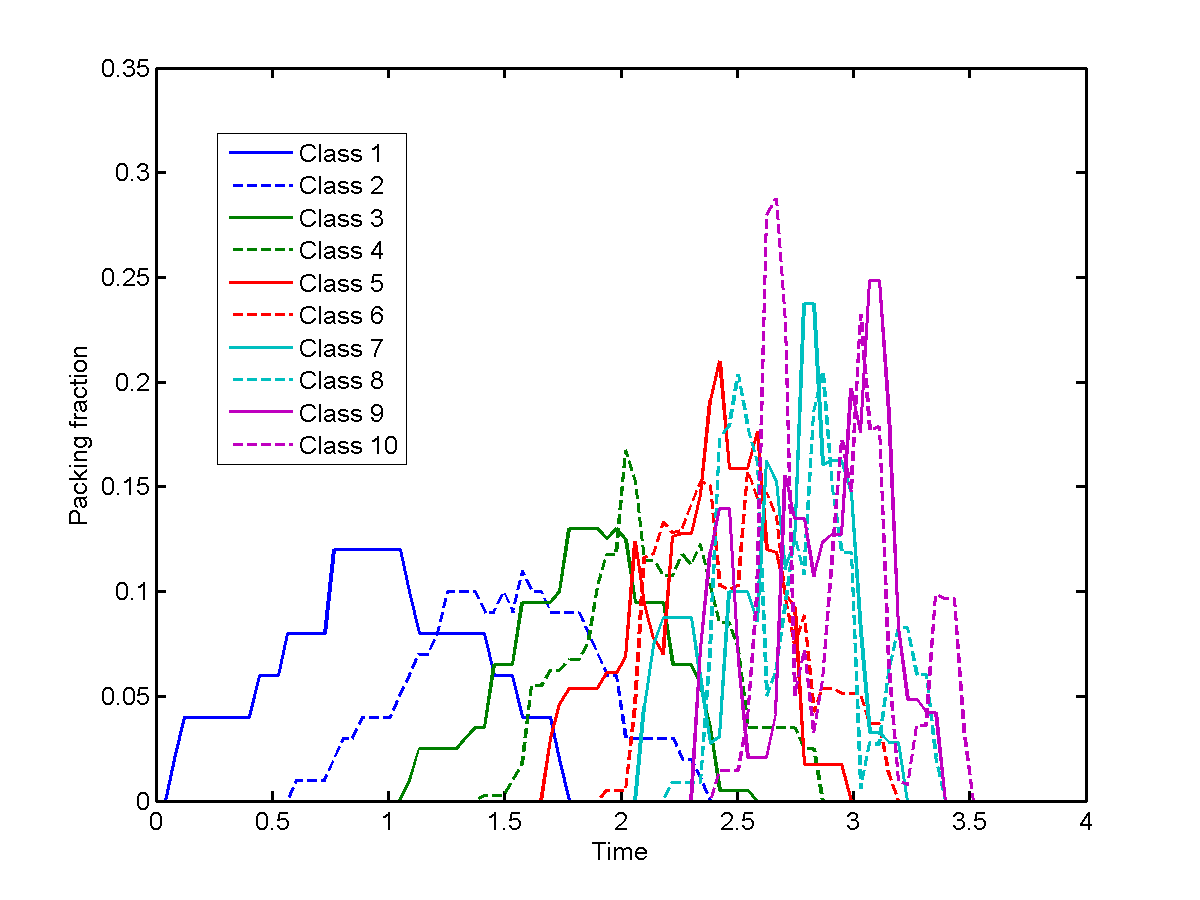}
\caption{Evolution of the empirical packing fractions for each of the five generations of the non-steady cascade process shown in Figs.~\ref{fig:NonSteadyState} and \ref{fig:NonSteadyStateCut}.}
\label{fig:NonSteadyPack2}
\end{figure}

Figure~\ref{fig:NonSteadyPack2} is similar to Fig.~\ref{fig:NonSteadyPack}, except that the model has been densified to $K=2$. The temporal shifting of the additional size class ($i=3/2,5/2,\ldots,11/2$, indicated by dashed lines), relative to their integer counterparts, is evident.

\subsection{Spectral Model}

Let us next consider development of a spectral model for a non-steady cascade process, analogous to the one for a static or steady-state cascade process, as developed in Sec.~\ref{sec:stat}. Now, however, the spectral content will evolve with time. In the following discussion, we assume that each size class represents a single generation of the process, and that the cascade is initialized through a Poisson process creating objects of size $a_1$ at a rate $\mathcal{R}_1=\mathcal{N}_1/\tau_1$ during the interval $[t_1,t_1+T)$. The rate of energy production (per unit volume) is thus $\mathcal{N}_1 q_1^2 a_1^D/\tau_1$. A total (per unit volume) of $\mathcal{N}_1 T/\tau_1$ objects of size $a_1$, with total energy (per unit volume) of $\mathcal{N}_1 q_1^2 a_1^D T/\tau_1=\phi_1 q_1^2 T/\tau_1$, are thus produced by the end of the interval. This same amount of energy must subsequently pass through each generation in the cascade.

At time $t=t_2=\tau_1$, the objects of size $a_1$ begin to decay into objects of size $a_2$. This process continues in linear fashion (on average) until $t=T+\tau_1$. Hence the energy in size class $a_1$ is lost over this interval (with length $T$) at a rate of $(\phi_1 q_1^2 T/\tau_1)/T=\phi_1 q_1^2/\tau_1$. During this same time interval, the loss of energy at $a_1$ is going into production at $a_2$. A total of $\mathcal{N}_1 M T/\tau_1=\mathcal{N}_2 T/\tau_2$ objects of size $a_2$, with energy $\mathcal{N}_1 q_1^2 a_1^D T/\tau_1$$=\mathcal{N}_2 q_2^2 a_2^D T/\tau_2 $$=\phi_2 q_2^2 T/\tau_2$, are created. Hence creation of $a_2$ occurs at a rate $\phi_2 q_2^2/\tau_2$.

Consider next the general case of production and destruction of objects of size $a_i$. A total of $(\mathcal{N}_1 T/\tau_1)M^{i-1}$ objects, with total energy $(\mathcal{N}_1 T/\tau_1)M^{i-1}q_i^2 a_i^D=\phi_i q_i^2 T/\tau_i$, are created during the interval $[t_i,t_i+T)$. The rate of energy increase during this interval is $\phi_i q_i^2/\tau_i$. By the end of this interval, all objects of size $a_i$ that will be created have been created. This motivates the definition of the \emph{cumulative production function} for the $i$th size class, namely
\begin{equation}
P_i(t)=\phi_i q_i^2 
\left[ R\left(\frac{t-t_i}{\tau_i} \right) -
R\left(\frac{t-t_i-T}{\tau_i} \right)\right] .
\label{def:production}
\end{equation}
where $R(t)$ is the unit ramp function ($=0$ for $t<0$, and $=t$ for $t\geq 0$). The function $P_i(t)$ is zero for $t<t_i$, then increases linearly until $t=T+t_i$, and thereafter equals the sum of all energy produced at scale $i$. Balancing the production, the energy of objects of size $a_i$ begins decaying at a rate $\phi_i q_i^2/\tau_i$ at $t=t_{i+1}$, which continues until they have disappeared entirely by $t=T+t_{i+1}$. This motivates the following definition of the \emph{cumulative destruction function}:
\begin{equation}
D_i(t)=\phi_i q_i^2
\left[ R\left(\frac{t-t_{i+1}}{\tau_i} \right) -
R\left(\frac{t-t_{i+1}-T}{\tau_i} \right)\right] 
\label{def:destruction}
\end{equation}
Note that, by Eq.~\ref{def:phiGen}, $q_i^2 a_i^D/\tau_i=q_{i+1}^2 a_{i+1}^D\tau_{i+1}$. It follows that $D_i(t)=P_{i+1}(t)$; that is, the destruction of size class $i$ is balanced by the production of class $i+1$. 

\begin{figure}[tbh]
\centering
\includegraphics[width=0.6\linewidth,natwidth=885,natheight=916]{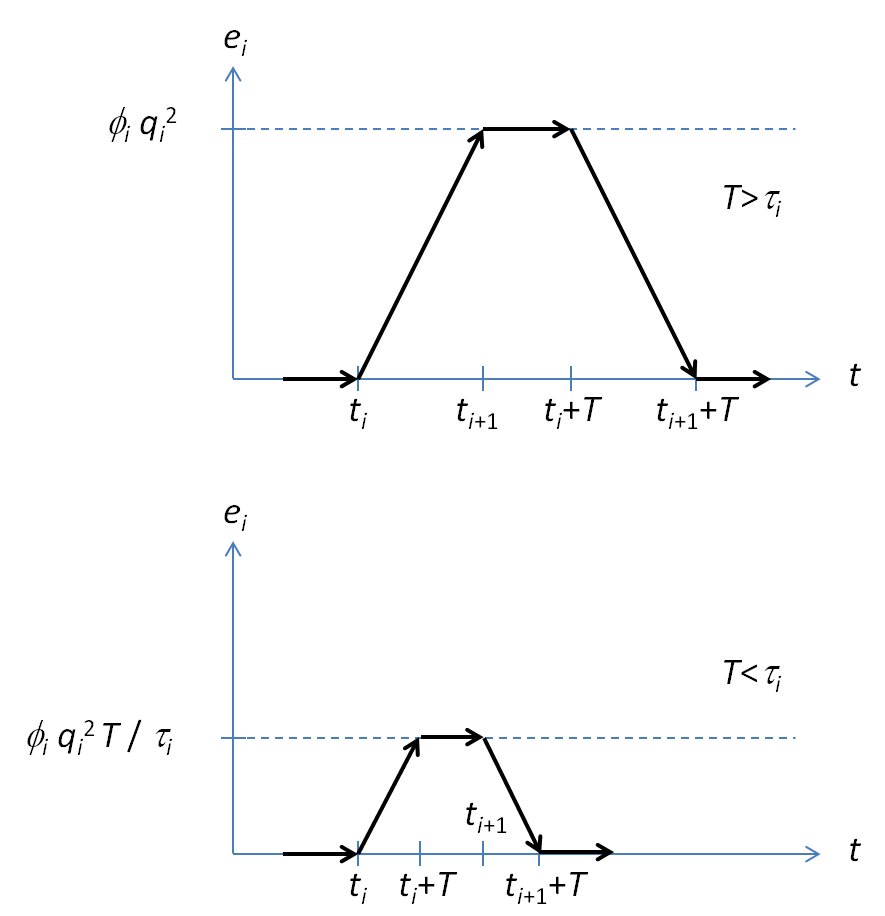}
\caption[Evolution of energy in the $i$th size class.]
{Evolution of energy in the $i$th size class. Top shows the case $\tau_1>T$; bottom shows the case $\tau_1<T$. This schematic assumes that each size class $i$ represents a new generation of the cascade process.}
\label{fig:dynamic}
\end{figure}

The difference between production and destruction for size class $i$, $P_i(t)-D_i(t)$, exhibits two distinct behaviors depending on the relative values of $T$ and $\tau_i=t_{i+1}-t_i$. These are illustrated in Fig.~\ref{fig:dynamic}. In either case, $P_i(t)-D_i(t)=0$ outside the range $[t_i,t_{i+1}+T)$. Within this range, production initially dominates, followed by a time interval at which production and destruction are balanced, and then followed by a time interval in which destruction dominates.
\begin{enumerate}
\item If $T>\tau_i$, during the interval $[t_{i+1},t_i+T)$ we have $P_i(t)-D_i(t)=\phi_i q_i^2$. Production of class $i$ continues to occur, but is balanced by its destruction. The spectrum stabilizes at its equilibrium (steady-state cascade) value.

\item If $T<\tau_i$, during the interval $[t_i+T,t_{i+1})$ we have $P_i(t)-D_i(t)=(T/\tau_i)\phi_i q_i^2$. Production of class $i$ has ceased, but destruction has not yet begun. The spectrum stabilizes at fraction of $T/\tau_i$ of its equilibrium (steady-state cascade) value.
\end{enumerate}

An alternative viewpoint of the production/destruction process follows by pairing the first term in the production function with the first term in the destruction function, i.e.,
\begin{equation}
\Psi_i(t)=\phi_i q_i^2 
\left[ R\left(\frac{t-t_i}{\tau_i} \right) -
R\left(\frac{t-t_{i+1}}{\tau_i} \right)\right] .
\label{def:Psii}
\end{equation}
We then have
\begin{equation}
P_i(t)-D_i(t)=\Psi_i(t)-\Psi_i(t-T).
\label{eq:PsiiRel}
\end{equation}
The function $\Psi_i(t)$ represents the \emph{step} response to an introduction of QWs at the rate $\mathcal{R}_1$, beginning at $t=t_1$. The second term on the right side of Eq.~\ref{eq:PsiiRel} represents the effect of turning off the production at $t=t_1+T$. The step function behaves as follows:
\begin{equation}
\Psi_i\left(t\right)=\left\lbrace \begin{array}{ll}
0, & t<t_i, \\
({\phi_i q_i^2}/{\tau_i})(t-t_i), & t_i\leq t<t_{i+1}, \\
\phi_i q_i^2, & t_{i+1}\leq t. 
\end{array}\right. 
\end{equation}
The step function is zero until $t=t_i$, and then increases linearly to a value $\phi_i q_i^2$ at $t=t_{i+1}=t_i+\tau_i$, after which it remains constant. 

Let us now return to the spectral model. In Eq.~\ref{eq:specalpha}, which gives the spectrum for size class $i$ in the steady-state model, the energy per unit volume is $(N_i/V)q_i^2 a_i^D=\phi_i q_i^2$. For the non-steady model, we simply replace $\phi_i q_i^2$ by its dynamic value, $\Psi_i(t)-\Psi_i(t-T)$. Hence
\begin{equation}
\Phi_{i}\left(k,t\right)=\left(2\pi a_{i}\right)^{D} F^{2}\left(ka_{i}\right)
\left[\Psi_i(t)-\Psi_i(t-T) \right] .
\end{equation}
Including the sum over all size classes yields
\begin{equation}
\Phi\left(k\right)= 
\left(2\pi\right)^{D} \sum_i a_{i}^{D} F^{2}\left(ka_{i}\right)
\left[ \Psi_i(t)-\Psi_i(t-T)\right].
\label{eq:specT}
\end{equation}
Before attempting to convert the summation to an integral, we must devise an approach to scale densification in the non-steady model, which is the topic of the next section.

Eq.~\ref{def:Psii} generalizes by recognizing that the energy in size class $i$ is transferred to class $i+K$. Hence, in the cumulative destruction function, Eq.~\ref{def:destruction}, $t_{i+1}$ is replaced by $t_{i+K}$. This results in the following replacement for Eq.~\ref{def:Psii}:
\begin{equation}
\Psi_i(t)=\frac{\phi_i q_i^2}{K}
\left[ R\left(\frac{t-t_i}{\tau_i} \right) -
R\left(\frac{t-t_{i+K}}{\tau_i} \right)\right] .
\label{def:Psii2}
\end{equation}
Note the introduction of the factor $1/K$ to account for the densification, as in Eq.~\ref{eq:specalpha}.

To seed the densified cascade, we adopt the procedure described in Sec.~\ref{sec:densify} based on generating random QWs for all production classes $i\leq K$. As before, the rate at which these seed QWs are generated must be $\bar{r}_i=\bar{N}_i/\tau_i$, although now $N_i$ is interpreted as the number of QWs of class $i$ that would exist if the cascade were in a steady state. To mimic the general behavior shown in Fig.~\ref{fig:dynamic}, the onset of production for class $i$ should occur at $t=t_i$, rather than $t=t_1$, and continue until $t=t_i+T$. While this may initially seem inconsistent with confining production to the fixed interval $[t_1,t_1+T)$, it must be kept in mind that the cascade process delays the smaller sizes relative to $a_1$. This delay must be accounted for in the numerical procedure used to densify the cascade.

With this approach to scale densification, we are now ready to convert summations over size classes, such as Eq.~\ref{eq:specT}, to integrals. We begin by writing the ramp functions in an equivalent form using the Heaviside step function $H(t)$ (which is $0$ for $t<0$, and $1$ for $t>=0$). Specifically, $R(t/a)=(t/a)H(t)$, where $a$ is a constant. Since ${\phi_1 q_1^2}/{\tau_1}={\phi_i q_i^2}/{\tau_i}$, we have
\begin{equation}
\Psi_i\left(t\right)=
\frac{\phi_i q_i^2}{K\tau_i}\left[(t-t_i)H\left(t-t_i\right)
-(t-t_{i+K})H\left(t-t_{i+K}\right)\right].
\end{equation}

Note that as we increase the densification of scales, the linear interval, $t_{i-1}\leq t<t_i$, becomes shorter. Hence, in a highly densified model, we may make the approximation
\begin{equation}
\Psi_i\left(k,t\right)\simeq
\left(2\pi a_{i}\right)^{D}
 F^{2}\left(ka_{i}\right)(t_i-t_{i-1})H\left(t-t_{i-1}\right).
\end{equation}
From Eq.~\ref{def:ti3}, and setting $a_{i+1}=a_i+\Delta a_i$, we have
\begin{equation}
t_i-t_{i-1}\simeq -\frac{\Delta a_i}{a_i}
\frac{\tau_1 (\beta+2\lambda)}{1-\tau^{1/K}}
\left( \frac{a_i}{a_1}\right)^{\beta+2\lambda}.
\end{equation}
(Since $\Delta a_i$ is negative, $t_i-t_{i-1}$ is positive.) We then have
\begin{equation}
\Psi_i\left(k,t\right)\simeq -\frac{\left(2\pi\right)^{D}\tau_1 (\beta+2\lambda)}
{a_1^{\beta+2\lambda}(1-\tau^{1/K})}\sum_i
 a_{i}^{\beta+2\lambda+D-1} F^{2}\left(ka_{i}\right)\Delta a_i,
\end{equation}
where the summation is to be carried out over all size classes such that $t_{i-1}\leq t$. Taking the limit $\Delta a_i\rightarrow 0$, we have the integral
\begin{equation}
\Psi_i\left(k,t\right)\simeq \frac{\tau_1 
\left(2\pi a_{1}\right)^{D} (\beta+2\lambda)}
{1-\overline{\tau}^{1/K}}(k a_1)^{-\beta-2\lambda-D}
\int_{k a(0)}^{k a(t)}
F^{2}\left(y\right) y^{\beta+2\lambda+D-1} \, dy.
\label{eq:nsint}
\end{equation}
In this result, $a(t)$ is the scale of the size class emerging at $t$, as calculated from Eq.~\ref{eq:aiemerge}, in the following continuous form:
\begin{equation}
a(t)=\left\lbrace\begin{array}{ll}
a_1, & t<0, \\
a_1\left[1-{t}(1-\overline{\tau}^{1/K})/{\tau_1}\right]^{1/(\beta+2\lambda)}, 
  & 0\leq t < t_{N-1}, \\
a_N, & t_{N-1}\leq t.
\end{array} \right. 
\end{equation}

The solution could, in principle, be completed by solving the integral, Eq.~\ref{eq:nsint}. Such a solution appears to be very challenging, however, and is not provided here.

\section{Conclusion}

This report described a wavelet-like model for distributions of objects in natural and man-made terrestrial environments. The model was constructed in a self-similar fashion, with the sizes, amplitudes, and numbers of objects occurring at a constant ratios between adjacent size classes. The objects were represented using \emph{quasi-wavelets}, which are similar to ordinary wavelets, except that they are generally spherically symmetric, and some conditions such as possessing a zero mean are not enforced.

The primary goal of the report was to introduce realistic \emph{intermittency} into models for random media. Three types of intermittency were specifically mentioned: \emph{global intermittency}, in which the largest objects and activity initiated by them is uneven, \emph{intrinsic intermittency}, in which the smaller objects are confined to less space than the larger ones, and \emph{sorting by size}, in which like-size objects tend to occur together. Fractal supports and a cascade process were used to introduce such intermittency features into the self-similar, quasi-wavelet model. 

The report began with a description of relevant concepts from fractal theory, which served to motivate many of the modeling choices made later on. The presentation then progressed through static (time-invariant), steady-state, and non-steady models. A two-way cascade model and spectral equations for the non-steady cascade were partially developed. Further development of these ideas could be a productive topic for future research.

The model developed in this report can be applied to such diverse phenomena as turbulence, geologic distributions, urban buildings, vegetation, and arctic ice floes. In particular, development of the model was motivated by the desire to provide a rigorous basis for synthesizing realistic terrestrial scenes, and for predicting the performance of sensing and communication systems in operating environments with complex, intermittent distributions of scattering objects. Some initial research on such applications has been done \citep{wilson2008asc,wilson2008sound}, although many possibilities for further improvements remain.

\section*{Acknowledgements}

Many of the authors' colleagues at CRREL contributed indirectly to this report by sharing their extensive knowledge of the structure of ice and geologic formations. In particular, the authors benefited from insights of Daniel Lawson and David Finnegan on the volcanic processes underlying the structure of the Amboy Crater site, which helped to shape some of the ideas in this report. Author K. Wilson is grateful to John Wyngaard (Pennsylvania State University) for introducing him to turbulent intermittency and related concepts, and to George Goedecke (New Mexico State University) and Vladimir Ostashev (Cooperative Institute for Research in Environmental Sciences), for many previous collaborations on quasi-wavelets which provided a foundation for this effort. Permission to publish was granted by Director, Cold Regions Research and Engineering Laboratory.

\appendix
\section{Appendix: QW Modeling Software Package}

Software was written in \textsc{Matlab} to conveniently implement the quasi-wavelet models described in this report. The software was used to create the example images. This appendix describes the overall software design and its usage.

The design uses object-oriented programming principles. Three types of objects are defined:
\begin{enumerate}
\item \emph{QWModel}: This object contains the parameters defining the QW model, such as the parent function, ratios between adjacent size classes, and values of the initial generation.
\item \emph{QWEnsemble}: This object contains an array of QWs, including their positions, amplitudes, generations, and birth times. The ensemble is generated from a QW model.
\item \emph{QWField}: This object contains a realization (calculated field) for a particular time and region of space, as calculated from a QW ensemble.
\end{enumerate}

The initial step in using the software is to define the parameters for the QW model. This is done by creating an instance of (i.e., calling the constructor method of) QWModel. Next, an ensemble is created by calling one of several methods available in QWModel, as will be described shortly. Thereafter, realizations of the ensemble are created using the QWField constructor method. These realizations can then be viewed using the visualizeField and visualizeCut methods, which are for continuously varying and level-cut fields, respectively.

\subsection{The QWModel Class}
The QWModel class defines the parent function for the QWs, as well as the placement function, which governs where offspring QWs are placed relative to the center of parent. (The placement function is not used in the creation of static ensembles.) 

The constructor for QWModel, besides setting the two just-described functions, also sets the spatial limits of the domain in which the QWs are to be placed. These are each two-element arrays, representing the minimum and the maximum. The limits for the $z$-direction may also be specified with a one-element array, in which case the model is two-dimensional (rather than three-dimensional). The power-law exponents $\beta$ and $\lambda$ (defined by Eqs.~\ref{def:beta} and ~\ref{def:lambda}, respectively) can also be specified. (If these exponents are not specified explicitly, they default to $\beta=0$ and $\lambda=1/3$.) The constructor initializes the generational length scale ratio ($\ell$) to 0.5. An energy-conserving cascade is also assumed, which fixes the value of $\overline{M}$ according to Eq.~\ref{eq:Mgen}. The values of $\ell$ and $\overline{M}$ can be overriden after calling the constructor, however, by setting the fields ScaleRat and OffspringRatio to alternative values. Note that the many other ratios appearing in the notes are not adjustable, as they are constrained by various relationships described previously in this report. Finally, values for the first generation of the process (packing fraction $\phi_1$, length scale $a_1$, energy $E_1$, amplitude $q_1$, and lifetime $\tau_1$) can all be set by the constructor. If desired, many of these parameters need not be specified, or an empty matrix can be passed, in which case the value is set to a reasonable default.

A couple other QWModel parameters can also be set by the user, although not through the constructor. These are the number of generations in the process (default 5), and the size of the spatial buffer around the domain (default 3).  For example, after instantiating a QWModel object, obj, the number of generations could be set to 6 by setting obj.NumGen=6. The buffer is specified in units of QW radii. For example, if the radius of the QW is $0.4$, the buffer will be $1.2$ in spatial units. The purpose of the buffer is to mitigate edge effects.

QWModel contains a number of methods to retrieve the generational properties of the QWModel. For example, getScale(n) and getAmplitude(n) returns the length scale and amplitude, respectively, associated with generation $n$. The various model ratios can also be retrieved.

QWModel includes several methods for generating new ensembles based on the model parameters. The available methods are:
\begin{itemize}
\item \emph{getStaticEnsemble()}: This method generates a static QW ensemble, i.e., one that does not vary in time. The ensemble is created in such a manner that the expected values of the packing fractions all equal their equilibrium (steady-state) values. The QWs are assumed to be uniformly dispersed in space, i.e., fully disorganized. The static ensemble should be viewed at $t=0$. Note that the settings of the placement function and $\overline{M}$ do not affect the static realizations.

\item \emph{getOrganizedStaticEnsemble()}: This method is the same as getStaticEnsemble, except that the positions of the QWs are organized using a random fractal support. This imposes intermittency properties on the realization.

\item \emph{getSteadyStateEnsemble()}: This method generates a steady-state QW ensemble, i.e., one that began at some time in the past and has now obtained an equilibrium. This is very similar to the static ensemble, except that the steady-state ensemble incorporates a cascade process which organizes the positions of offspring relative to their parent. On average, the packing fractions of the steady-state ensemble equal the static ensemble. The steady-state ensemble should be viewed at $t=0$.

\item \emph{getNonSteadyEnsemble(tlims, stopTime)}: This method generates an ensemble which is initialized during the time interval tlims (a two-element array, specifying the beginning and end of the time interval). The initialization consists of random generation of the first-generation (largest) QWs through a Poisson process. The rate of the Poisson process is such that if the length of the time interval, tlims(2)-tlims(1), is equal to the lifetime of the first generation, the equilibrium packing fraction will be reached at tlims(2). The cascade process is simulated up until stopTime. The non-steady ensemble can be viewed at any desired time.
\end{itemize}

\subsection{The QWEnsemble Class}
The QWEnsemble class defines a collection of QWs of multiple generations. The ensemble does not necessarily represent a particular time or region in space. Internally, the ensemble is represented by the positions, signed amplitudes, generation, birth time, and orientation of the QW.  When this information is combined with the QWModel, all properties of the ensemble are known. For example, the expiration time for the QW is known by adding its lifetime, as calculated from QWModel and its generation, to the birth time.

Generally, users have no need to invoke the constructor method of QWEnsemble. Rather, ensembles are created using one of the three methods in the QWModel class described above. The QWEnsemble class contains convenient methods for retrieving the numbers of QWs and empirical packing fractions associated with the ensemble. These quantities may be retrieved for the entire ensemble, or as a function of QW generation, time, and region in space.

A key set of capabilities provided by QWEnsemble consists of methods for filtering. The filtering can be performed by generation, size, time, and spatial position. Generally, limits can be specified (i.e., extract all members of the ensemble between the limits), or at a particular scalar value (i.e., extract generation n or all the QWs existing at time t).

\subsection{The QWField Class}
The main function of QWField is to use the positions, sizes, and parent function, as extracted from the QW ensemble and associated model, to calculate the value of the field at each point in space and time. This realization is calculated by the constructor method of QWField, which takes as input the QW model, ensemble, desired time, and limits on the spatial region. If the spatial limits are omitted, they are set to the default values from the QW model. Depending on the size of the ensemble, and the complexity of the parent function, the realization may be time consuming to calculate. 

Once the realization has been calculated (i.e., the QWField object has been instantiated), the getField and getCut methods can be recalled to retrieve it. These are intended for continuous and two-phase media, respectively. The methods visualizeField and visualizeCut methods produce corresponding graphical images. Alternative version of all four these methods, with the names appended by ``Ext'', retrieve the realization including the buffer region around the edges. 

\subsection{Dispersion Models}
Three dispersion models, which control the positions of the offspring relative to the parent, as described in Sec.~\ref{sec:placement}, are currently available:

\begin{itemize}
\item \emph{UniformDisp} randomly distributes the offspring over the entire QW model domain. When UniformDisp is used in the generation of a steady-state model, the result is statistically equivalent to a static model. 
\item \emph{NormalDisp} places the offspring using a normal (Gaussian) distribution centered on the position of the parent. Hence the probability peaks at the location of the parent, and decays with increasing distance.
\item \emph{Chi2Disp} places the offspring using a fourth degree $\chi^2$ distribution centered on the position of the parent. The probability is zero at the location of the parent, increases to a peak around the edge of the parent, and then decays with increasing distance.
\end{itemize}

\bibliographystyle{apalike}
\bibliography{QWmodel}

\end{document}